\documentclass[sigconf,10pt,nonacm]{acmart}
\usepackage{graphicx}
\usepackage{balance}  
\usepackage{amsmath,amsfonts,amsthm}
\usepackage{xcolor}
\usepackage{paralist}
\usepackage{weiwMath}
\usepackage{lipsum}
\usepackage{multirow}
\usepackage[scriptsize,tight]{subfigure}
\usepackage{url}
\usepackage{footnote}

\usepackage[all=normal, floats, bibnotes, wordspacing, charwidths, indent, lists]{savetrees}

\makeatletter%
\skip\footins 8.1pt plus 4pt minus 2pt 
\newcommand{\mycomment}[1]{}

\newcommand{\vae}{\textsf{VAE}\xspace}
\newcommand{\gph}{\textsf{GPH}\xspace}

\newcommand{\repr}{$\Gamma$\xspace}

\newcommand{\pubchem}{\textsf{HM-PubChem}\xspace}

\newcommand{\gloveone}{$\textsf{EU-Glove}_{50}$\xspace}
\newcommand{\glovetwo}{$\textsf{EU-Glove}_{300}$\xspace}

\newcommand{\imagenet}{\textsf{HM-ImageNet}\xspace}

\newcommand{\aminer}{\textsf{ED-AMiner}\xspace}

\newcommand{\youtube}{\textsf{HM-Youtube}\xspace}
\newcommand{\video}{\textsf{HM-UQVideo}\xspace}
\newcommand{\fasttext}{\textsf{HM-fastText}\xspace}

\newcommand{\mnist}{\textsf{HM-EMNIST}\xspace}
\newcommand{\gistlarge}{$\textsf{HM-GIST}_{2048}$\xspace}

\newcommand{\wikijacc}{\textsf{JC-Wikipedia}\xspace}

\newcommand{\bmsjacc}{\textsf{JC-BMS}\xspace}

\newcommand{\dblped}{\textsf{ED-DBLP}\xspace}

\newcommand{\dblpjacclong}{$\textsf{JC-DBLP}_{q3}$\xspace}

\newcommand{\mse}{\textsf{MSE}\xspace}
\newcommand{\mape}{\textsf{MAPE}\xspace}
\newcommand{\msle}{\textsf{MSLE}\xspace}

\newcommand{\hist}{\textsf{Histogram}\xspace}
\newcommand{\mean}{\textsf{Mean}\xspace}
\newcommand{\exact}{\textsf{Exact}\xspace}

\newcommand{\xgbexp}{\textsf{TL-XGB}\xspace}
\newcommand{\lightgbmexp}{\textsf{TL-LGBM}\xspace}
\newcommand{\hierdnnexp}{\textsf{DL-RMI}\xspace}
\newcommand{\dlnexp}{\textsf{DL-DLN}\xspace}
\newcommand{\moeexp}{\textsf{DL-MoE}\xspace}

\newcommand{\usexp}{\textsf{DB-US}\xspace}
\newcommand{\spestexp}{\textsf{DB-SE}\xspace}
\newcommand{\kdeexp}{\textsf{TL-KDE}\xspace}

\newcommand{\dnnexp}{\textsf{DL-DNN}\xspace}
\newcommand{\dnnsexp}{$\textsf{DL-DNNs}_{\tau}$\xspace}
\newcommand{\lstmexp}{\textsf{DL-BiLSTM}\xspace}
\newcommand{\lstmaexp}{\textsf{DL-BiLSTM-A}\xspace}

\newcommand{\modelone}{\textsf{CardNet}\xspace}
\newcommand{\modeltwo}{\textsf{CardNet-A}\xspace}
\newcommand{\modelpart}{\textsf{CardNet\{-A\}}_{\text{-C}}\xspace}
\newcommand{\modelcom}[1]{\textsf{CardNet\{-A\}}_{-#1}\xspace}

\newcommand{\single}{single uniform sample\xspace}
\newcommand{\multiple}{multiple uniform samples\xspace}
\newcommand{\skewed}{single skewed sample\xspace}

\newtheorem{lemma}{Lemma}
\newtheorem{example}{Example}

\newtheorem{problem}{Problem}

\usepackage{mathtools}

\DeclarePairedDelimiter\floor{\lfloor}{\rfloor}

\newcommand{\confversion}[1]{}
\newcommand{\fullversion}[1]{#1}
\newcommand{\reviseR}[1]{#1}
\newcommand{\eat}[1]{}
\newcommand{\revise}[1]{#1}

\setcopyright{rightsretained}
\confversion{
\copyrightyear{2020}
\acmYear{2020}
\setcopyright{acmlicensed}
\acmConference[SIGMOD'20]{Proceedings of the 2020 ACM SIGMOD International Conference on Management of Data}{June 14--19, 2020}{Portland, OR, USA}
\acmBooktitle{Proceedings of the 2020 ACM SIGMOD International Conference on Management of Data (SIGMOD'20), June 14--19, 2020, Portland, OR, USA}
\acmPrice{15.00}
\acmDOI{10.1145/3318464.3380570}
\acmISBN{978-1-4503-6735-6/20/06}
}
\fullversion{
\acmConference[]{}{}{}
\acmYear{}
\copyrightyear{}
}

\begin{document}

\fancyhead{}
\pagenumbering{arabic}
\pagestyle{plain}


\title{Monotonic Cardinality Estimation of Similarity Selection: A Deep Learning Approach}

\settopmatter{authorsperrow=4}

\author{Yaoshu Wang}
\affiliation{%
  \institution{Shenzhen Institute of Computing Sciences, Shenzhen University}
}
\email{yaoshuw@sics.ac.cn}
\author{Chuan Xiao} 
\affiliation{%
  \institution{Osaka University\\\& Nagoya University}
}
\email{chuanx@ist.osaka-u.ac.jp}
\author{Jianbin Qin} 
\affiliation{%
  \institution{Shenzhen Institute of Computing Sciences, Shenzhen University}
}
\authornote{Corresponding author.}
\email{jqin@sics.ac.cn}
\author{Xin Cao}
\affiliation{%
  \institution{The University of New South Wales}
}
\email{xin.cao@unsw.edu.au}
\author{Yifang Sun}
\affiliation{%
  \institution{The University of New South Wales}
}
\email{yifangs@cse.unsw.edu.au}
\author{Wei Wang}
\affiliation{%
  \institution{The University of New South Wales}
}
\email{weiw@cse.unsw.edu.au}
\author{Makoto Onizuka}
\affiliation{%
  \institution{Osaka University}
}
\email{onizuka@ist.osaka-u.ac.jp}


\begin{abstract}
Due to the outstanding capability of capturing underlying data distributions, 
deep learning techniques have been recently utilized for a series of traditional 
database problems. 
In this paper, we investigate the possibilities of utilizing 
deep learning for cardinality estimation of similarity selection. 
Answering this problem accurately and efficiently is essential to many data 
management applications, especially for query optimization. 
Moreover, in some applications the estimated cardinality is supposed to be 
consistent and interpretable. Hence a monotonic estimation w.r.t. the query 
threshold is preferred. 
We propose a novel 
and generic method that can be applied to any data type and distance function. 
Our method consists of a feature extraction model and a regression model. The 
feature extraction model transforms original data and 
threshold to a Hamming space, in which a deep learning-based regression model 
is utilized to exploit the incremental property of cardinality w.r.t. the threshold 
for both accuracy and monotonicity. We develop a training strategy tailored 
to our model as well as techniques for fast estimation. We also discuss how to handle 
updates. We demonstrate the accuracy and the efficiency of our method through 
experiments, and show how it improves the performance of a query optimizer. 
\end{abstract}



\maketitle

\section{Introduction} \label{sec:intro}
Deep learning has been recently utilized to deal with traditional database problems, 
such as indexing~\cite{kraska2018case}, query 
execution~\cite{kraska2019sagedb,DBLP:conf/sigmod/DingDM0CN19, treedeepmodel, surveycarddeeplearning}, 
and database 
tuning~\cite{DBLP:conf/sigmod/ZhangLZLXCXWCLR19}. Compared to traditional 
database methods and non-deep-learning models (logistic regression, random 
forest, etc.), deep learning exhibits outstanding capability of reflecting 
the underlying patterns and correlations of data as well as exceptions and 
outliers that capture the extreme anomalies of data instances~\cite{kraska2019sagedb}. 

In this paper, we explore in the direction of applying deep learning techniques 
for a data management problem -- cardinality estimation of similarity selection, 
i.e., given a set of records $\mathcal{D}$, a query record $x$, a 
distance function and a threshold $\theta$, to estimate the number of records 
in $\mathcal{D}$ whose distances to $x$ are no greater than $\theta$. It is 
an essential procedure in many data management tasks,  
such as search and retrieval, 
data integration, data exploration, and query optimization. 
\revise{For example: 
\begin{inparaenum} [(1)]
  \item In image retrieval, images are converted to binary vectors (e.g., by 
  a HashNet~\cite{cao2017hashnet}), and then the 
  vectors whose Hamming distances to the query are within 
  a threshold of 16 are identified as candidates~\cite{DBLP:conf/mm/ZhangGZL11} 
  for further image-level verification (e.g, by a CNN). Since the image-level 
  verification is costly, estimating the cardinalities of similarity selection 
  yields the number of candidates, and thus helps estimate the overall running 
  time in an end-to-end system to create a service level agreement.
  \item In query optimization, estimating cardinalities for similarity selection 
  benefits the computation of operation costs and the choice of execution orders 
  of query plans that involve multiple similarity predicates; e.g., hands-off 
  entity matching systems~\cite{DBLP:conf/sigmod/GokhaleDDNRSZ14,DBLP:conf/sigmod/DasCDNKDARP17} 
  extract paths from random forests and take each path (a conjunction of 
  similarity predicates over multiple attributes) as a blocking rule. Such query 
  was also studied in \cite{DBLP:conf/sigmod/LiHDL15} for sets and strings. 
\end{inparaenum}
}

\begin{figure} [t]
  \centering
  \subfigure[Cardinality v.s. threshold.]{
    \includegraphics[width=0.46\linewidth]{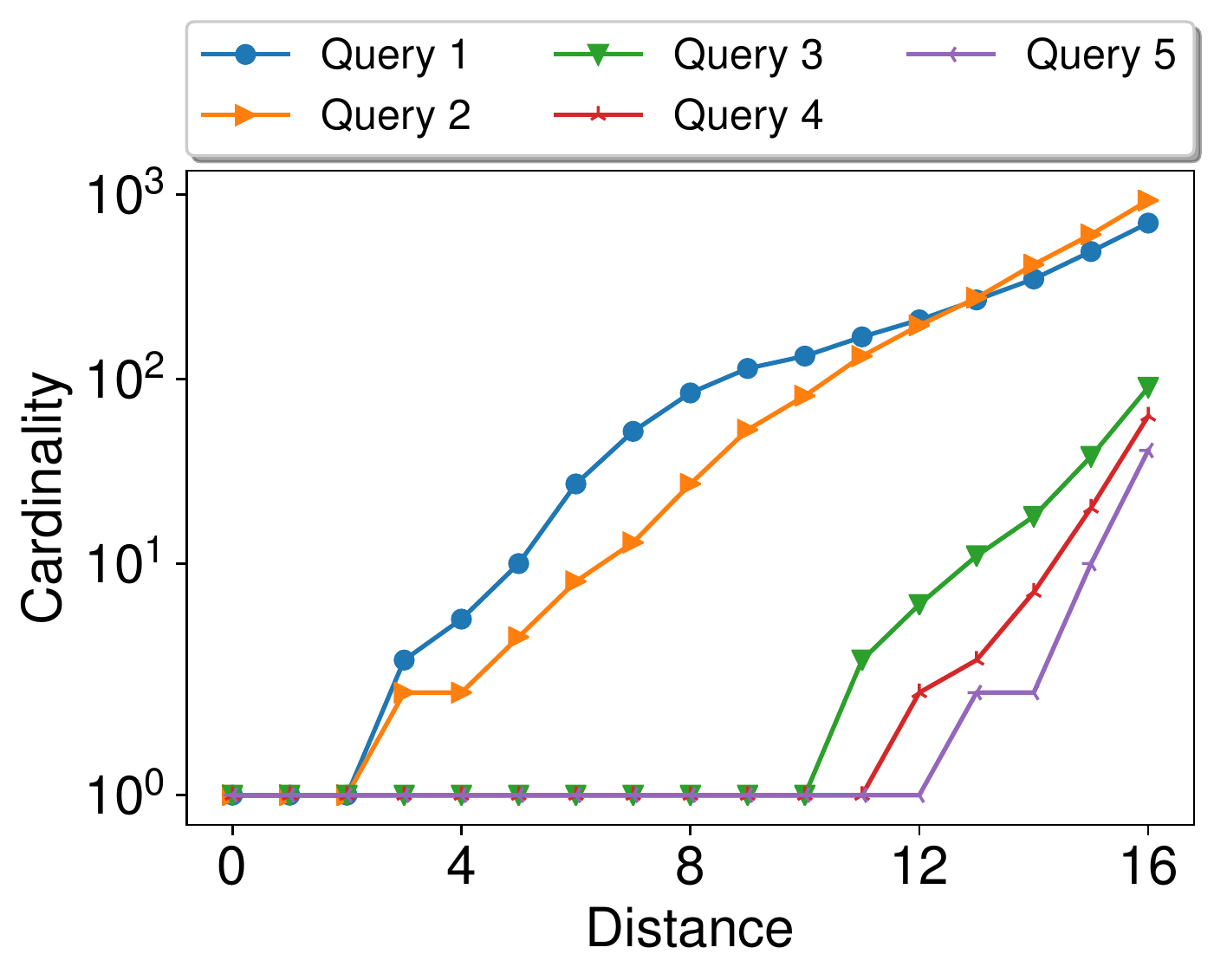}
    \label{fig:card-dist}
  }
  \subfigure[Percentage of queries v.s. cardinality.]{
    \includegraphics[width=0.46\linewidth]{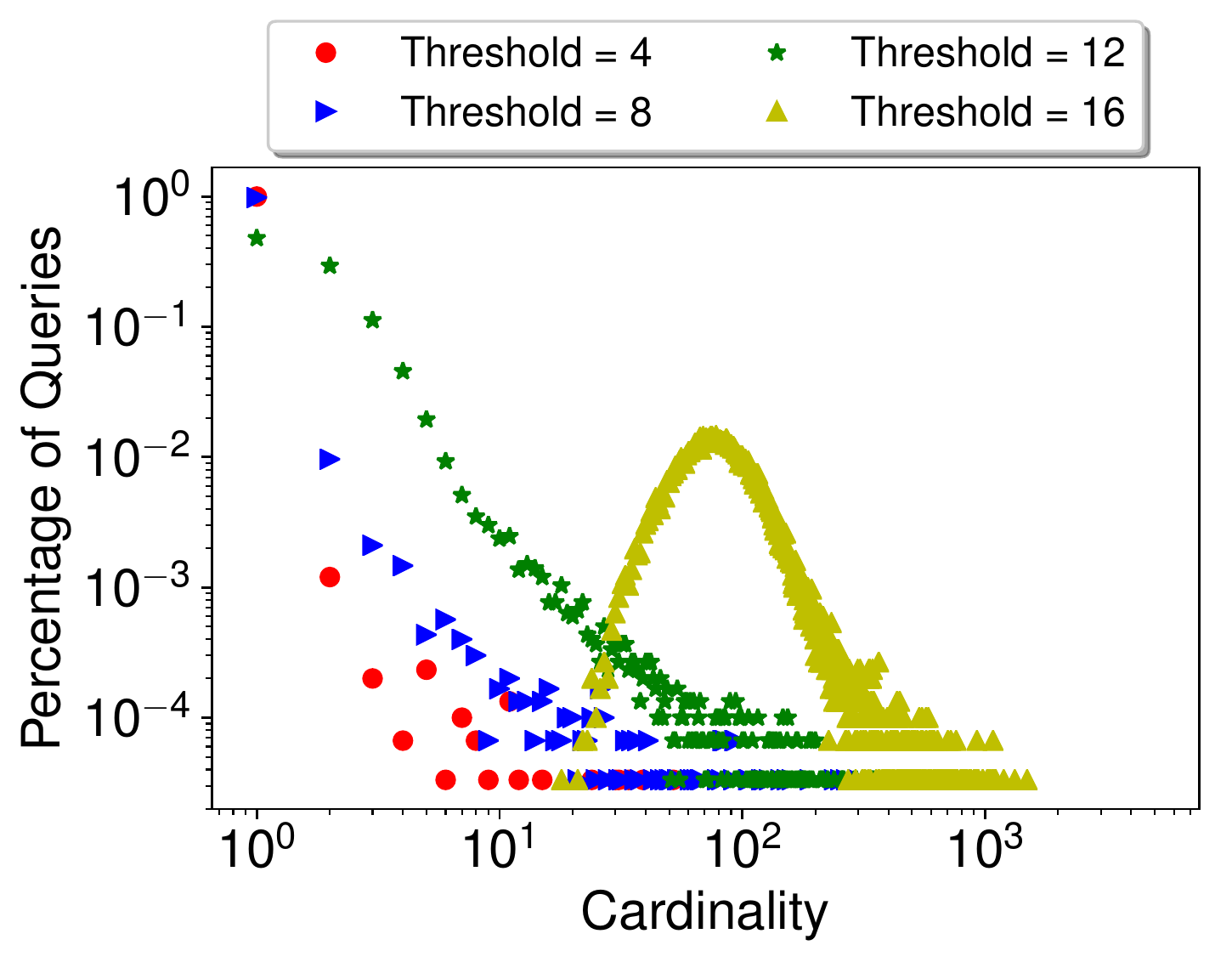}
    \label{fig:pct-card}
  }
  \caption{Cardinality distribution on ImageNet.}
\end{figure}

The reason why deep learning approaches may outperform other options for cardinality 
estimation of similarity selection can be seen from the following example: 
\begin{inparaenum}
  \item Figure~\ref{fig:card-dist} shows the cardinalities of five randomly chosen 
  queries on the ImageNet dataset~\cite{URL:imagenet} by varying Hamming distance 
  threshold. The cardinalities keep unchanged at some thresholds but surge at others. 
  \item Figure~\ref{fig:pct-card} shows the percentage of queries (out of 30,000) for 
  each cardinality value, under Hamming distance thresholds 4, 8, 12, and 16. The 
  cardinalities are small or moderate for most queries, yet exceptionally large for 
  long-tail queries (on the right side of the figure). 
\end{inparaenum}
Both facts cause considerable difficulties for traditional database methods which require 
large samples to achieve good accuracy and traditional learning methods which 
are incapable to learn such complex underlying distributions. In contrast, deep learning 
is a good candidate to capture such data patterns and generalizes well on queries that are 
not covered by training data, thereby delivering better accuracy. Another reason for 
choosing deep learning is that the training data -- though large training sets are usually 
needed for deep learning -- are easily acquired by running similarity selection algorithms 
(without producing label noise when exact algorithms are used). 

In addition to \textbf{accuracy}, there are several other technical issues for cardinality 
estimation of similarity selection: 
\begin{inparaenum} [(1)]
  \item A good estimation is supposed to be \textbf{fast}. 
  \item A \textbf{generic} method that applies to a variety of data types and 
  distance functions is preferred. 
  \item Users may want the estimated 
  cardinality to be consistent and interpretable in applications like data exploration. Since the actual cardinality is 
  \textbf{monotonically} increasing with the threshold, when a greater threshold 
  is given, a larger or equal number of results is preferable, so the user is able 
  to interpret the cardinality for better analysis. 
\end{inparaenum}

To cope with these technical issues, we propose a novel method that 
separates data modelling and cardinality estimation into two components: 
\begin{itemize}
  \item A \emph{feature extraction} component transforms original data and thresholds to a 
  Hamming space such that the semantics of the input distance function is exactly or 
  approximately captured by 
  Hamming distance. As such, our method becomes \textbf{generic} and applies to any data 
  type and distance. 
  \item A \emph{regression} component models the estimation as a regression problem and 
  estimates the cardinality on the transformed vectors and threshold using deep learning. 
\end{itemize}



\revise{To achieve good accuracy of regression, rather than feeding a deep neural network 
with training data in a straightforward manner, we devise a novel approach based on 
\emph{incremental prediction} to exploit the incremental property of cardinality;} i.e., 
when the threshold is increasing, the increment of cardinality is only caused by the records 
in the increased range of distance. 
Since our feature extraction maps original distances to 
discrete distance values, we can use multiple regressors, each dealing with one distance 
value, and then sum up the individual results to get the total cardinality. In 
doing so, we are able to learn the cardinality distribution for each distance value, so 
the overall estimation becomes more \textbf{accurate}. Another benefit of incremental 
prediction is that it guarantees the \textbf{monotonicity} w.r.t. the threshold, and thus 
yields more interpretability of the estimated results. To estimate the cardinality of each 
distance value, we utilize an \emph{encoder-decoder} model through careful neural network 
design: 
\begin{inparaenum} [(1)]
  \item To cope with the sparsity in Hamming space, as output by the feature extraction, 
  we employ a variational auto-encoder to \revise{embed the binary vector in Hamming space} 
  to a dense representation. 
  \item To generalize for queries and thresholds not covered by the training data, we also 
  \revise{embed (Hamming) distance values}. The distance embeddings are concatenated to the 
  binary vector and its dense representation, and then fed to a neural network to produce 
  final embeddings. The decoders takes the final embeddings as input and outputs the 
  estimated cardinality. 
\end{inparaenum}

\revise{We design a loss function and a dynamic training strategy, both 
tailored to our incremental prediction model. The loss function adds more loss to the 
distance values that tend to cause more estimation error. The impact of such loss is 
dynamically adjusted through training to improve the accuracy and the generalizability. 
For \textbf{fast} online estimation, optimizations are developed on 
top of our regression model by reducing multiple encoders to one.} As we are applying 
machine learning on a traditional database problem, an important 
issue is whether the solution works when \textbf{update} exists. For this reason, we 
discuss incremental learning to handle updates in the dataset. 

\revise{Extensive experiments were carried out on four common distance functions using 
real datasets. 
\fullversion{We took a uniform sample of records from each dataset as a query workload 
for training, validation, and testing, and computed labels by running exact similarity 
selection algorithms.}
} The takeaways are: 
\begin{inparaenum} [(1)]
  \item The proposed deep learning method is more accurate than existing methods 
  while also running faster with a moderate model size. 
  \item Incremental prediction guarantees monotonicity and at the same time achieves 
  high accuracy, substantially outperforming the method that simply feeding a deep 
  neural network with training data. 
  \item The components in our model are all useful to improve accuracy and speed. 
  \item Incremental learning is fast and effective against updates. 
  \item Our method delivers excellent performance on long-tail queries having 
  exceptionally large cardinalities and generalizes well on out-of-dataset queries 
  that significantly differ from the dataset. 
  \item A case study shows that query processing performance is improved by 
  integrating our method into a query optimizer. 
\end{inparaenum}

Our contributions are summarized as follows. 
\begin{itemize}
  \item We develop a deep learning method for cardinality estimation of 
  similarity selection (Section~\ref{sec:framework}). Our method guarantees 
  the monotonicity of cardinality w.r.t. the threshold. 
  \item Through feature extraction (Section~\ref{sec:case-study}) and regression 
  (Section~\ref{sec:model}), our method is generic to any data type and distance 
  function, and exploits the incremental property of cardinality to achieve accuracy 
  and monotonicity. The training 
  techniques that favor our method are developed (Section~\ref{sec:train}). 
  \item We accelerate our model for online estimation (Section~\ref{sec:accelerate}) 
  and propose incremental learning for updates (Section~\ref{sec:update}). 
  \item We conduct extensive experiments to demonstrate the superiority and the 
  generalizability of our method, as well as how it works in a query optimizer 
  (Section~\ref{sec:exp}). 
\end{itemize}


\confversion{Due to the page limitation, we provide lemma proofs, model complexity analysis, 
details of experiment setup, and additional experiments in the extended version of 
the paper~\cite{wang2020monotonic}.}

\section{Preliminaries}
\label{sec:pre}

\subsection{Problem Definition and Notations}
Let $\mathcal{O}$ be a universe of records. $x$ and $y$ are two records in $\mathcal{O}$. 
$f: \mathcal{O} \times \mathcal{O} \to \mathbb{R}$ is a 
function which evaluates the distance (similarity) of a pair of records. 
Common distance (similarity) functions include Hamming distance, Jaccard similarity, 
edit distance, Euclidean distance, etc. Without loss of generality, we 
assume $f$ is distance function. Given a collection of records $\mathcal{D} \subseteq \mathcal{O}$, 
a query record $x \in \mathcal{O}$, and a threshold $\theta$, a similarity selection 
is to find all the records $y \in \mathcal{D}$ such that $f(x, y) \leq \theta$. 
We formally define our problem.
\begin{problem} [Cardinality Estimation of Similarity Selection]
  Given a collection $\mathcal{D}$ of records, a query record $x \in \mathcal{O}$, a 
  distance function $f$, and a threshold $\theta \in [0, \theta_{\max}]$, our task 
  is to estimate the number of records that satisfy 
  the similarity constraint, i.e., $\size{\set{ y \mid f(x, y) \leq \theta, y \in 
  \mathcal{D}}}$. 
\end{problem}
$\theta_{\max}$ is the maximum threshold (reasonably large for 
similarity selection to make sense) we are going to support. 
A good estimation is supposed to be close to the actual cardinality. Mean squared 
error (\textsf{MSE}) and mean absolute percentage error (\textsf{MAPE}) are two 
widely used evaluation metrics in the cardinality 
estimation problem~\cite{DBLP:conf/sigmod/DasguptaJJZD10,wu2018local,DBLP:conf/edbt/MattigFBS18,DBLP:conf/sigmod/HeimelKM15,liu2015cardinality}. 
Given $n$ similarity selection queries, let $c_i$ denote the actual cardinality of 
the $i$-th selection and $\widehat{c}_i$ denote the estimated cardinality. 
\textsf{MSE} and \textsf{MAPE} are computed as
\begin{align*}
  \textsf{MSE} = \frac{1}{n}\sum_{i = 1}^{n} (c_i - \widehat{c}_i)^2, \quad
  \textsf{MAPE} = \frac{1}{n}\sum_{i = 1}^{n} \left|{\frac{c_i-\widehat{c}_i}{c_i}}\right|. \\
\end{align*}
Smaller errors are preferred. We adopt these two metrics to evaluate the estimation 
accuracy. We focus on evaluating the cardinality estimation as stand-alone (in contrast 
to in an RDBMS) and 
only consider in-memory implementations. 

Table~\ref{tab:notation} lists the notations frequently used in this paper.
We use bold uppercase letters (e.g., $\mathbf{A}$) for matrices; bold lowercase 
letters (e.g., $\mathbf{a}$) for vectors; and non-bold lowercase letters (e.g., 
$a$) for scalars and other variables. Uppercase Greek symbols (e.g., $\Phi$) 
are used to denote neural networks. $\mathbf{A}[i, *]$ 
and $\mathbf{A}[*, i]$ denote the $i$-th row and the $i$-th column of 
$\mathbf{A}$, respectively. $\mathbf{a}[i]$ denotes the $i$-th dimension of 
$\mathbf{a}$. Semicolon represents the concatenation of vectors; e.g., given 
an $a$-dimensional vector $\mathbf{a}$ and a $b$-dimensional vector $\mathbf{b}$, 
$\mathbf{c} = [\mathbf{a}; \mathbf{b}]$ means that $\mathbf{c}[1 \twoldots a] = \mathbf{a}$ 
and $\mathbf{c}[a + 1 \twoldots a + b] = \mathbf{b}$. Colon represents the 
construction of a matrix by column vectors or matrices; e.g., $\mathbf{C} = 
[\mathbf{a}: \mathbf{b}]$ means that $\mathbf{C}[*, 1]  = \mathbf{a}$ and 
$\mathbf{C}[*, 2]  = \mathbf{b}$. 

\begin{table} [t]
  \caption{Frequently used notations.}
  \small
  \label{tab:notation}
  \centering
  \begin{tabular}[b]{| l | l |}
    \hline%
    Symbol & Definition \\
    \hline%
    $\mathcal{O}$, $\mathcal{D}$ & a record universe, a dataset \\
    \hline%
    $x, y$ & records in $\mathcal{O}$ \\
    \hline%
    $f$ & a distance function \\
    \hline%
    $\theta$, $\theta_{\max}$ & a distance threshold and its maximum value\\    
    \hline%
    $c$, $\widehat{c}$ & cardinality and the estimated value \\
    \hline%
    $g$, $h$ & regression function and feature extraction function \\
    \hline%
    $\mathbf{x}$, $d$ & the binary representation of $x$ and its dimensionality\\
    \hline%
    $\tau$, $\tau_{\max}$ & a threshold in Hamming space and its maximum value \\
    \hline%
    $\mathbf{e}^i$ & the embedding of distance $i$ \\
    \hline%
    $\mathbf{z}_x^{i}$ & the embedding of $\mathbf{x}$ and distance $i$ \\
    \hline%
                                                                   
  \end{tabular}
\end{table}

\subsection{Related Work}

\subsubsection{Database Methods}
Auxiliary data structure is one of the main types of database methods 
for the cardinality estimation of similarity selection. 
For binary vectors, histograms~\cite{qin2018gph} can be constructed to 
count the number of records by partitioning dimensions into buckets and 
enumerating binary vectors and thresholds. 
For strings and sets, semi-lattice structures~\cite{lee2007extending,lee2009power} 
and inverted indexes~\cite{jin2008sepia,mazeika2007estimating} 
are utilized for estimation. 
The major drawback of auxiliary structure methods is that they only 
perform well on low dimensionality and small thresholds. Another type 
of database methods is based on sampling, e.g., uniform sampling, 
adaptive sampling~\cite{lipton1990query}, and sequential sampling~\cite{haas1992sequential}. 
State-of-the-art sampling strategies~\cite{wu2016sampling, DBLP:conf/sigmod/ZhaoC0HY18,DBLP:conf/icml/Jiang17a,leis2017cardinality} 
focus on join size estimation in query optimization, and are difficult 
to be adopted to our problem defined on distance functions.
Sampling methods are often combined with 
sketches~\cite{DBLP:conf/vldb/GionisIM99,lee2011similarity} to improve 
the performance. 
A state-of-the-art method was proposed in \cite{wu2018local} for 
high-dimensional data. 
In general, sampling methods need a large set of samples to achieve good 
accuracy, and thus become either slow or inaccurate when applied on large 
datasets. As for the cardinalities of SQL queries, recent studies proposed 
a tighter bound for intermediate join cardinalities~\cite{pesscardinality}
and adopted inaccurate cardinalities to generate optimal query plans~\cite{cardqueryopt}.

\subsubsection{Traditional Learning Models}
A prevalent traditional learning method for the cardinality estimation of similarity 
selection is to train a kernel-based estimator~\cite{DBLP:conf/sigmod/HeimelKM15,DBLP:conf/edbt/MattigFBS18} 
using a sample. Monotonicity is guaranteed when the sample is deterministic 
w.r.t. the query record (i.e., the sample does not change if only the threshold 
changes). 
Kernel methods require large number of instances 
for accurate estimation, hence resulting in low estimation speed. Moreover, 
they impose strong assumptions on kernel functions, e.g., only diagonal 
covariance matrix for Gaussian kernels. 
Other traditional learning models~\cite{ivanov2017adaptive}, such as support 
vector regression, logistic regression, and gradient boosting tree, are 
also adopted to solve the cardinality estimation problem. A query-driven 
approach~\cite{anagnostopoulos2017query} was proposed to learn several query 
prototypes (i.e., interpolation) that are differentiable. These approaches 
deliver comparable performance to kernel methods. 
A recent study~\cite{DBLP:journals/pvldb/DuttWNKNC19} explored the application 
of two-hidden-layer neural networks and tree-based ensembles (random forest and 
gradient boosted trees) on cardinality estimation of multi-dimensional range 
queries. It targets numerical attributes but does not 
apply to similarity selections. 

\subsubsection{Deep Learning Models}
Deep learning models are recently adopted to learn the best join order~\cite{krishnan2018learning,marcus2018deep,neoqueryoptimizer} or 
estimate the join size~\cite{kipf2018learned}. Deep reinforcement 
learning is also explored to generate query plans~\cite{ortiz2018learning}. 
Recent studies also adopted local-oriented approach~\cite{cardlocaldeepmodel}, 
tree or plan based learning model~\cite{treedeepmodel,plandeeppredict,neoqueryoptimizer}, 
recurrent neural network~\cite{surveycarddeeplearning}, and autoregressive 
model~\cite{autogressivemultiattr,autoregressivelikelihood}. 
The method that can be adapted to solve our problem is the mixture of expert 
model~\cite{shazeer2017outrageously}. It utilizes a 
sparsely-gated mixture-of-experts layer and assigns good experts (models) to 
these inputs. 
Based on this model, the recursive-model index~\cite{kraska2018case} was 
designed to replace the traditional B-tree index for range queries. The two 
models deliver good accuracy but do not guarantee monotonicity.

For monotonic methods, an early attempt used a min-max 4-layer neural network for 
regression~\cite{daniels2010monotone}. Lattice 
regression~\cite{garcia2009lattice,gupta2016monotonic,fard2016fast,you2017deep} 
is recent monotonic method. It adopts a lattice structure to construct all 
combinations of monotonic interpolation values. To handle high-dimensional 
monotonic features, ensemble of lattices~\cite{fard2016fast} splits lattices 
into several small pieces using ensemble learning. To improve the performance
of regression, deep lattice network (DLN) was proposed~\cite{you2017deep}. 
It consists of multiple calibration layers and an ensemble of lattices layers. 
However, lattice regression does not directly target 
our problem, and our experiments show that DLN is rather inaccurate. 
\section{Cardinality Estimation Framework}
\label{sec:framework}


\subsection{Basic Idea}
Let $c(x, \theta)$ denote the cardinality of a query $x$ with threshold $\theta$. 
We model the estimation as a regression problem with a unique framework designed to
alleviate the challenges mentioned in Section~\ref{sec:intro}.
We would like to find a function $\widehat{c}$ within a function family, such that
$\widehat{c}$ returns an approximate value of $c$ for any input $x$, i.e., 
$\widehat{c}(x, \theta) \approx c(x, \theta), \forall x \in \mathcal{O}$ and $\theta 
\in [0, \theta_{\max}]$.
We consider $\widehat{c}$ that belongs to the following family: 
$\widehat{c} \definedas g \concat h$, 
where $h(x, \theta) = (\mathbf{x}, \tau)$, $g(\mathbf{x}, \tau) \in \mathbb{Z}_{\geq 0}$, 
$\mathbf{x} \in \set{0, 1}^d$, and $\tau \in \mathbb{Z}_{\geq 0}$.
Intuitively, we can deem $h$ as a \emph{feature extraction} function, which maps
an object $x$ and a threshold $\theta$ to a fixed-dimensional binary vector $\mathbf{x}$
and an integer threshold $\tau$. Then, the function $g$ essentially performs the
\emph{regression} using the transformed input, i.e., the $(\mathbf{x}, \tau)$ pair. 
The rationales of such design are analyzed as follows: 
\begin{itemize}
  \item This design separates data modelling and cardinality estimation into two
  functions, $h$ and $g$, respectively. On one hand, this allows the system to cater for
  different data types, and distance functions. On the other hand, it
  allows us to choose the best models for the
  estimation problem. To decouple the two components, some common interface needs
  to be established. 
  We pose the constraints that
  \begin{inparaenum}[(1)]
    \item $\mathbf{x}$ belonging to a Hamming space, and 
    \item $\tau$ is an non-negative integer. 
  \end{inparaenum}
  \fullversion{For (1), many DB applications deal with discrete objects (e.g., sets and strings) 
  or discrete object representations (e.g., binary codes from learned hash functions). 
  For (2), since there is a finite number of thresholds that make a difference in cardinality, 
  in theory we can always map them to integers, albeit a large number. Here, we take the 
  approximation to limit it to a fixed number. Other learning models also make similar 
  modelling choice (e.g., most regularizations adopt the smoothness bias in the function 
  space).}
  We will leave other interface design choices to future work due to their added
  complexity. For instance, it is entirely possible to restrict $\mathbf{x}$ to
  $\mathbb{R}^d$ and $\tau \in \mathbb{R}$. While the modelling power of the framework 
  gets increased, this will inevitably result in more complex models that are 
  potentially difficult to train. 
  \item The design can be deemed as an instance of the encoder-decoder 
  model, where two functions $h$ and $g$ are used for some prediction tasks. As a 
  close analog, Google translation~\cite{johnson2017google} trains an $h$ that maps 
  inputs in the source language to a latent representation, and then train a $g$ 
  that maps the latent representation to the destination language. As such, it can 
  support translation between $n$ languages by training only $2n$ functions, instead 
  of $n(n-1)$ direct functions. 
\end{itemize}


By this model design, the function $\widehat{c} = g \circ h(x, \theta)$ is monotonic 
if it satisfies the condition in the following lemma. 
\begin{lemma} \label{lm:overall-monotonicity}
  Consider a function $h(x, \theta)$ monotonically increasing with $\theta$ and a 
  function $g(\mathbf{x}, \tau)$ monotonically increasing with $\tau$, our framework 
  $g \circ h(x, \theta)$ is monotonically increasing with $\theta$.
\end{lemma}

\subsection{Feature Extraction} \label{sec:feature-extraction}
The process of feature extraction is to transfer any data type and distance 
threshold into binary representation and integer threshold. 
Formally, we have a function $h_{rec}: \mathcal{O} \to \set{0, 1}^d$; 
i.e., given any record $x \in \mathcal{O}$, $h_{rec}$ maps $x$ to a $d$-dimensional 
binary vector, denoted by $\mathbf{x}$, called $x$'s binary representation. We can 
plug in any user-defined functions or neural networks for feature extraction. For 
the sake of estimation accuracy, the general criteria is that the Hamming distance 
of the target $d$-dimensional binary vectors can equivalently or 
approximately capture the semantics of the original distance function. We will show 
some example feature extraction methods and a series of case studies in 
Section~\ref{sec:case-study}. 

Besides the transformation to binary representations, 
we also have a monotonically increasing (as demanded by Lemma~\ref{lm:overall-monotonicity}) 
function $h_{thr}: [0, \theta_{\max}] 
\to [0, \tau_{\max}]$ to transform the threshold. $\tau_{\max}$ is a tunable 
parameter to control the model size (as introduced later, there are $(\tau + 1)$ decoders 
in our model, $\tau \leq \tau_{\max}$). Given a $\theta \in [0, \theta_{\max}]$, $h_{thr}$ 
maps $\theta$ to an integer between $0$ and $\tau_{\max}$, denoted by $\tau$. The 
purpose of threshold transformation is: for real-valued distance functions, it makes the 
distances countable; for integer-valued distance functions, it can reduce the threshold to a 
small number, hence to prevent the model growing too big when the input threshold is large. 
As such, we are able to use finite number of estimators to predict the cardinality for each 
distance value. 
The design of threshold transformation depends on how original data are transformed to 
binary representations. In general, a transformation with less skew leads to better 
performance. 
Using the threshold of the Hamming distance between binary representations is not 
necessary, but would be a preferable option. 
A few case studies will be given in Section~\ref{sec:case-study}. 

\subsection{Regression (in a Nutshell)} \label{sec:regression-brief}
Our method for the regression is based on the following observation: given 
a binary representation $\mathbf{x}$ and a threshold $\tau$~\footnote{Note that the 
threshold is $\tau$ not $\tau_{\max}$ here because $\theta$ is mapped to $\tau$.}, 
the cardinality can be divided into $(\tau + 1)$ parts, each 
representing the cardinality of a Hamming distance $i$, $i \in [0, \tau]$. 
This suggests that we can learn $(\tau + 1)$ functions $g_0(\mathbf{x}), 
\ldots, g_{\tau}(\mathbf{x})$, each $g_{i}(\mathbf{x})$ estimating the cardinality 
of the set of records whose Hamming distances to $\mathbf{x}$ are exactly $i$. 
So we have 
\begin{align} 
  \label{eq:model}
  g(\mathbf{x}, \tau) = \sum_{i = 0}^{\tau}{g_{i}(\mathbf{x})}. 
\end{align}

This design has the following advantages: 
\begin{itemize}
  \item As we have shown in Figure~\ref{fig:card-dist}, 
  the cardinalities for different distance values may differ significantly. 
  By using individual estimators, the distribution of each distance value can be 
  learned separately to achieve better overall accuracy. 
  \item Our method exploits the \emph{incremental property} of cardinality: when the 
  threshold increases from $i$ to $i + 1$, the increased cardinality is the 
  cardinality for distance $i + 1$. This incremental prediction can guarantee the 
  monotonicity of cardinality estimation: 
  \begin{lemma} \label{lm:model-monotonicity}
    $g(\mathbf{x}, \tau)$ is monotonically increasing with $\tau$, if each 
    $g_i(\mathbf{x})$, $i \in [0, \tau]$ is deterministic and non-negative. 
  \end{lemma}
  The lemma suggests that a deterministic and non-negative model satisfies the requirement 
  in Lemma~\ref{lm:overall-monotonicity}, hence leading to the overall monotonicity. 
  \item We are able to control the size of the model by setting the maximum number of 
  estimators. Thus, working with the feature extraction, the regression achieves fast 
  speed even if the original threshold is large. 
\end{itemize}

We employ a deep encoder-decoder model to process each regression $g_i$. The 
reasons for choosing deep models are: 
\begin{inparaenum} [(1)]
  \item Cardinalities may significantly differ across queries, as shown in 
  Figure~\ref{fig:pct-card}. 
  Deep models are able to learn a variety 
  of underlying distributions and deliver salient performance for general 
  regression tasks~\cite{shazeer2017outrageously,kraska2018case,you2017deep}. 
  \item Deep models generalize well on queries that are not covered by the 
  training data. 
  \item Although deep models usually need large training sets for good accuracy, 
  the training data here can be easily and efficiently acquired by running 
  state-of-the-art similarity selection algorithms (and producing no label noise 
  when exact algorithms are used). 
  \item 
  Deep 
  models can be accelerated by modern hardware (e.g., GPUs/TPUs) or software 
  (e.g., Tensorflow) that optimizes batch strategies or matrix 
  manipulation~\footnote{Despite such possibilities for acceleration, we only 
  use them for training but not inference (estimation) in our experiments 
  for the sake of fair comparison.}.   
\end{inparaenum}

We employ a deep learning model to process each regression $g_i$. 
By carefully choosing encoders and decoders, we can meet the requirement in 
Lemma~\ref{lm:model-monotonicity} to guarantee the monotonicity. The details will 
be given in Section~\ref{sec:model}. Before that, we show some options and case studies of 
feature extraction.

\section{Case Studies for Feature Extraction} \label{sec:case-study}

As stated in Section~\ref{sec:feature-extraction}, for good accuracy, a desirable 
feature extraction is that the Hamming distance between the binary vectors can 
capture the semantics of the original distance function. We discuss a few example 
options. 
\begin{itemize} 
  \item \textbf{Equivalency}: Some distances can be equivalently 
  expressed in a Hamming space, e.g., $L_1$ distance
  on integer values~\cite{DBLP:conf/vldb/GionisIM99}. 
  \item \textbf{LSH}: We use $d$ hash functions in the locality sensitive hashing (LSH) 
  family~\cite{DBLP:conf/vldb/GionisIM99}, each hashing a record to a bit.  
  $\mathbf{x}$ and $\mathbf{y}$ agree on a bit with high probability if $f(x, y) \leq \theta$, 
  thus yielding a small Hamming distance between $\mathbf{x}$ and $\mathbf{y}$. 
  \item \textbf{Bounding}: We may derive a necessary condition of the original 
  distance constraint; e.g., $f(x, y) \leq \theta \implies H(\mathbf{x}, \mathbf{y}) \leq \tau$, 
  where $H(\cdot, \cdot)$ denotes the Hamming distance. 
\end{itemize}
For the equivalency method, since the conversion to Hamming distance is lossless, 
it can be used atop of the other two. This is useful when the output of the hash function 
or the bound is not in a Hamming space. Note that our model is not limited to these 
options. Other feature extraction methods, such as embedding~\cite{DBLP:conf/kdd/Zhang017}, 
can be also used here. 

As for threshold transformation, we have two parameters: $\theta_{\max}$, the maximum threshold 
we are going to support, and $\tau_{\max}$, a tunable parameter to control the size of 
our model. 
Any threshold $\theta \in [0, \theta_{\max}]$ is monotonically mapped to an integer $\tau 
\in [0, \tau_{\max}]$. In our case studies, we consider using a transformation proportional to 
the (expected/bounded) Hamming distance between binary representations. Note that 
$\theta_{\max}$ is not necessarily mapped to $\tau_{\max}$, because for integer-valued 
distance functions, the number of available thresholds is smaller than $\tau_{\max} + 1$ when 
$\theta_{\max} < \tau_{\max}$, meaning that only $(\theta_{\max} + 1)$ decoders are 
useful. In this case, $\theta_{\max}$ is mapped to a value smaller than $\tau_{\max}$. 
Next we show four case studies of some common data types and distance functions.

\subsection{Hamming Distance}
We consider binary vector data and Hamming distance as the input distance function. 
The original data are directly fed to our regression model. 
Since the function is already Hamming distance, 
we use the original threshold $\theta$ as $\tau$, if $\theta_{\max} \leq \tau_{\max}$. 
Otherwise, we map $\theta_{\max}$ to $\tau_{\max}$, and other thresholds are mapped 
proportionally: $\tau = \floor{\tau_{\max} \cdot \theta / \theta_{\max}}$. 
\revise{Although multiple thresholds may map to the same $\tau$, we can increase the number 
of decoders to mitigate the imprecision.}

\subsection{Edit Distance}
The (Levenshtein) edit distance measures the minimum number of operations, including insertion, 
deletion, and substitution of a character, to transform one string to another. 

The feature extraction is based on bounding. The basic idea is to map each character to 
$(2\tau_{\max} + 1)$ bits, hence to cover the effect of insertion and deletion. 
Let $\Sigma$ denote the alphabet of strings, and $l_{\max}$ denote the maximum string length 
in $\mathcal{D}$. Each binary vector has $d = ((l_{\max} + 2\tau_{\max}) \cdot \size{\Sigma})$ 
bits. 
They are divided into $\size{\Sigma}$ groups, each group representing a character in $\Sigma$. 
For ease of exposition, we assume the subscript of a string start from $0$, and the subscript of 
each group of the binary vector start from $-\tau_{\max}$. All the bits are initialized as $0$. 
Given a string $x$, for each character $\sigma$ at position $i$, 
we set the $j$-th bit in the $\sigma$-th group to $1$, where $j$ iterates through $i - \tau_{\max}$ 
to $i + \tau_{\max}$. For example, 
given a string $x = \texttt{abc}$, $\Sigma = \set{\texttt{a}, \texttt{b}, \texttt{c}, \texttt{d}}$, 
$l_{\max} = 4$, and $\tau_{max} = 1$, the binary vector is 
\texttt{111000, 011100, 001110, 000000} (groups separated by comma).

It can be proved that an edit operation causes at most $(4\tau_{\max} + 2)$ different bits. 
Hence $f(x, y)$ edit operations yield a Hamming distance no greater than $f(x, y) \cdot 
(4\tau_{\max} + 2)$. Since it is proportional to $f(x, y)$ and thresholds are integers, we 
use the same threshold transformation as for Hamming distance. 

\subsection{Jaccard Distance}
Given two sets $x$ and $y$, the Jaccard similarity is defined as 
$\size{x \intersect y} / \size{x \union y}$. For 
ease of exposition, we use its distance form: $f(x, y) = 1 - 
\size{x \intersect y} / \size{x \union y}$. 

We use $b$-bit minwise hashing~\cite{li2010b} (LSH) for feature extraction. Given a record 
$x$, $\pi(x)$ orders the elements of $x$ by a permutation on the record universe $\mathcal{O}$. 
We uniformly choose a set of $k$ permutations $\set{\pi_1, \ldots, \pi_k}$. 
Let $bmin(\pi(x))$ denote the last $b$ bits of the smallest element of $\pi(x)$. We regard 
$bmin(\pi(x))$ as an integer in $[0, 2^b - 1]$ and transform it to a Hamming space. 
Let $set\_bit(i, j)$ produce a one-hot binary vector such that only the $i$-th bit is $1$ 
out of $j$ bits. $x$ is transformed to a $d$-dimensional ($d = 2^{b}k$) binary vector: 
$[set\_bit(bmin(\pi_1(x)), 2^b); \ldots; set\_bit(bmin(\pi_k(x)), 2^b)]$. For example: 
$x = \set{1, 2, 4}$. $\mathcal{O} = \set{1, 2, 3, 4, 5}$. 
$\pi_1 = 12345$, $\pi_2 = 54321$, and $\pi_3 = 21453$. $b = 2$. 
We have $bmin(\pi_1(x)) = 1$, $bmin(\pi_2(x)) = 0$, and $bmin(\pi_3(x)) = 2$. Suppose 
$set\_bit$ counts from the lowest bit, starting from $0$. The binary vector is 
\texttt{0010, 0001, 0100} (permutations separated by comma). 

Given two sets $x$ and $y$, the probability that $bmin(\pi(x)) = bmin(\pi(y))$ equals to  
$1 - f(x, y)$~\cite{li2010b}. 
The expected Hamming distance between $\mathbf{x}$ and $\mathbf{y}$ is thus 
$f(x, y) \cdot d$. Since it is proportional to $f(x, y)$, we use the following threshold 
transformation: 
$\tau = \floor{\tau_{\max} \cdot \theta / \theta_{\max}}$. 

\subsection{Euclidean Distance} \label{sec:featext-eu}
We use LSH based on $p$-stable distribution~\cite{datar2004locality} to handle 
Euclidean distance on real-valued vectors. The hash function is $h_{\mathbf{a}, b}(x) = 
\floor{\frac{\mathbf{a}x + b}{r}}$, where $\mathbf{a}$ is a $\size{x}$-dimensional vector 
with each element independently drawn by a normal distribution $\mathcal{N}(0, 1)$, $b$ is a real 
number chosen uniformly from $[0, r]$, and $r$ is a predefined constant value. Let $v$ denote the 
maximum hash value. We use the aforementioned $set\_bit$ function to transform hash values to a 
Hamming space. Given $k$ hash 
functions, $x$ is transformed to a $d$-dimensional ($d = k(v + 1)$) binary 
vector: $[set\_bit(h_{\mathbf{a}_1, b_1}(x), v + 1); \ldots; set\_bit(h_{\mathbf{a}_k, b_k}(x), 
v + 1)]$. For example: $x = [0.1, 0.2, 0.4]$. $v = 4$. 
$h_{\mathbf{a}_1, b_1}(x) = 1$. $h_{\mathbf{a}_2, b_2}(x) = 3$. 
$h_{\mathbf{a}_3, b_3}(x) = 4$. Suppose $set\_bit$ counts from the lowest bit, starting from $0$. 
The binary vector is \texttt{00010, 01000, 10000} (hash functions separated by comma). 

Given two records $x$ and $y$ such that $f(x, y) = \theta$, the probability that two hash 
values match is $Pr \set{h_{\mathbf{a}, b}(x) = h_{\mathbf{a}, b}(y)} = \epsilon(\theta)$ 
$ = 1 - 2\cdot norm(-r/\theta) - \frac{2}{\sqrt{2\pi} r / \theta} (1 - e^{-(r^2/2\theta^2)})$, 
where $norm(\cdot)$ is the cumulative distribution function of a random variable 
with normal distribution $\mathcal{N}(0, 1)$~\cite{datar2004locality}. Hence the 
expected Hamming distance between their binary representations is $(1 - \epsilon(\theta)) 
\cdot d$. The threshold transformation is  
$\tau = \floor{\tau_{\max} \cdot \frac{1 - \epsilon(\theta)}{1 - \epsilon(\theta_{\max})}}$. 

\section{Regression} \label{sec:model}
\begin{figure*} [t]
  \centering
  \includegraphics[width=\linewidth]{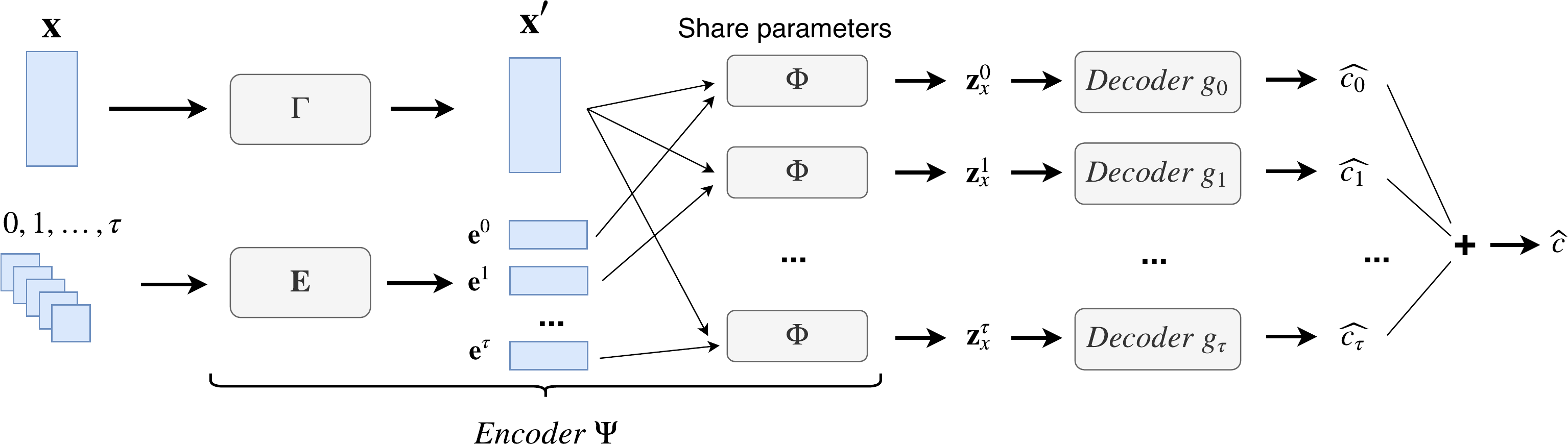}
  \caption{The regression model.}
  \label{fig:model:mono}
\end{figure*}

We present the detailed regression model in this section. Figure~\ref{fig:model:mono} 
shows the framework of our model. 
\begin{inparaenum} [(1)]
  \item $\mathbf{x}$ and $\tau$ are input to the encoder $\Psi$, which returns 
  $\tau + 1$ embeddings $\mathbf{z}_x^i$. Specifically, $\mathbf{x}$ is embedded 
  to a dense vector space. Each distance $i$ is also embedded, concatenated to the 
  embedding of $\mathbf{x}$, and fed to a neural network $\Phi$ to produce 
  $\mathbf{z}_x^i$. 
  \item Each of the $\tau + 1$ decoders $g_i$ takes an embedding $\mathbf{z}_x^i$ as 
  input and returns the cardinality of distance $i$, $i \in [0, \tau]$. 
  \item The $\tau + 1$ cardinalities are summed up to get the final cardinality.
\end{inparaenum}

\subsection{Encoder-Decoder Model} 
Our solution to the regression is to embed the binary vector $\mathbf{x}$ and distance $i$
to a dense real-valued vector $\mathbf{z}^1_x$ 
by an encoder $\Psi$, and then model $g_{i}(\mathbf{x})$ as a 
decoder that performs an affine transformation and applies an \textsf{ReLU} activation 
function: 
\begin{align*}
  g_i(\mathbf{x}) = \textsf{ReLU}(\mathbf{w}_i^{\rm T}
  \Psi(\mathbf{x}, i) + b_i) = \textsf{ReLU}(\mathbf{w}_i^{\rm T}
  \mathbf{z}^i_x + b_i).
\end{align*}
$\mathbf{w}_{i}$ and $b_{i}$ are parameters of the mapping from
the embedding $\mathbf{z}^i_x$ 
to the cardinality estimation of distance $i$. 
From the machine learning perspective, if a representation of input features 
is well learned through an encoder, then a linear model (affine 
transformation) is capable of making final decisions~\cite{bengio2013representation}. 
\textsf{ReLU} is chosen here because cardinality is non-negative and matches 
the range of \textsf{ReLU}. 
The reason why we also embed distance $i$ is as follows. Consider only $\mathbf{x}$ 
is embedded. If the cardinalities of two records $x_1$ and $x_2$ are close for 
distance values in a range $[\tau_1, \tau_2]$ covered by the training examples, 
their embeddings are likely to become similar after training, because the encoder may 
mistakenly regard $x_1$ and $x_2$ as similar. This may cause $g_i(\mathbf{x}_1) \approx 
g_i(\mathbf{x}_2)$ for $i \notin [\tau_1, \tau_2]$, i.e., the distance values not 
covered by the training examples, even if their actual cardinalities significantly 
differ. 

By Equation~\ref{eq:model}, the output of the $\tau$ decoders 
are summed up to obtain the cardinality. $g_{i}(\mathbf{x})$ is deterministic 
if we use a deterministic $\Psi(\cdot, \cdot)$. Hence the model can satisfy the requirement 
in Lemma~\ref{lm:model-monotonicity} to guarantee the monotonicity. 

\subsection{Encoder in Detail}
To encode both $\mathbf{x}$ and distance $i$ to embedding $\mathbf{z}^i_x$, 
$\Psi$ includes a representation network \repr that maps $\mathbf{x}$ to a 
dense vector space, a distance embedding layer $\mathbf{E}$, and a shared 
neural network $\Phi$ that outputs the embedding $\mathbf{z}^i_x$. Next we 
introduce the details. 

\subsubsection{Representation Network}
Given a binary representation $\mathbf{x}$ generated by feature extraction 
function $h(\cdot, \cdot)$, we design a neural network \repr that maps 
$\mathbf{x}$ to another vector space: 
$\mathbf{x}' = \Gamma(\mathbf{x})$, 
because the correlations of sparse high-dimensional binary vectors 
are difficult to learn. 
Variational auto-encoder (\vae)~\cite{kingma2013auto} is a generative
model to estimate data distribution by unsupervised learning. 
We can view auto-encoders (AEs) as non-linear PCAs to reduce dimensionality and 
extract meaningful and robust features, and \vae enforces continuity in the latent 
space. 
\fullversion{\vae improves upon other types of AEs (such as denoising AEs and sparse 
AEs) by imposing some regularization condition on the embeddings. This results in 
embeddings that are robust and disentangled, and hence have been widely used in 
various models~\cite{DBLP:journals/corr/abs-1812-05069}.} 
We use the latent layer of \vae to produce a dense representation, denoted 
by $VAE(\mathbf{x}, \epsilon)$. $\epsilon$ is a random noise 
generated by normal distribution $\mathcal{N}(0, \mathbf{I})$. 
\repr concatenates $\mathbf{x}$ and the output of \vae; i.e., 
$\mathbf{x}' = [\mathbf{x}; VAE(\mathbf{x}, \epsilon)]$. 
The reason for such concatenation (i.e., not using only the output of \vae 
as $\mathbf{x}'$) is that the (cosine) distance in the $VAE(\mathbf{x}, \epsilon)$ 
space captures less semantics of the original distance than does the Hamming 
distance between binary vectors. 
Due to the noise $\epsilon$, the output of \vae becomes nondeterministic. 
Since we need a deterministic output to guarantee the monotonicity, we 
choose the following option: for training, we still use the nondeterministic 
$\mathbf{x}' = [\mathbf{x}; VAE(\mathbf{x}, 
\epsilon)]$, because this makes our model generalize 
to unseen records and thresholds; for inference (online estimation), we 
set $\mathbf{x}' = 
[\mathbf{x}; \mathrm{E}_{\epsilon \sim \mathcal{N}(0, \mathbf{I})}[VAE(\mathbf{x}, 
\epsilon)]]$, where $\mathrm{E}[\cdot]$ denotes the expected value, so 
it becomes deterministic~\cite{sohn2015learning}. 

\begin{figure} [t]
  \centering
  \includegraphics[width=\linewidth]{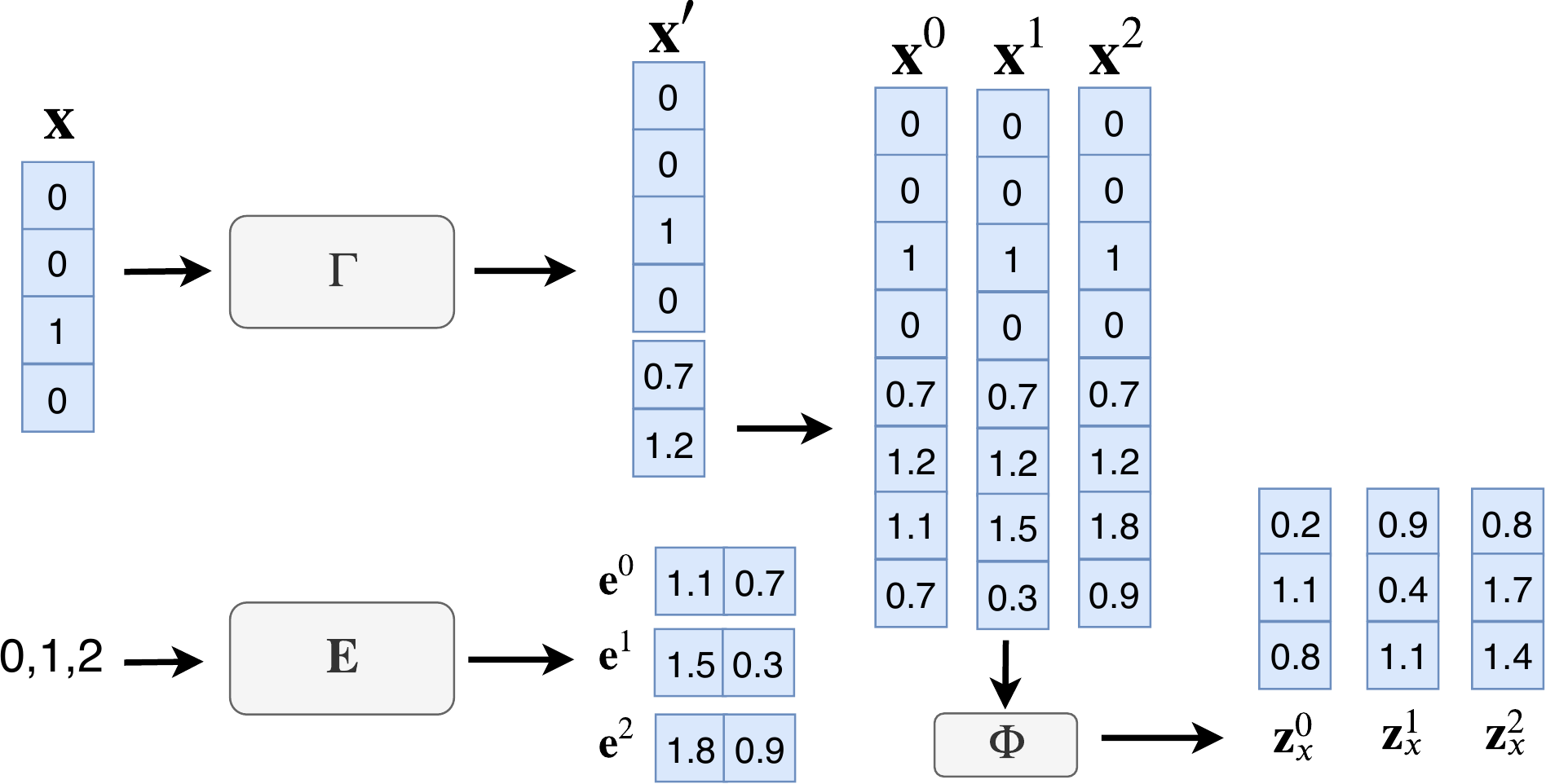}
  \caption{Example of encoder $\Psi$.}
  \label{fig:encoder-example}
\end{figure}

\begin{example}
  \label{ex:dense-representation}
  Figure~\ref{fig:encoder-example} shows an example of $\mathbf{x}$ and its 
  embedding $\mathbf{x}'$. Suppose $\mathbf{x} = \texttt{0010}$. The \vae 
  takes $\mathbf{x}$ as input and output a dense vector, say $[0.7, 1.2]$. 
  Then they are concatenated to obtain $\mathbf{x}' = [\texttt{0010}, 0.7, 1.2]$.
\end{example}

\subsubsection{Distance Embeddings}
In order to embed $\mathbf{x}$ and distance $i$ into the same vector, 
we design a distance embedding layer (a matrix) $\mathbf{E}$ 
to embed each distance $i$. Each column in $\mathbf{E}$ 
represents a distance embedding; i.e., $\mathbf{e}^i = \mathbf{E}[*, i]$.
$\mathbf{E}$ is initialized randomly, following standard normal 
distribution. 

\subsubsection{Final Embeddings}
The distance embedding $\mathbf{e}^{i}$ is concatenated with $\mathbf{x}'$; i.e., 
$\mathbf{x}^{i} = [\mathbf{x}' ; \mathbf{e}^{i}]$. 
Then we use a feedforward neural network (FNN) $\Phi$ to generate embeddings 
$\mathbf{z}^i_x = \Phi(\mathbf{x}^{i})$.

\begin{example}
  We follow Example~\ref{ex:dense-representation}. Suppose $\tau = 2$. Then 
  we have three distance embeddings $\mathbf{e}^0 = [1.1, 0.7]$, 
  $\mathbf{e}^1 = [1.5, 0.3]$, and $\mathbf{e}^2 = [1.8, 0.9]$. 
  By concatenating $\mathbf{x}'$ and each $\mathbf{e}^i$, 
  $\mathbf{x}^0 = [\texttt{0010}, 0.7,1.2,1.1,0.7]$, 
  $\mathbf{x}^1 = [\texttt{0010}, 0.7,1.2,1.5,0.3]$, 
  and $\mathbf{x}^2 = [\texttt{0010}, 0.7, 1.2, 1.8,0.9]$. 
  They are sent to neural network $\Phi$, whose output is 
  $\mathbf{z}^0_x = [0.2, 1.1, 0.8]$, 
  $\mathbf{z}^1_x = [0.9, 0.4, 1.1]$, 
  and $\mathbf{z}^2_x = [0.8, 1.7, 1.4]$. 
\end{example}

\section{Model Training} \label{sec:train}

\subsection{Data Preparation} \label{sec:traindata}
\revise{Consider a query workload $\mathcal{Q}$ of records (see 
Section~\ref{sec:exp-setup-datasets-queries} for the choice of $\mathcal{Q}$ in our experiments). 
We split data in $\mathcal{Q}$ for training, validation, and testing sets. Then 
we uniformly generate a set of thresholds in $[0, \theta_{\max}]$, denoted by 
$S$. For each  record $x$ in the training set, we iterate through all the 
thresholds $\theta$ in $S$ and compute the cardinality $c$ w.r.t. $\mathcal{D}$ 
using an exact similarity selection algorithm. Then $\pair{x, \theta, c}$ is 
used as a training example. We uniformly choose thresholds in $S$ for 
validation and in $[0, \theta_{\max}]$ for testing.}

\subsection{Loss Function \& Dynamic Training}
The loss function is defined as follows. 
\begin{align} \label{eq:monomodel1:loss}
\mathcal{L}(\widehat{\mathbf{c}}, \mathbf{c}) = \mathrm{E}_{\mathbf{\tau} \sim
  P(\cdot)}[\mathcal{L}_{g}(\widehat{\mathbf{c}}, \mathbf{c})] +
  \lambda \mathcal{L}_{vae}(\mathbf{x}),
\end{align}
where $\mathcal{L}_{g}(\cdot, \cdot)$ is the loss of regression model, and
$\mathcal{L}_{vae}(\cdot)$ is the loss of  
\vae. $\widehat{\mathbf{c}}$ and $\mathbf{c}$ are two 
vectors, each dimension representing the estimated and the real cardinalities 
of a set of training examples, respectively. 
$\lambda$ is a positive hyperparameter for the importance of \vae. A caveat 
is that although we uniformly sample thresholds in 
$[0, \theta_{\max}]$ for training data, it does not necessarily mean the 
threshold $\tau$ after feature extraction is uniformly distributed in 
$[0, \tau_{\max}]$, e.g., for Euclidean distance (Section~\ref{sec:featext-eu}). 
To take this factor into account, we approximate the probability of $\tau$ using 
the empirical probability of thresholds after running feature extraction on the 
validation set; i.e., 
$P(\tau) \approx \frac{\sum_{\pair{x, i, c} \in \mathcal{T}^{valid}}\mathbf{1}_{h_{thr}(i) = \tau}}{|\mathcal{T}^{valid}|}$, 
where $\mathcal{T}^{valid}$ is the validation set, and $\mathbf{1}$ is the 
indicator function.


For $\mathcal{L}_{g}$, instead of using \textsf{MSE} or 
\textsf{MAPE}, 
we resort to the mean
squared logarithmic error (\msle) 
for the following reason: \msle is an approximation of 
\textsf{MAPE}~\cite{park1998relative} and narrows down the large output space to 
a smaller one, thereby decreasing the learning difficulty.

Then we propose a training strategy for better accuracy. Given a set 
of training examples, let $\mathbf{c}_0$, $\ldots$, $\mathbf{c}_{\tau}$ and 
$\widehat{\mathbf{c}_0}$, $\ldots$, $\widehat{\mathbf{c}_{\tau}}$ denote the 
cardinalities and the estimated values for distance $0$, $\ldots$, $\tau$ in 
these training examples, respectively. 
As we have shown in 
Figure~\ref{fig:card-dist}, the cardinalities may vary significantly for different 
distance values. Some of them may result in much worse estimations than others and 
compromise the overall performance. The training procedure should gradually focus 
on training these bad estimations. Thus, we consider the loss caused by the 
estimation for distance $i$, combined with \msle:  
\begin{align}
  \label{eq:monomodel1:lossnn}
  \mathcal{L}_{g}(\widehat{\mathbf{c}}, \mathbf{c}) = 
  \msle(\widehat{\mathbf{c}}, \mathbf{c}) + 
  \lambda_{\Delta} \cdot \bigg{(}
  \sum_{i=0}^{\tau_{max}}{\omega_i \cdot \msle(\widehat{\mathbf{c}_i}. 
  \mathbf{c}_i)}  \bigg{)}. 
\end{align} 
Each $\omega_i$ is a hyperparameter automatically adjusted during the training 
procedure. Hence we call it dynamic training. It controls the loss of each estimation 
for distance $i$. $\sum_{i=0}^{\tau_{max}}{\omega_i} = 1$. 
$\lambda_{\Delta}$ is a hyperparameters to control the impact of the losses 
of all the estimations for $i \in [0, \tau_{\max}]$.

Due to the non-convexity of $\mathcal{L}_{g}$, it is difficult to find 
the correct direction of gradient that reaches the global or a good local 
optimum. Nonetheless, we can adjust its direction by considering 
the loss trend of the estimation for distance $i$, hence to encourage the 
model to generalize rather than to overfit the training data. 
Let $\ell_i(t) = \msle(\widehat{\mathbf{c}_i}, \mathbf{c}_i)$ denote the 
loss of the estimation for distance $i$ in the $t$-th iteration of 
validation. The loss trend $\Delta \ell_i(t) = \ell_i(t) - \ell_i(t - 1)$ is 
calculated, and then after each validation we adjust $\omega_i$ by adding 
more gradients to where the loss occurs: 
\begin{inparaenum} [(1)]
  \item If $\Delta \ell_i(t) > 0$, $\omega_i = \frac{\Delta \ell_i(t)}{\sum_{i \in A}{\Delta \ell_i(t)}, \,
  A = \set{i \mid \Delta \ell_i(t) > 0, 0 \leq i \leq \tau_{\max}}}$; 
  \item otherwise, $\omega_i = 0$. 
\end{inparaenum}

\section{Accelerating Estimation}\label{sec:accelerate}

\begin{figure} [t]
  \centering
  \includegraphics[width=\linewidth]{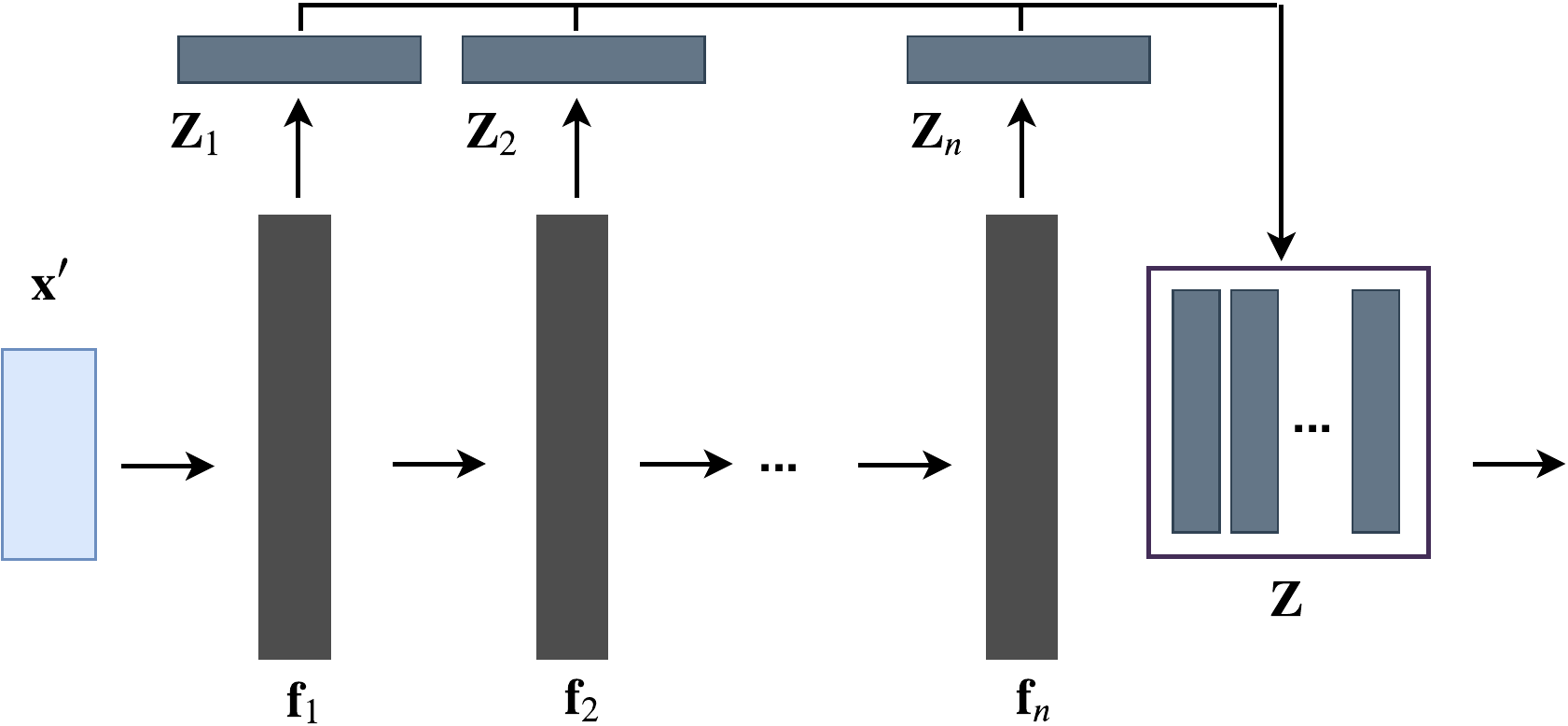}
  \caption{$\Phi'$ in the accelerated regression model.}
  \label{fig:model:mono2}
\end{figure}

Recall in the regression model, we pair $\mathbf{x}'$ and $(\tau + 1)$ 
distance embeddings in encoder $\Psi$ to produce embedings $\mathbf{z}_x^i$. 
This leads to 
high computation cost for online cardinality estimation when $\tau$ is 
large. To reduce the cost, we propose an accelerated model, using a neural 
network $\Phi'$ to replace $\Phi$ and the distance embedding layer 
$\mathbf{E}$ to output $(\tau_{\max} + 1)$
$\mathbf{z}_x^i$ embeddings together~\footnote{We output $(\tau_{\max} + 1)$ 
embeddings since it is constant for all queries, hence to favor implementation. 
Only the first $(\tau + 1)$ embedding are used for $\tau$.}. 
$\Phi'$ only takes an input $\mathbf{x'}$ and reduces the computation cost 
from $O((\tau + 1)\size{\Phi})$ to $O(\size{\Phi'})$, where $\size{\Phi}$ 
and $\size{\Phi'}$ denote the number of parameters in the two neural networks. 

Figure~\ref{fig:model:mono2} shows the framework of $\Phi'$, 
an FNN comprised of $n$ hidden layers 
$\mathbf{f}_1$, $\mathbf{f}_2$, $\ldots$, $\mathbf{f}_n$, each 
outputting some dimensions of the embeddings $\mathbf{z}_x^i$. 
Each $\mathbf{z}^i_x$ 
is partitioned into $n$ regions, denoted by 
$\mathbf{z}^i_x[r_0, r_1]$, $\mathbf{z}^i_x[r_1, r_2]$, $\ldots$, 
$\mathbf{z}^i_x[r_{n-1}, r_n]$, where $0 = r_0 \leq r_1 \leq \ldots \leq r_n$ 
and $r_n$ equals to the dimensionality of $\mathbf{z}_x^i$. 
A hidden layer $\mathbf{f}_j$ outputs the $j$-th region of 
$\mathbf{z}_x^i$, $i \in [0, \tau_{\max}]$; i.e., 
$\mathbf{Z}_j = [\mathbf{z}^0_x[r_{j-1}, r_j]: \mathbf{z}^1_x[r_{j-1}, r_j]: \ldots : \mathbf{z}^{\tau_{max}}_x[r_{j-1}, r_j]]$.
Then we concatenate all the regions: 
$\mathbf{Z} = [\mathbf{Z}_1 : \mathbf{Z}_2 : \ldots : \mathbf{Z}_n]$. 
Each row of $\mathbf{Z}$ is an embedding: 
$\mathbf{z}_x^i = \mathbf{Z}[i, *]$.

\fullversion{
In addition to fast estimation, the model $\Phi'$ has the following advantages: 
\begin{inparaenum} [(1)]
  \item 
  The parameters of each hidden layer $\mathbf{f}_j$ are updated from 
  the following layer $\mathbf{f}_{j+1}$ and the final embedding matrix 
  $\mathbf{Z}$ through backpropagation, hence increasing the ability to learn 
  good embeddings. 
  \item Since each embedding is affected by all the hidden 
  layers, the model is more likely to reach a better local optimum through training.
  \item In contrast to one output layer, all the hidden layers output 
  embeddings, so gradient vanishing and overfitting can be prevented. 
\end{inparaenum}
}


\fullversion{We analyze the complexities of our models. 
Assume an FNN has hidden layers $\mathbf{a}_1, \ldots, \mathbf{a}_n$. 
Given an input $\mathbf{x}$ and an output $\mathbf{y}$, the complexity of an FNN is 
$\size{\textsf{FNN}(\mathbf{x}, \mathbf{y})} = \size{\mathbf{x}} \cdot \size{\mathbf{a}_1} 
+ \sum_{i=1}^{n-1}{\size{\mathbf{a}_i} \cdot \size{\mathbf{a}_{i+1}}} 
+ \size{\mathbf{a}_n} \cdot \size{\mathbf{y}}$. 
Our model has the following components: $\Phi, \Gamma, \mathbf{E}$, and $g_i$. 
The complexities of $\Phi$ and $\Gamma$ are 
$\size{\textsf{FNN}([\mathbf{x}';\mathbf{e}^i], \mathbf{z}_x^i)}$ and 
$\size{\textsf{FNN}(\mathbf{x}, \mathbf{x})}$, respectively. 
$\size{\mathbf{E}} = (\tau_{\max} + 1) \size{\mathbf{e}^i}$. 
$\size{g_i} = (\tau_{\max} + 1) \size{\mathbf{z}_x^i} + \tau_{\max} + 1$. 
Thus, the complexity of our model without acceleration is 
$\size{\textsf{FNN}([\mathbf{x}'; \mathbf{e}^i], \mathbf{z}_x^i)} 
+ \size{\textsf{FNN}(\mathbf{x}, \mathbf{x})} + (\tau_{\max} + 1) \size{\mathbf{e}^i} 
+ (\tau_{\max} + 1) \size{\mathbf{z}_x^i} + \tau_{\max} + 1$.
With acceleration for FNN (AFNN), the complexity is 
$\size{\textsf{AFNN}(\mathbf{x}', \mathbf{Z})} 
+ \size{\textsf{FNN}(\mathbf{x}, \mathbf{x})} 
+ (\tau_{\max} + 1)\size{\mathbf{z}_x^i} + \tau_{\max} + 1$, 
where $\size{\textsf{AFNN}(\mathbf{x}', \mathbf{Z})} = \size{\mathbf{x}'} \cdot \size{\mathbf{a}_1} + \sum_{i=1}^{n-1}{\size{\mathbf{a}_i} \cdot \size{\mathbf{a}_{i+1}}} + (\tau_{\max} + 1) \size{\mathbf{a}_n} \cdot \size{\mathbf{Z}[i, *]}$.
} 

\section{Dealing with Updates} \label{sec:update} 
\revise{When the dataset is updated, the labels of the validation data are first updated 
by running a similarity selection algorithm on the updated dataset. Then we monitor the 
error (\msle) in the validated data by running our model. 
If the error increases,
we 
train our model with incremental learning:} 
First, the labels of the training data are updated by running a similarity selection 
algorithm on the updated dataset. Then the model is trained with the updated training 
data until the validation error does not change for three consecutive epochs. 
\revise{Note that 
\begin{inparaenum} [(1)]
  \item the training does not start from scratch but from the current model, and it is 
  processed on the entire training data to prevent catastrophic 
  forgetting~\cite{mccloskey1989catastrophic,kemker2018measuring}; and
  \item we always keep the original queries and only update their labels.
\end{inparaenum}
}
\confversion{
\begin{table*} [t]
  \small
  \caption{Statistics of datasets.}
  \label{tab:dataset}
  \centering
  \begin{tabular}[b]{| l | c | c | c | c | c | c | c | c | c |}
    \hline%
    \texttt{Dataset} & \texttt{Source} & \texttt{Process} & \texttt{Data Type} &
    \texttt{Domain} & \texttt{\# Records} & $\ell_{max}$ & $\ell_{avg}$ &
      \texttt{Distance} & $\theta_{max}$ \\
    \hline%
  \imagenet & \cite{URL:imagenet} 
  & HashNet~\cite{cao2017hashnet} 
  & binary vector & image & 1,431,167 & 64 & 64 & Hamming & 20 \\ 
   \hline%
  \aminer & \cite{URL:aminer} 
  & - & string & author name & 1,712,433 & 109 & 13.02
  & edit & 10 \\
   \hline%
  \bmsjacc & \cite{URL:bms-pos} 
  & - & set & product entry & 515,597 & 164 & 6.54 
  & Jaccard & 0.4 \\
   \hline%
  \glovetwo & \cite{URL:glove} 
  & normalize & real-valued vector & word embedding & 1,917,494 & 300 & 300 
  & Euclidean & 0.8 \\
   \hline%
  \end{tabular}
\end{table*}
\begin{table} [t]
  \small
  \caption{\textsf{MSE}.}
  \label{tab:hm:errors-mse}
  \centering
  \resizebox{\linewidth}{!}{
  \begin{tabular}[b]{| l | c | c | c | c |}
    \hline%
    Model & \imagenet & \aminer & \bmsjacc & \glovetwo\\  
    \hline%
    \spestexp & 41563 & 8219583 & 4725 & 116820\\
    \usexp & 27776 & 159572 & 6090 & 78552\\
    \xgbexp & 12082 & 4147509 & 3784 & 821937\\
    \lightgbmexp & 14132 & 4830965 & 4011 & 844301\\
    \kdeexp & 279782 & 3412627 & 7236 & 102200\\
    \dlnexp & 7307 & 1285010 & 2892 & 1063687\\
    \moeexp & 7096 & 265257 & 1503 & 988918\\
    \hierdnnexp & 6774 & 93158 & 264 & 45165\\
    \dnnexp & 10075 & 207286 & 5281 & 1192426\\
    \dnnsexp & 4236 & 217193 & 7500 & 1178239\\
    \lstmexp & - & 104152 & - & -\\
    \lstmaexp & - & 115111 & - & -\\
    \textbf{\modelone} & \textbf{2871} & \textbf{52101} & 75 & \textbf{6822}\\
    \textbf{\modeltwo} & 3044 & 64831 & \textbf{64} & 16809\\
    \hline%
  \end{tabular}
  }
\end{table}

\begin{table} [t]
  \small
  \caption{\textsf{MAPE} (in percentage).}
  \label{tab:hm:errors-mape}
  \centering
  \resizebox{\linewidth}{!}{  
  \begin{tabular}[b]{| l | c | c | c | c |}
    \hline%
    Model & \imagenet & \aminer & \bmsjacc & \glovetwo\\
    \hline%
    \spestexp & 56.14 & 80.15 & 59.38 & 41.91\\
    \usexp & 62.51 & 61.98 & 63.84 & 112.06\\
    \xgbexp & 13.87 & 113.68 & 34.76 & 14.46\\
    \lightgbmexp & 14.12 & 115.88 & 39.01 & 17.49\\
    \kdeexp & 85.57 & 105.17 & 42.01 & 52.38\\
    \dlnexp & 20.72 & 73.48 & 49.67 & 21.67\\
    \moeexp & 11.93 & 57.79 & 19.73 & 11.94\\
    \hierdnnexp & 12.36 & 52.81 & 23.89 & 5.48\\
    \dnnexp & 14.24 & 51.36 & 43.44 & 7.24\\
    \dnnsexp & 13.00 & 53.82 & 21.25 & 9.19\\
    \lstmexp & - & 43.44 & - & -\\
    \lstmaexp & - & 45.25 & - & -\\
    \textbf{\modelone} & \textbf{8.41} & \textbf{42.26} & \textbf{11.25} & \textbf{4.04}\\
    \textbf{\modeltwo} & 9.63 & 44.78 & 13.94 & 4.58\\
    \hline%
  \end{tabular}
  }
\end{table} 

}
\fullversion{
\begin{table*} [t]
  \small
  \caption{Statistics of datasets.}
  \label{tab:dataset}
  \centering
  \begin{tabular}[b]{| l | c | c | c | c | c | c | c | c | c |}
    \hline%
    \texttt{Dataset} & \texttt{Source} & \texttt{Process} & \texttt{Data Type} &
    \texttt{Domain} & \texttt{\# Records} & $\ell_{max}$ & $\ell_{avg}$ &
      \texttt{Distance} & $\theta_{max}$ \\
    \hline%
  \textbf{\imagenet} & \cite{URL:imagenet} 
  & \textsf{HashNet}~\cite{cao2017hashnet} 
  & binary vector & image & 1,431,167 & 64 & 64 & Hamming & 20 \\ 
   \hline%
  \pubchem & \cite{URL:pubchem} 
  & - & binary vector & biological sequence & 1,000,000 & 881 & 881
  & Hamming & 30 \\
   \hline%
  \textbf{\aminer} & \cite{URL:aminer} 
  & - & string & author name & 1,712,433 & 109 & 13.02
  & edit & 10 \\
   \hline%
  \dblped & \cite{URL:dblp} 
  & - & string & publication title & 1,000,000 & 199 & 72.49
  & edit & 20 \\
   \hline%
  \textbf{\bmsjacc} & \cite{URL:bms-pos} 
  & - & set & product entry & 515,597 & 164 & 6.54 
  & Jaccard & 0.4 \\
   \hline%
  \dblpjacclong & \cite{URL:dblp} 
  & 3-gram & set & publication title & 1,000,000 & 197 & 70.49
  & Jaccard & 0.4 \\
   \hline%
  \textbf{\glovetwo} & \cite{URL:glove} 
  & normalize & real-valued vector & word embedding & 1,917,494 & 300 & 300 
  & Euclidean & 0.8 \\
   \hline%
  \gloveone & \cite{URL:glove} 
  & normalize & real-valued vector & word embedding & 400,000 & 50 & 50 
  & Euclidean & 0.8 \\
   \hline%
  \end{tabular}
\end{table*}

\eat{
\begin{table*} [t]
  \small
  \caption{\textsf{MSE}, best values highlighted in boldface.}
  \label{tab:hm:errors-mse}
  \centering
  \begin{tabular}[b]{| l | c | c | c | c | c | c | c | c |}
    \hline%
    Model & \imagenet & \pubchem & \aminer & \dblped & \bmsjacc & \dblpjacclong & \glovetwo & \gloveone\\  
    \hline%
    \spestexp & 41563 & 445182 & 8219583 & 1681 & 4725 & 177 & 116820 & 45631\\
    \usexp & 27776 & 66255 & 159572 & 1095 & 6090 & 427 & 78552 & 16249\\  
    \xgbexp & 12082 & 882206 & 4147509 & 1657 & 3784 & 23 & 821937 & 557229\\
    \lightgbmexp & 14132 & 721609 & 4830965 & 2103 & 4011 & 49 & 844301 & 512984\\
    \kdeexp & 279782 & 112952 & 3412627 & 2097 & 7236 & 100 & 102200 & 169604\\
    \dlnexp & 7307 & 189743 & 1285010 & 1664 & 2892 & 50 & 1063687 & 49389\\
    \moeexp & 7096 & 95447 & 265257 & 1235 & 1503 & 23 & 988918 & 315437\\ 
    \hierdnnexp & 6774 & 42186 & 93158 & 928 & 264 & 15 & 45165 & 6791\\    
    \dnnexp & 10075 & 231167 & 207286 & 1341 & 5281 & 138 & 1192426 & 27892\\
    \dnnsexp & 4236 & 51026 & 217193 & 984 & 7500 & 207 & 1178239 & 87991\\
    \lstmexp & - & - & 104152 & 1034 & - & - & - & -\\
    \lstmaexp & - & - & 115111 & 1061 & - & - & - & -\\    
    \textbf{\modelone} & \textbf{2871} & 12809 & \textbf{52101} & 446 & 75 & \textbf{2} & \textbf{6822} & \textbf{3245}\\
    \textbf{\modeltwo} & 3044 & \textbf{11598} & 64831 & \textbf{427} & \textbf{64} & 3 & 16809 & 3269\\
    \hline%
  \end{tabular}
\end{table*}
}

\begin{table*} [t]
  \small
  \caption{\textsf{MSE}, best values highlighted in boldface.}
  \label{tab:hm:errors-mse}
  \centering
  \begin{tabular}[b]{| l | c | c | c | c | c | c | c | c |}
    \hline%
    Model & \imagenet & \pubchem & \aminer & \dblped & \bmsjacc & \dblpjacclong & \glovetwo & \gloveone\\  
    \hline%
    \spestexp & 41563 & 445182 & 8219583 & 1681 & \reviseR{1722} & 177 & 116820 & 45631\\
    \usexp & 27776 & 66255 & 159572 & 1095 & \reviseR{3052} & 427 & 78552 & 16249\\  
    \xgbexp & 12082 & 882206 & 4147509 & 1657 & \reviseR{2031} & 23 & 821937 & 557229\\
    \lightgbmexp & 14132 & 721609 & 4830965 & 2103 & \reviseR{1394} & 49 & 844301 & 512984\\
    \kdeexp & 279782 & 112952 & 3412627 & 2097 & \reviseR{873} & 100 & 102200 & 169604\\
    \dlnexp & 7307 & 189743 & 1285010 & 1664 & \reviseR{1349} & 50 & 1063687 & 49389\\
    \moeexp & 7096 & 95447 & 265257 & 1235 & \reviseR{425} & 23 & 988918 & 315437\\ 
    \hierdnnexp & 6774 & 42186 & 93158 & 928 & \reviseR{151} & 15 & 45165 & 6791\\    
    \dnnexp & 10075 & 231167 & 207286 & 1341 & \reviseR{1103} & 138 & 1192426 & 27892\\
    \dnnsexp & 4236 & 51026 & 217193 & 984 & \reviseR{1306} & 207 & 1178239 & 87991\\
    \lstmexp & - & - & 104152 & 1034 & - & - & - & -\\
    \lstmaexp & - & - & 115111 & 1061 & - & - & - & -\\    
    \textbf{\modelone} & \textbf{2871} & 12809 & \textbf{52101} & 446 & 75 & \textbf{2} & \textbf{6822} & \textbf{3245}\\
    \textbf{\modeltwo} & 3044 & \textbf{11598} & 64831 & \textbf{427} & \textbf{64} & 3 & 16809 & 3269\\
    \hline%
  \end{tabular}
\end{table*}

\eat{
\begin{table*} [t]
  \small
  \caption{\textsf{MAPE} (in percentage), best values highlighted in boldface.}
  \label{tab:hm:errors-mape}
  \centering
  \begin{tabular}[b]{| l | c | c | c | c | c | c | c | c |}
    \hline%
    Model & \imagenet & \pubchem & \aminer & \dblped & \bmsjacc & \dblpjacclong & \glovetwo & \gloveone\\
    \hline%
    \spestexp & 56.14 & 79.74 & 80.15 & 57.23 & 59.38 & 10.42 & 41.91 & 47.12\\
    \usexp & 62.51 & 141.04 & 61.98 & 56.80 & 63.84 & 50.52 & 112.06 & 98.23\\
    \xgbexp & 13.87 & 152.20 & 113.68 & 33.26 & 34.76 & 6.70 & 14.46 & 33.87\\
    \lightgbmexp & 14.12 & 110.22 & 115.88 & 30.29 & 39.01 & 10.71 & 17.49 & 37.54\\
    \kdeexp & 85.57 & 179.39 & 105.17 & 60.23 & 42.01 & 37.79 & 52.38 & 59.84\\
    \dlnexp & 20.72 & 174.69 & 73.48 & 39.10 & 49.67 & 6.54 & 21.67 & 33.95\\    
    \moeexp & 11.93 & 49.47 & 57.79 & 31.81 & 19.73 & 4.10 & 11.94 & 26.82\\
    \hierdnnexp & 12.36 & 50.57 & 52.81 & 32.24 & 23.89 & 4.78 & 5.48 & 15.03\\    
    \dnnexp & 14.24 & 198.36 & 51.36 & 34.12 & 43.44 & 5.11 & 7.24 & 17.57\\
    \dnnsexp & 13.00 & 46.43 & 53.82 & 30.91 & 21.25 & 5.81 & 9.19 & 21.95\\
    \lstmexp & - & - & 43.44 & 40.95 & - & - & - & -\\
    \lstmaexp & - & - & 45.25 & 41.12 & - & - & - & -\\    
    \textbf{\modelone} & \textbf{8.41} & \textbf{35.66} & \textbf{42.26} & \textbf{22.53} & \textbf{11.25} & 3.18 & \textbf{4.04} & \textbf{11.85}\\
    \textbf{\modeltwo} & 9.63 & 36.57 & 44.78 & 23.07 & 13.94 & \textbf{3.05} & 4.58 & 12.71\\
    \hline%
  \end{tabular}
\end{table*} 
}

\begin{table*} [t]
  \small
  \caption{\textsf{MAPE} (in percentage), best values highlighted in boldface.}
  \label{tab:hm:errors-mape}
  \centering
  \begin{tabular}[b]{| l | c | c | c | c | c | c | c | c |}
    \hline%
    Model & \imagenet & \pubchem & \aminer & \dblped & \bmsjacc & \dblpjacclong & \glovetwo & \gloveone\\
    \hline%
    \spestexp & 56.14 & 79.74 & 80.15 & 57.23 & \reviseR{45.12} & 10.42 & 41.91 & 47.12\\
    \usexp & 62.51 & 141.04 & 61.98 & 56.80 & \reviseR{60.06} & 50.52 & 112.06 & 98.23\\
    \xgbexp & 13.87 & 152.20 & 113.68 & 33.26 & \reviseR{26.52} & 6.70 & 14.46 & 33.87\\
    \lightgbmexp & 14.12 & 110.22 & 115.88 & 30.29 & \reviseR{21.39} & 10.71 & 17.49 & 37.54\\
    \kdeexp & 85.57 & 179.39 & 105.17 & 60.23 & \reviseR{27.59} & 37.79 & 52.38 & 59.84\\
    \dlnexp & 20.72 & 174.69 & 73.48 & 39.10 & \reviseR{42.50} & 6.54 & 21.67 & 33.95\\    
    \moeexp & 11.93 & 49.47 & 57.79 & 31.81 & \reviseR{16.37} & 4.10 & 11.94 & 26.82\\
    \hierdnnexp & 12.36 & 50.57 & 52.81 & 32.24 & \reviseR{15.02} & 4.78 & 5.48 & 15.03\\    
    \dnnexp & 14.24 & 198.36 & 51.36 & 34.12 & \reviseR{28.41} & 5.11 & 7.24 & 17.57\\
    \dnnsexp & 13.00 & 46.43 & 53.82 & 30.91 &\reviseR{19.67} & 5.81 & 9.19 & 21.95\\
    \lstmexp & - & - & 43.44 & 40.95 & - & - & - & -\\
    \lstmaexp & - & - & 45.25 & 41.12 & - & - & - & -\\    
    \textbf{\modelone} & \textbf{8.41} & \textbf{35.66} & \textbf{42.26} & \textbf{22.53} & \textbf{11.25} & 3.18 & \textbf{4.04} & \textbf{11.85}\\
    \textbf{\modeltwo} & 9.63 & 36.57 & 44.78 & 23.07 & 13.94 & \textbf{3.05} & 4.58 & 12.71\\
    \hline%
  \end{tabular}
\end{table*} 


}



\section{Experiments}
\label{sec:exp}
\subsection{Experiment Setup} 
\label{sec:exp-setup}
\subsubsection{Datasets and Queries} \label{sec:exp-setup-datasets-queries}
\confversion{
We use one dataset for each of the four distance functions: Hamming distance (\textsf{HM}), 
edit distance (\textsf{ED}), Jaccard distance (\textsf{JC}), and Euclidean distance 
(\textsf{EU}). The datasets and the statistics are shown in Table~\ref{tab:dataset}. 
}
\fullversion{
We use eight datasets for four distance functions: Hamming distance (\textsf{HM}), 
edit distance (\textsf{ED}), Jaccard distance (\textsf{JC}), and Euclidean distance 
(\textsf{EU}). 
The datasets and the statistics are shown in Table~\ref{tab:dataset}. Boldface indicates 
default datasets.
}
\texttt{Process} indicates how 
we process the dataset; e.g., 
HashNet~\cite{cao2017hashnet} is adopted to convert images in the ImageNet 
dataset~\cite{ILSVRC15} to hash codes. 
$\ell_{max}$ and $\ell_{avg}$ are the maximum and average lengths or dimensionalities 
of records, respectively. 
\revise{We uniformly sample 10\% data from dataset $\mathcal{D}$ as the query workload 
$\mathcal{Q}$. Then we follow the method in Section~\ref{sec:traindata} to split 
$\mathcal{Q}$ in $80:10:10$ to create training, validation, and testing instances. 
\fullversion{Multiple uniform sampling is not considered here because even if our models 
are trained on skewed data sampled with equal chance from each cluster of $\mathcal{D}$, 
we observe only moderate change of accuracy of our models (up to 48\% \mse) when testing 
over multiple uniform samples and they still perform significantly better than the other 
competitors.} 
Thresholds and labels (cardinalities w.r.t. $\mathcal{D}$) are generated 
using the method in Section~\ref{sec:traindata}.} 

\subsubsection{Models} \label{sec:exp-setup-models}
Our models are referred to as \modelone and \modeltwo. The latter is equipped 
with the acceleration (Section~\ref{sec:accelerate}). 
We compare with the following categories of methods: 
\begin{inparaenum} [(1)]
  \item Database methods: \spestexp, a specialized estimator for each distance function  
  (histogram~\cite{qin2018gph} for \textsf{HM}, inverted index~\cite{jin2008sepia} 
  for \textsf{ED}, semi-lattice~\cite{lee2009power} for \textsf{JC}, and 
  LSH-based sampling~\cite{wu2018local} for \textsf{EU})  
  and \usexp, which uniformly samples 1\% records from $\mathcal{D}$ 
  and estimates cardinality using the sample. We do not consider higher sample ratios 
  because 1\% samples are already very slow (Table~\ref{tab:hm:estitime}). 
  \item Traditional learning methods: \xgbexp (XGBoost)~\cite{chen2016xgboost}, 
  \lightgbmexp (LightGBM)~\cite{DBLP:conf/nips/KeMFWCMYL17}, and 
  \kdeexp~\cite{DBLP:conf/edbt/MattigFBS18}.  
  \item Deep learning methods: \dlnexp~\cite{you2017deep}; 
  \moeexp~\cite{shazeer2017outrageously}; \hierdnnexp~\cite{kraska2018case}; 
  \dnnexp, a vanilla FNN with four hidden layers; and 
  \dnnsexp, a set of $(\tau_{max} + 1)$ independently learned deep neural networks, 
  each against a threshold range (computed using the threshold transformation in 
  Section~\ref{sec:case-study}). 
  For \textsf{ED}, we also have a method that replaces $\Gamma$ in \modelone or  
  \modeltwo with a character-level bidirectional LSTM~\cite{chiu2016named}, referred to 
  as \lstmexp or \lstmaexp.
\end{inparaenum}
Since the above learning models need vectors (except for \kdeexp which is fed with 
original input) as input, we use the same feature extraction as our models on \textsf{ED} 
and \textsf{JC}. On \textsf{HM} and \textsf{EU}, they are fed with original 
vectors. As for other deep learning 
models~\cite{kipf2018learned,ortiz2018learning,cardlocaldeepmodel,treedeepmodel,plandeeppredict,neoqueryoptimizer,surveycarddeeplearning,autogressivemultiattr,autoregressivelikelihood}, 
when adapted for our problem, \cite{kipf2018learned} becomes a feature extraction by deep 
set~\cite{zaheer2017deep} plus a regression by \dnnexp, while the others become exactly 
\dnnexp. We will show that deep set is outperformed by our feature extraction 
(Table~\ref{tab:components}, also observed when applying to other methods). Hence these 
models are not repeatedly compared. 
Among the compared models, \spestexp, \xgbexp, \lightgbmexp, \kdeexp, \dlnexp, and our 
models are monotonic. 
\fullversion{Among the models involving FNNs, \moeexp and \hierdnnexp are more complex than 
ours in most cases, depending on the number of FNNs and other hyperparameter tuning. 
\dnnexp is less complex than ours. \dnnsexp is more complex.} 


\subsubsection{Hyperparameter Tuning} \label{sec:exp-setup-tuning}
\confversion{We use 256 hash functions for Jaccard 
distance and 512 hash functions for Euclidean distance.} 
\fullversion{We use 256 hash functions for Jaccard 
distance, 256 (on \gloveone) and 512 (on \glovetwo) hash functions 
for Euclidean distance.} 
The \vae is a fully-connected neural network, with three hidden 
layers of 256, 128, and 128 nodes for both encoder and decoder. The activation function 
is \textsf{ELU}, in line with \cite{kingma2013auto}. The dimensionality of the \vae's 
output is 40, 128, 128, 128, 64, 64, 64, 32 as per the order in 
Table~\ref{tab:dataset}. We use a fully-connected neural network with four hidden layers 
of 512, 512, 256, and 256 nodes for both $\Phi$ and $\Phi'$. 
The activation function is \textsf{ReLU}. The dimensionality of distance embeddings is 5. 
The dimensionality of $\mathbf{z}_x^i$ is 60. 
We set $\lambda$ in Equation~\ref{eq:monomodel1:loss} and $\lambda_{\Delta}$ in 
Equation~\ref{eq:monomodel1:lossnn} to both 0.1. 
The \vae is trained for 100 epochs. Our models are trained for 800 epochs. 

\subsubsection{Environments} \label{sec:exp-setup-environments}
The experiments were carried out on a server with a Intel Xeon E5-2640 @2.40GHz 
CPU and 256GB RAM running Ubuntu 16.04.4 LTS. 
Non-deep models were implemented in C++. Deep models were trained in Tensorflow, 
and then the parameters were copied to C++ implementations for a fair comparison 
of estimation efficiency, in line with \cite{kraska2018case}.


\confversion{
\begin{figure} [t]
  \centering
  \subfigure[\textsf{MSE}, \imagenet]{
    \includegraphics[width=0.46\linewidth]{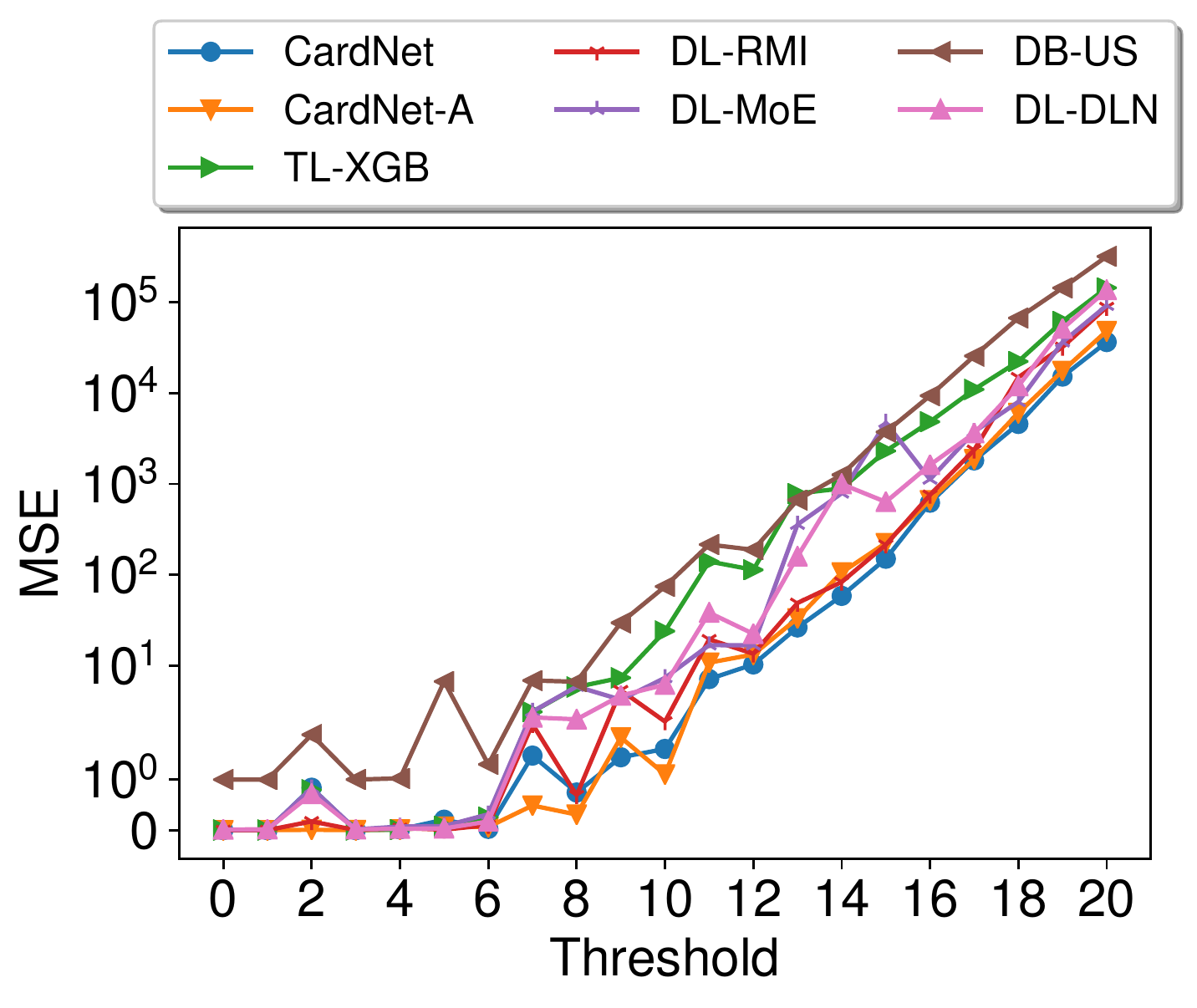}
    \label{fig:exp-mse-gist-tau}
  }
  \subfigure[\textsf{MSE}, \aminer]{
    \includegraphics[width=0.46\linewidth]{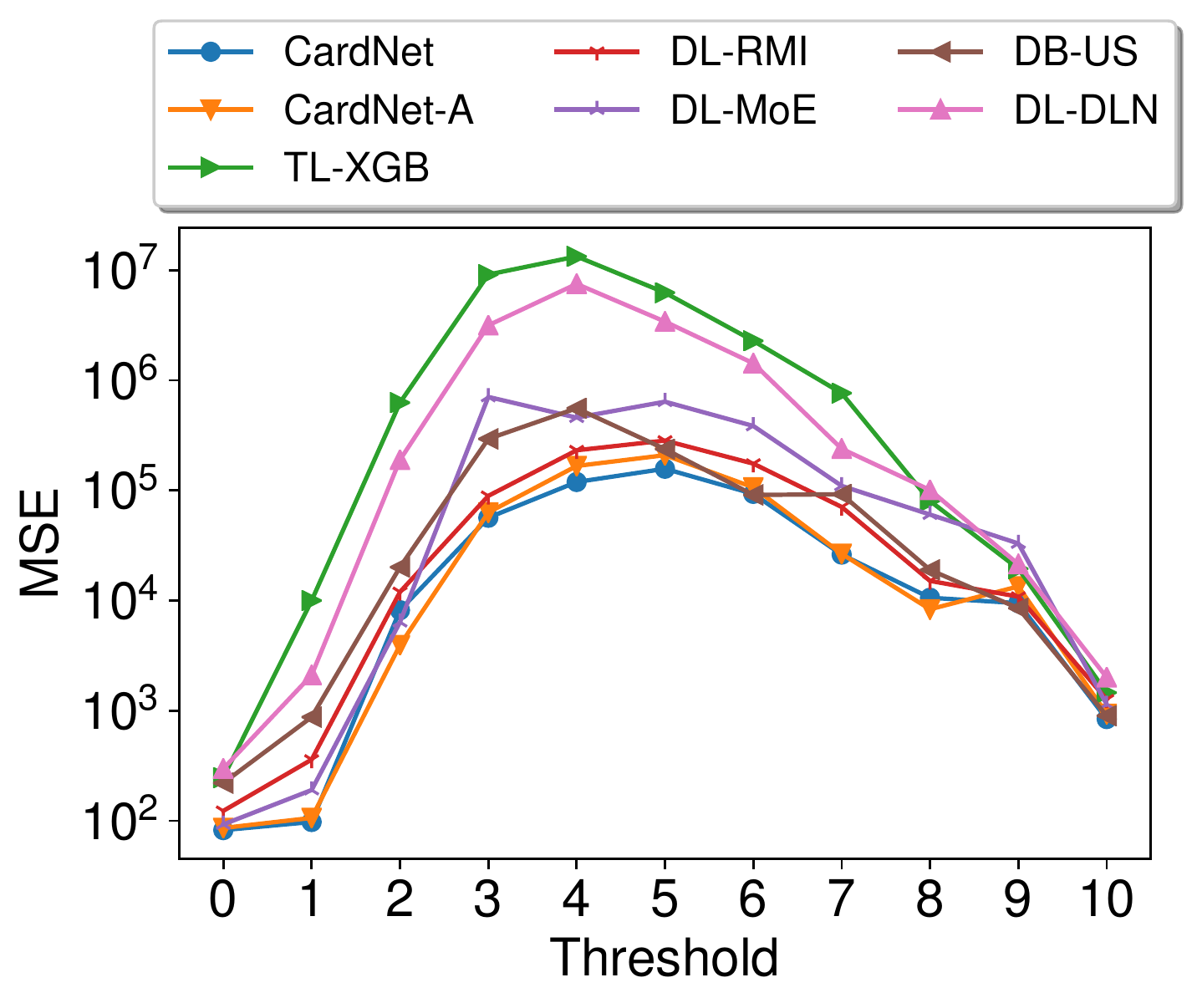}
    \label{fig:exp-mse-aminer-tau}
  }
  \subfigure[\textsf{MSE}, \bmsjacc]{
    \includegraphics[width=0.46\linewidth]{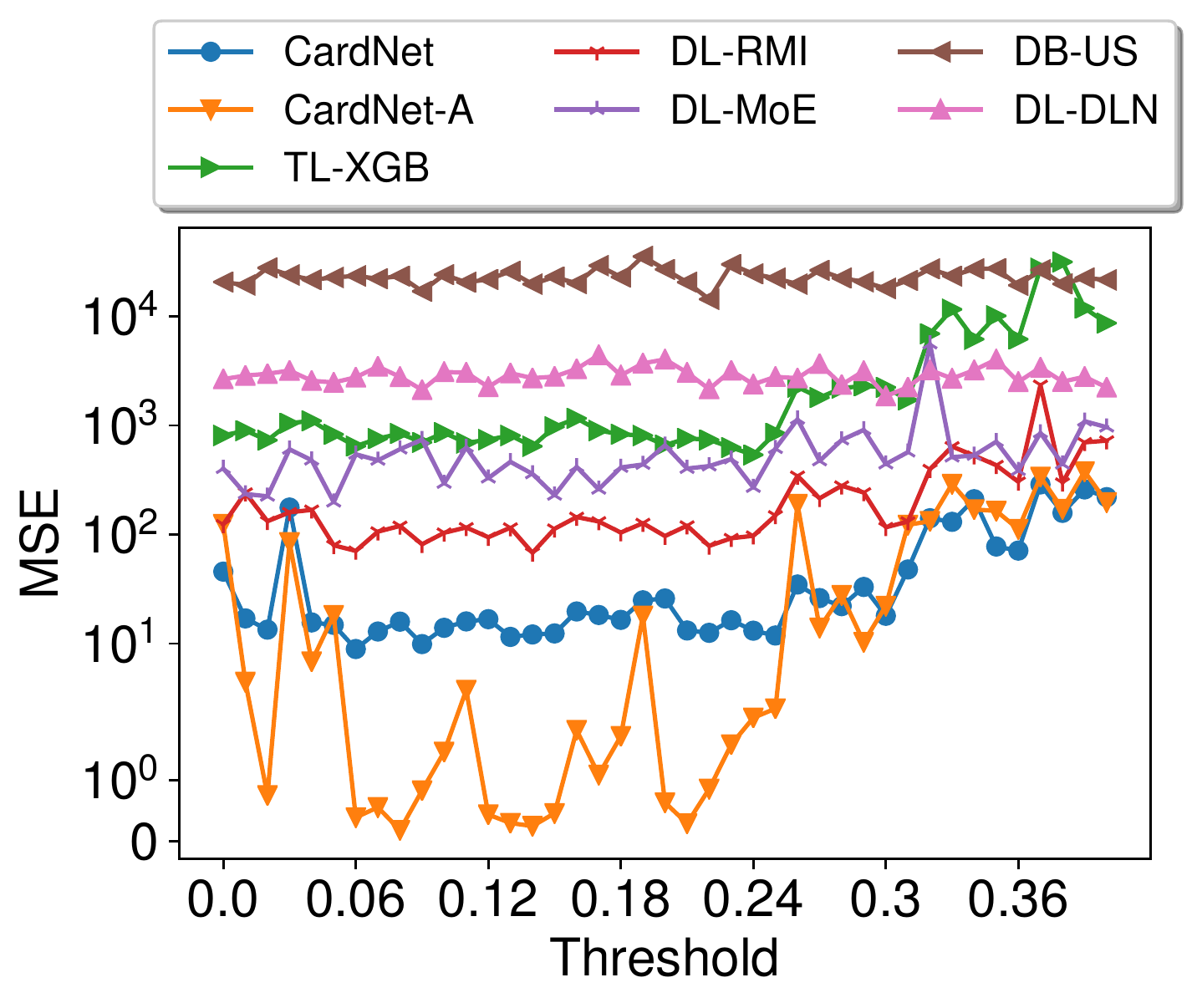}
    \label{fig:exp-mse-bmsjacc-tau}
  }
  \subfigure[\textsf{MSE}, \glovetwo]{
    \includegraphics[width=0.46\linewidth]{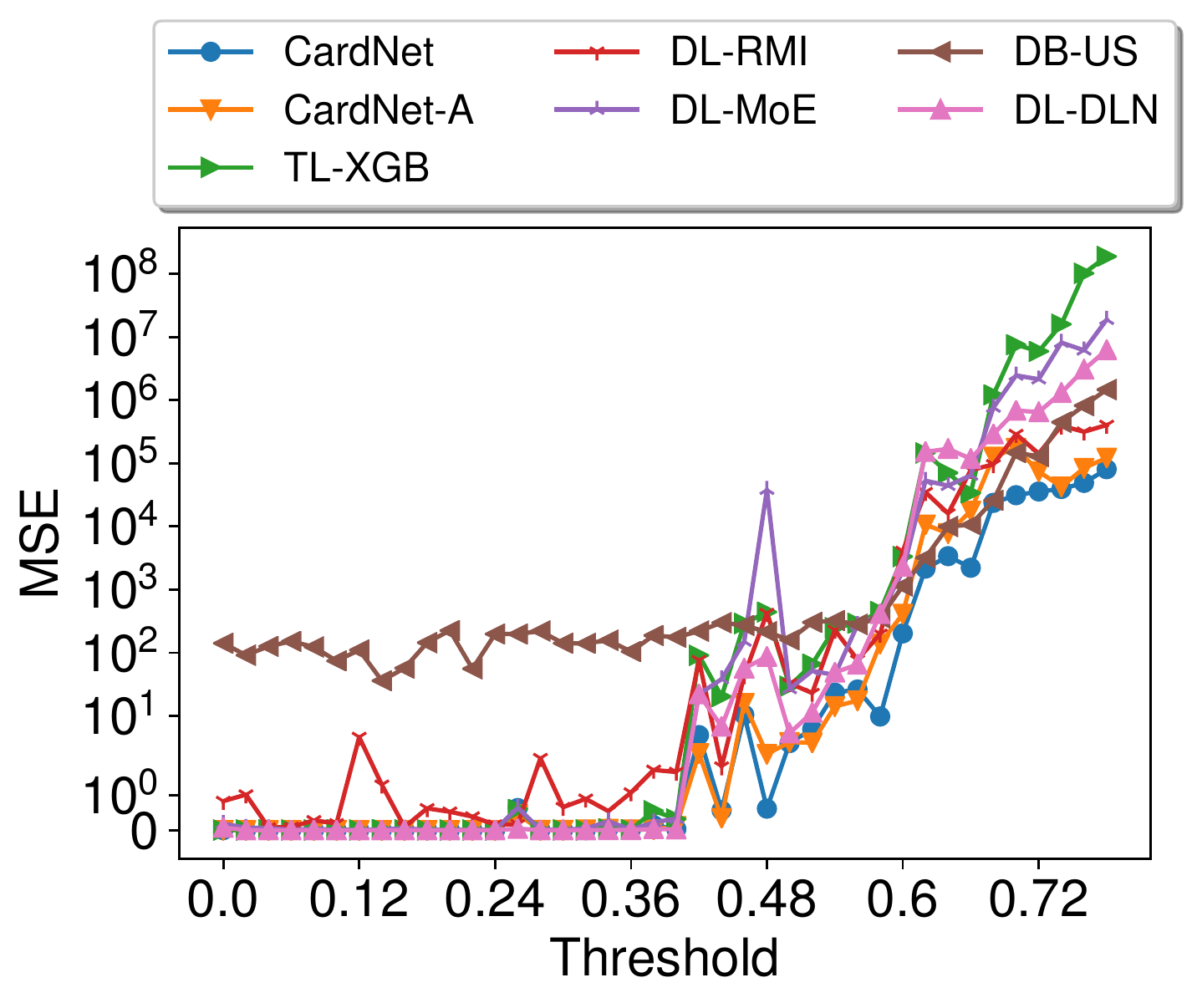}
    \label{fig:exp-mse-glove300-tau}
  }
  \caption{Accuracy v.s. threshold.}
  \label{fig:exp-error-tau}
\end{figure}
}

\fullversion{
\begin{figure} [t]
  \centering
  \subfigure[\textsf{MSE}, \imagenet]{
    \includegraphics[width=0.46\linewidth]{exp-figs/imagenet_mse_thresholds.pdf}
    \label{fig:exp-mse-gist-tau}
  }
  \subfigure[\textsf{MAPE}, \imagenet]{
    \includegraphics[width=0.46\linewidth]{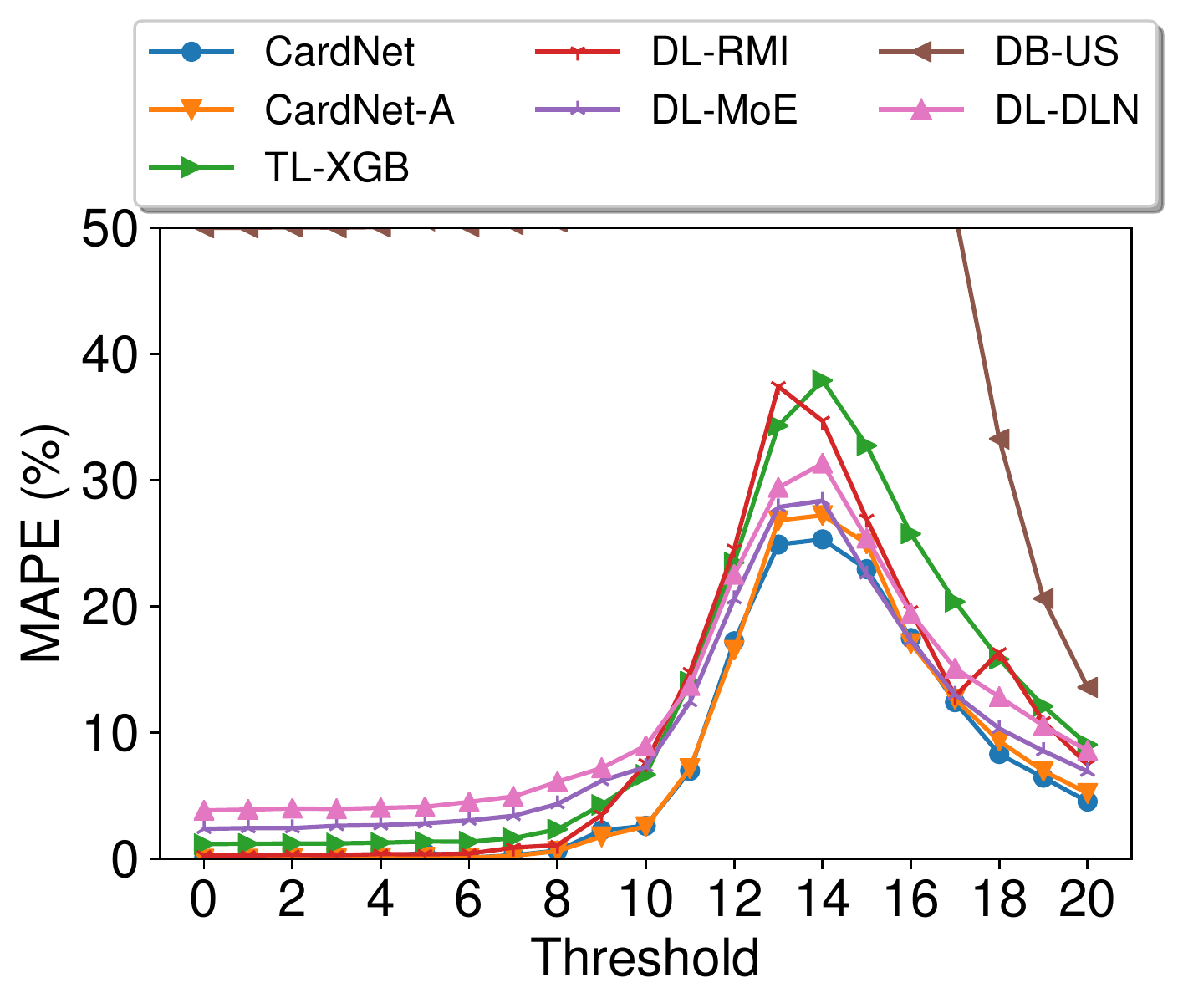}
    \label{fig:exp-mape-gist-tau}
  }  
  \subfigure[\textsf{MSE}, \aminer]{
    \includegraphics[width=0.46\linewidth]{exp-figs/aminer_mse_thresholds.pdf}
    \label{fig:exp-mse-aminer-tau}
  }
  \subfigure[\textsf{MAPE}, \aminer]{
    \includegraphics[width=0.46\linewidth]{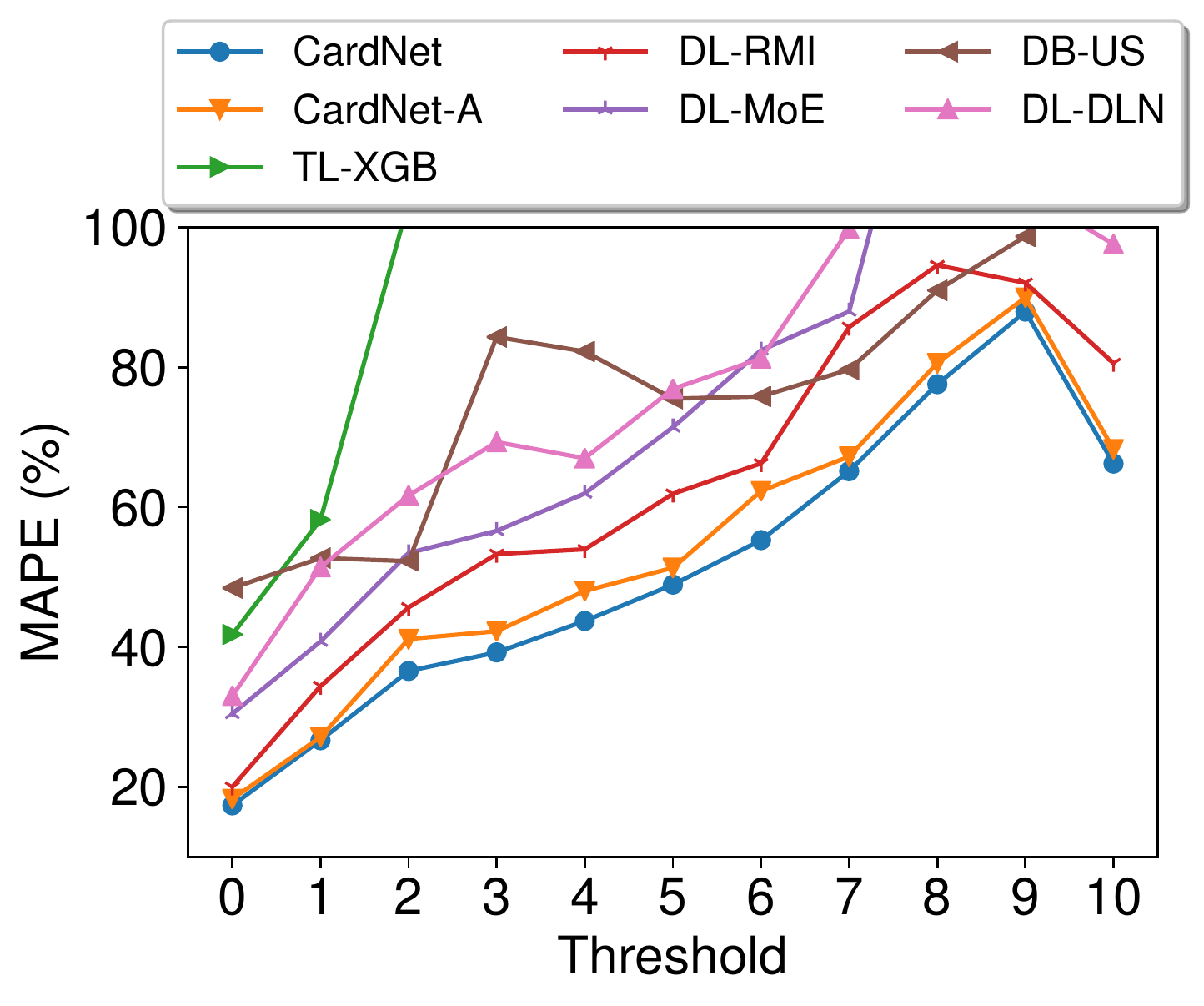}
    \label{fig:exp-mape-aminer-tau}
  }
  \subfigure[\textsf{MSE}, \bmsjacc]{
    \includegraphics[width=0.46\linewidth]{exp-figs/bms_jacc_mse_thresholds.pdf}
    \label{fig:exp-mse-bmsjacc-tau}
  }
  \subfigure[\textsf{MAPE}, \bmsjacc]{
    \includegraphics[width=0.46\linewidth]{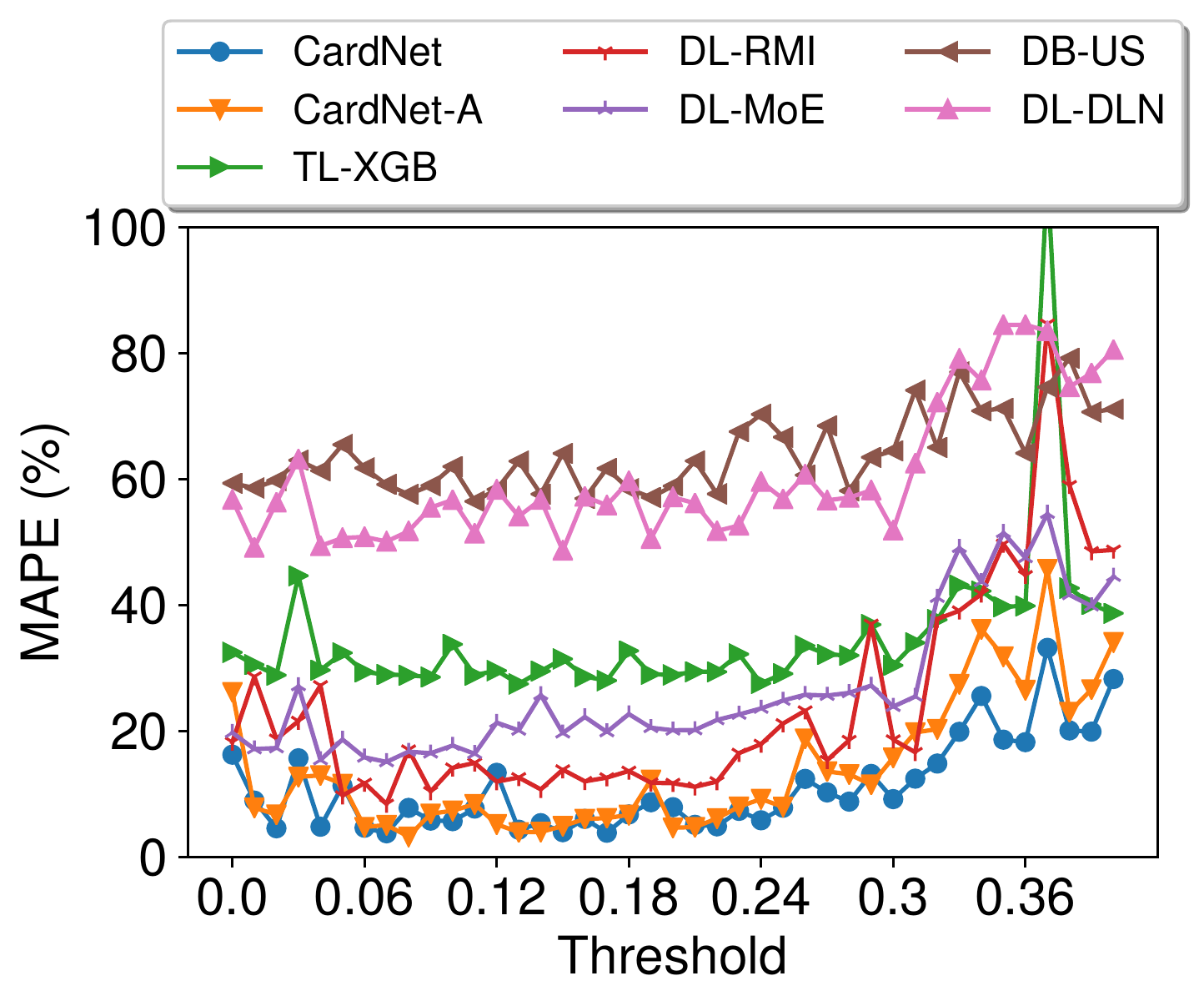}
    \label{fig:exp-mape-bmsjacc-tau}
  }
  \subfigure[\textsf{MSE}, \glovetwo]{
    \includegraphics[width=0.46\linewidth]{exp-figs/glove_300_mse_thresholds.pdf}
    \label{fig:exp-mse-glove300-tau}
  }
  \subfigure[\textsf{MAPE}, \glovetwo]{
    \includegraphics[width=0.46\linewidth]{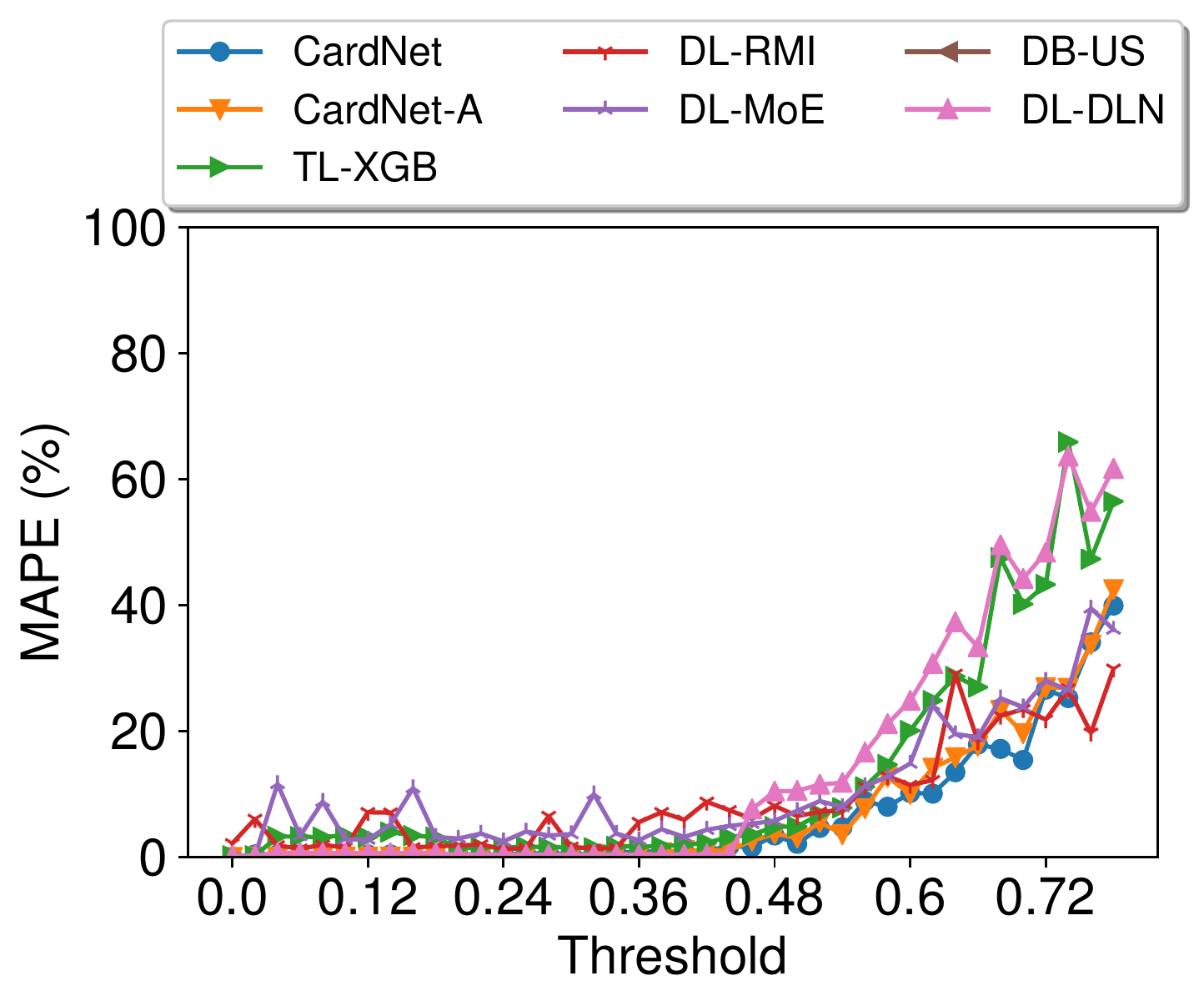}
    \label{fig:exp-mape-glove300-tau}
  }


  \caption{Accuracy v.s. threshold.}
  \label{fig:exp-error-tau}
\end{figure}
}

\subsection{Estimation Accuracy} \label{sec:exp-accuracy}

We report the accuracies of various models in Tables~\ref{tab:hm:errors-mse} 
and~\ref{tab:hm:errors-mape}, measured by \textsf{MSE} and \textsf{MAPE}. 
\modelone and \modeltwo report similar \mse and \mape. They achieve the best 
performance on almost all the datasets (except \modeltwo's \mape on \aminer), 
showcasing that the components in our model design collectively lead to better 
accuracy. 
For the four distance functions, the \mse (of the better one of our two models) 
is at least 1.5, 1.8, 4.1, and 2.1 times smaller than the best of the 
others, respectively. The \mape is reduced by at least 23.2\%, 2.7\%, 
25.6\%, and 21.2\% from the best of the others, respectively. 

In general, deep learning methods are more accurate than database and 
traditional learning methods. Among the deep learning methods, 
\dlnexp's performance is the worst, \moeexp is in the middle, and
\hierdnnexp is the runner-up to our models. 
The performance of \hierdnnexp relies on the models on upper levels. Although 
the neural networks on upper 
levels discretize output space into multiple regions, they tend to mispredict 
the cardinalities that are closest to the region boundaries. 
\dnnexp does not deliver good accuracy, and \dnnsexp is even worse than 
in some cases due to overfitting. This suggests that simply feeding deep 
neural networks with training data yields limited performance gain. 
\lstmexp and \lstmaexp exhibit small \mape on \aminer, but are outperformed by 
\hierdnnexp in the other cases, suggesting they do not learn the semantics of edit 
distance very well. 

\confversion{
\begin{table} [t]
  \small
  \caption{Average estimation time (milliseconds).}
  \label{tab:hm:estitime}
  \centering
  \resizebox{\linewidth}{!}{  
  \begin{tabular}[b]{| l | c | c | c | c |}
    \hline%
    Model & \imagenet & \aminer & \bmsjacc & \glovetwo\\
    \hline%
    \textsf{SimSelect} & 5.12 & 6.22 & 4.24 & 14.60\\
    \spestexp & 6.20 & 7.64 & 4.67 & 8.45\\
    \usexp & 1.17 & 1.26& 1.75& 6.23\\
    \xgbexp & 0.41 & 0.36 & 0.71 & 0.69\\
    \lightgbmexp & 0.32 & 0.31 & 0.52 & 0.49\\
    \kdeexp & 0.83 & 4.73 & 0.97& 1.28\\
    \dlnexp & 0.42 & 0.83 & 0.66 & 1.23\\
    \moeexp & 0.21 & 0.35 & 0.31 & 0.41\\
    \hierdnnexp & 0.37 & 0.41 & 0.39 & 0.68\\
    \dnnexp & 0.09 & 0.15 & 0.11 & 0.15\\
    \dnnsexp & 0.26 & 0.26 & 0.27 & 0.42\\
    \lstmexp & - & 3.11 & - & -\\
    \lstmaexp & - & 3.46 & - & -\\
    \textbf{\modelone} & 0.36 & 0.39 & 0.55 & 0.67\\
    \textbf{\modeltwo} & 0.13 & 0.21 & 0.18 & 0.24\\
    \hline%
  \end{tabular}
  }
\end{table}

\begin{table*} [t]
  \small
  \caption{Performance of model components.}
  \label{tab:components}
  \centering
  \begin{tabular}[b]{| l | l | c | c | c | c | c | c | c | c |}
    \hline%
    Metric & Dataset & \multicolumn{2}{|c|}{Feature Extraction} & \multicolumn{2}{|c|}{Incremental Prediction} & \multicolumn{2}{|c|}{Variational Auto-encoder} & \multicolumn{2}{|c|}{Dynamic Training} \\\cline{3-10}
    & & \modelone & \modeltwo & \modelone & \modeltwo & \modelone & \modeltwo & \modelone & \modeltwo\\\hline    
    \hline%
    \multirow{4}{*}{$\gamma_{\mse}$} &
    \imagenet & - & - & 84\% & 86\% & 13\% & 17\% & 20\% & 28\%\\ 
    & \aminer & 49\% & 44\% & 57\% & 62\% & 11\% & 14\% & 14\% & 21\%\\ 
    & \bmsjacc & 31\% & 34\% & 83\% & 76\% & 34\% & 26\% & 21\% & 28\%\\ 
    & \glovetwo & 5\% & 8\% & 93\% & 82\% & 18\% & 23\% & 26\% & 17\% \\
    \hline%
    \hline%
    \multirow{4}{*}{$\gamma_{\mape}$} &
    \imagenet & - & - & 47\% & 46\% & 14\% & 16\% & 15\% & 16\%\\ 
    & \aminer & 5\% & 6\% & 52\% & 51\% & 19\% & 18\% & 18\% & 22\%\\ 
    & \bmsjacc & 26\% & 16\% & 51\% & 60\% & 40\% & 29\% & 22\% & 34\%\\ 
    & \glovetwo & 32\% & 21\% & 54\% & 48\% & 23\% & 14\% & 32\% & 27\%\\ 
    \hline%
  \end{tabular}
\end{table*}
}
\fullversion{
\begin{table*} [t]
  \small
  \caption{Average estimation time (milliseconds).}
  \label{tab:hm:estitime}
  \centering
  \begin{tabular}[b]{| l | c | c | c | c | c | c | c | c |}
    \hline%
    Model & \imagenet & \pubchem & \aminer & \dblped & \bmsjacc & \dblpjacclong & \glovetwo & \gloveone\\
    \hline%
    \textsf{SimSelect} & 5.12 & 14.68 & 6.22 & 10.51 & 4.24 & 5.89 & 14.60 & 8.52\\
    \spestexp & 6.20 & 8.50 & 7.64 & 10.01 & 4.67 & 5.78 & 8.45 & 7.34\\
    \usexp & 1.17 & 3.60 & 1.26 & 6.08 & 1.75 & 1.44 & 6.23 & 1.05\\
    \xgbexp & 0.41 & 0.41 & 0.36 & 0.41 & 0.71 & 0.65 & 0.69 & 0.60\\
    \lightgbmexp & 0.32 & 0.34 & 0.31 & 0.33 & 0.52 & 0.48 & 0.49 & 0.47\\
    \kdeexp & 0.83 & 0.96 & 4.73 & 1.24 & 0.97 & 2.35 & 1.28 & 1.22\\
    \dlnexp & 0.42 & 0.84 & 0.83 & 6.43 & 0.66 & 0.57 & 1.23 & 0.46\\
    \moeexp & 0.21 & 0.32 & 0.35 & 0.59 & 0.31 & 0.36 & 0.41 & 0.28\\
    \hierdnnexp & 0.37 & 0.46 & 0.41 & 0.57 & 0.39 & 0.45 & 0.68 & 0.57\\
    \dnnexp & 0.09 & 0.11 & 0.15 & 0.25 & 0.11 & 0.14 & 0.15 & 0.12\\
    \dnnsexp & 0.26 & 0.58 & 0.26 & 0.62 & 0.27 & 0.34 & 0.42 & 0.38\\
    \lstmexp & - & - & 3.11 & 5.22 & - & - & - & -\\
    \lstmaexp & - & - & 3.46 & 5.80 & - & - & - & -\\
    \textbf{\modelone} & 0.36 & 0.45 & 0.39 & 0.69 & 0.55 & 0.48 & 0.67 & 0.50\\
    \textbf{\modeltwo} & 0.13 & 0.19 & 0.21 & 0.29 & 0.18 & 0.20 & 0.24 & 0.19\\
    \hline%
  \end{tabular}
\end{table*}

\begin{table*} [t]
  \small
  \caption{Performance of model components.}
  \label{tab:components}
  \centering
  \begin{tabular}[b]{| l | l | c | c | c | c | c | c | c | c |}
    \hline%
    Metric & Dataset & \multicolumn{2}{|c|}{Feature Extraction} & \multicolumn{2}{|c|}{Incremental Prediction} & \multicolumn{2}{|c|}{Variational Auto-encoder} & \multicolumn{2}{|c|}{Dynamic Training} \\\cline{3-10}
    & & \modelone & \modeltwo & \modelone & \modeltwo & \modelone & \modeltwo & \modelone & \modeltwo\\\hline    
    \hline%
    \multirow{4}{*}{$\gamma_{\mse}$} &
    \imagenet & - & - & 84\% & 86\% & 13\% & 17\% & 20\% & 28\%\\ 
    & \aminer & 49\% & 44\% & 57\% & 62\% & 11\% & 14\% & 14\% & 21\%\\ 
    & \bmsjacc & 31\% & 34\% & 83\% & 76\% & 34\% & 26\% & 21\% & 28\%\\ 
    & \glovetwo & 5\% & 8\% & 93\% & 82\% & 18\% & 23\% & 26\% & 17\% \\
    \hline%
    \hline%
    \multirow{4}{*}{$\gamma_{\mape}$} &
    \imagenet & - & - & 47\% & 46\% & 14\% & 16\% & 15\% & 16\%\\ 
    & \aminer & 5\% & 6\% & 52\% & 51\% & 19\% & 18\% & 18\% & 22\%\\ 
    & \bmsjacc & 26\% & 16\% & 51\% & 60\% & 40\% & 29\% & 22\% & 34\%\\ 
    & \glovetwo & 32\% & 21\% & 54\% & 48\% & 23\% & 14\% & 32\% & 27\%\\ 
    \hline%
  \end{tabular}
\end{table*}
}

In Figure~\ref{fig:exp-error-tau}, we evaluate the accuracy \confversion{(\mse)}
with varying thresholds\fullversion{ on the four default datasets}. 
We compare with the following models: \usexp, \xgbexp, \dlnexp, \moeexp, \hierdnnexp, 
the more accurate or monotonic models out of each category. The general trend 
is that the errors increase with the threshold, meaning that larger thresholds are harder. 
\confversion{The exception is the edit distance.}
\fullversion{The exceptions are \mape on Hamming distance and \mse on edit distance.} 
The reason is that the 
cardinalities of some large thresholds tend to resemble for different queries, 
and regression models are more likely to predict well.

\fullversion{
}

\subsection{Estimation Efficiency}

In Table~\ref{tab:hm:estitime}, we show the average estimation time. 
We also report the time of running a state-of-the-art 
similarity selection algorithm~\cite{DBLP:journals/pvldb/QinX18,fastercovertree} to 
process queries to obtain the cardinality 
(referred to as \textsf{SimSelect}). The estimation time of \modelone 
is close to \hierdnnexp and faster than the database methods and \kdeexp. Thanks to 
the acceleration technique, \modeltwo becomes the runner-up, 
and its speed is close to the fastest model \dnnexp. \modeltwo is faster than running 
the similarity selection algorithms by at least 24 times. 

\subsection{Evaluation of Model Components}
\label{sec:exp:comp}
We evaluate the following components in our models: feature extraction, incremental 
prediction, variational auto-encoder (\vae), and dynamic training strategy. We use 
the following radio to measure the improvement by each component: 
\begin{align*}
  \gamma_{\xi} = 
  \frac{\xi(\modelpart) - \xi(\textsf{CardNet\{-A\}})}{\xi(\modelpart)}, 
\end{align*}
\confversion{where $\xi \in \set{\mape, \mse}$, }
\fullversion{where $\xi \in \set{\mape, \mse\, \text{mean q-error}}$, }
\textsf{CardNet\{-A\}} is our model \modelone or \modeltwo, and $\modelpart$
is \textsf{CardNet\{-A\}} with component C replaced 
by other options; e.g., $\modelcom{\vae}$ 
is our model with \vae replaced. A positive $\gamma_{\xi}$ means 
the component has a positive effect in accuracy. 

We consider the following replacement options: 
\begin{inparaenum} [(1)]
  \item For feature extraction, we adopt a character-level bidirectional LSTM to 
  transfer a string to a dense representation for edit distance, a deep set 
  model~\cite{zaheer2017deep} to transfer a set to its representation for Jaccard 
  distance, and use the original record as input for Euclidean distance. Hamming 
  distance is not repeatedly tested as we use original vectors as input. 
  \item For incremental prediction, we compare it with a deep neural network 
  that takes as input the concatenation of $\mathbf{x'}$ and the embedding of $\tau$ 
  and outputs the cardinality. 
  \item For \vae, we compare it with an option that directly concatenates 
  the binary representation and distance embeddings. 
  \item For dynamic training, we compare it with using only \msle as loss, 
  i.e., removing the second term on the right side of Equation~\ref{eq:monomodel1:lossnn}. 
\end{inparaenum}
We report the $\gamma_{\xi}$ values \fullversion{on the four default datasets} in Table~\ref{tab:components}. 
The effects of the four components in our models are all positive, ranging from 5\% to 93\% 
improvement of \mse 
\confversion{and 5\% to 60\% improvement of \mape.} 
\fullversion{, 5\% to 60\% improvement of \mape, and 9\% to 64\% improvement of mean q-error.} 
The most useful component is 
incremental prediction, with \confversion{46\%} \fullversion{38\%} 
to 93\% performance improvement. This demonstrates 
that using incremental prediction on deep neural networks is significantly better than 
directly feeding neural networks with training data, in accord with what we have observed 
in Tables~\ref{tab:hm:errors-mse} and~\ref{tab:hm:errors-mape}.

\fullversion{
\subsection{Number of Decoders}
In Figure~\ref{fig:exp-granu}, we evaluate the accuracy by varying the number of 
decoders. In order to show the trend clearly, we use four datasets with large 
lengths or dimensionality, whose statistics is given in Table~\ref{tab:largedataset}. 
As seen from the experimental results, we discover that using the largest 
$\tau_{\max}$ setting does not always lead to the best performance. E.g., 
on \youtube, the best performance is achieved when $\tau_{max} = 326$ (327 decoders).
When there are too few decoders, the feature extraction becomes lossy and cannot 
successfully capture the semantic information of the original distance functions. 
As the number of decoders increases, the feature extraction becomes more effective 
to capture the semantics. On the other hand, the performance drops if we use an 
excessive number of decoders. This is because given a query, the cardinality only 
increases at a few thresholds (e.g., a threshold of 50 and 51 might produce the same 
cardinality). Using too many decoders will involve too many non-increasing points, 
posing difficulty in learning the regression model. 

\begin{table*} [t]
  \small
  \caption{Statistics of datasets with high dimensionality.}
  \label{tab:largedataset}
  \centering
  \resizebox{\linewidth}{!}{  
  \begin{tabular}[b]{| l | c | c | c | c | c | c | c | c | c |}
    \hline%
    \texttt{Dataset} & \texttt{Source} & \texttt{Process} & \texttt{Data Type} &
    \texttt{Attribute} & \texttt{\# Record }& $\ell_{max}$ & $\ell_{avg}$ &
    \texttt{Distance} & $\theta_{max}$ \\
    \hline%
    \youtube & \textsf{Youtube\_Faces}~\cite{URL:youtube-faces} 
    & normalize & real-valued vector & video & 346,194 & 1770 & 1770 & Euclidean & 0.8 \\     
    \hline%
    \gistlarge & \textsf{GIST}~\cite{URL:gist} 
    & \textsf{Spectral Hashing}~\cite{weiss2009spectral} 
    & binary vector & image & 982,677 & 2048 & 2048 & Hamming & 512 \\
    \hline%
    \dblped & \cite{URL:dblp} 
    & - & string & publication title & 1,000,000 & 199 & 72.49
    & edit & 20 \\
    \hline%
    \wikijacc & \textsf{Wikipedia}~\cite{URL:wikipedia} 
    & 3-gram & string & abstract & 1,150,842 & 732 & 496.06
    & Jaccard & 0.4 \\
    \hline%
  \end{tabular}
  }
\end{table*}

\begin{figure} [t]
  \centering
  \subfigure[\textsf{MSE}, \youtube]{
    \includegraphics[width=0.46\linewidth]{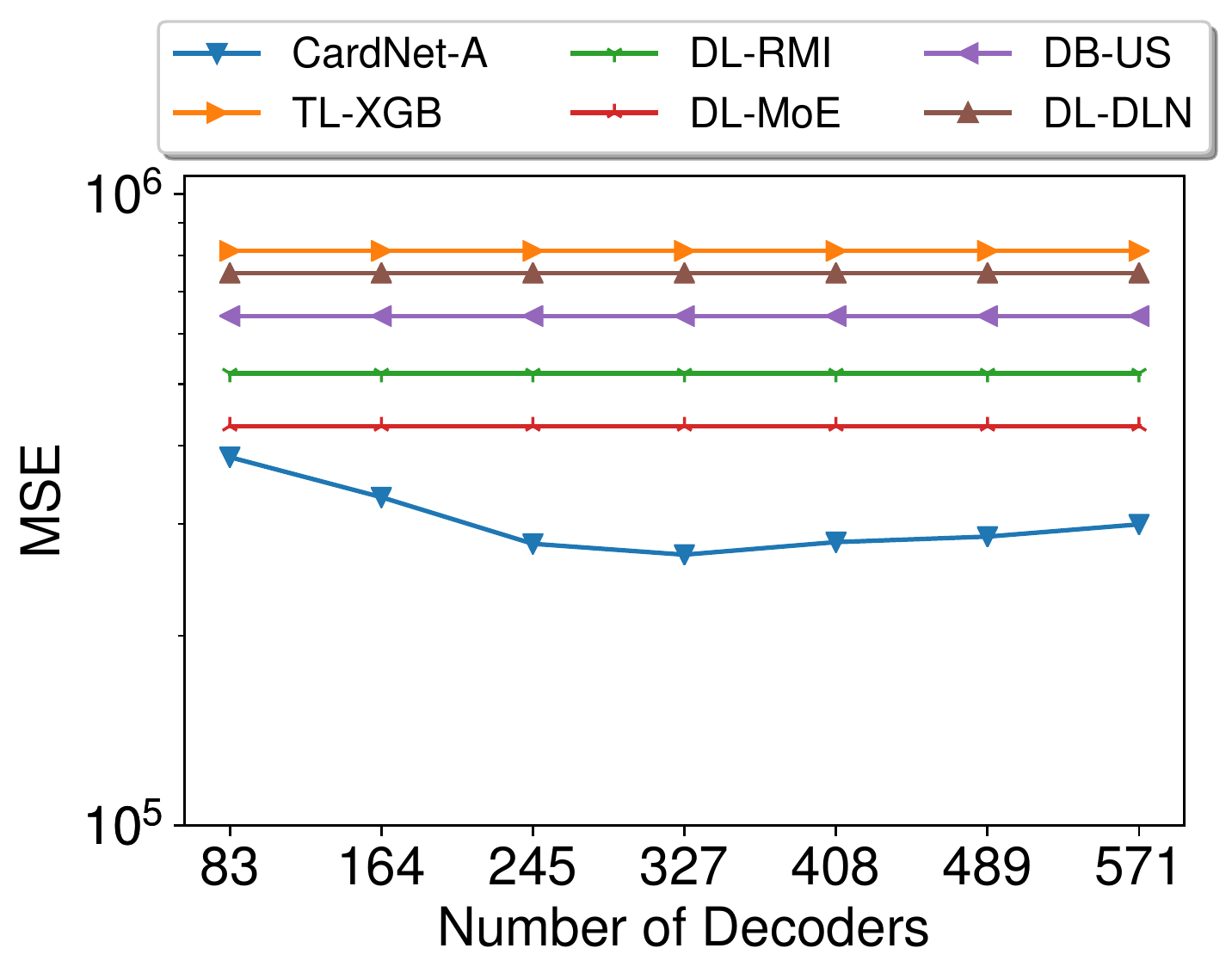}
    \label{fig:exp-granu-mse-youtube}
  }
  \subfigure[\textsf{MAPE}, \youtube]{
    \includegraphics[width=0.46\linewidth]{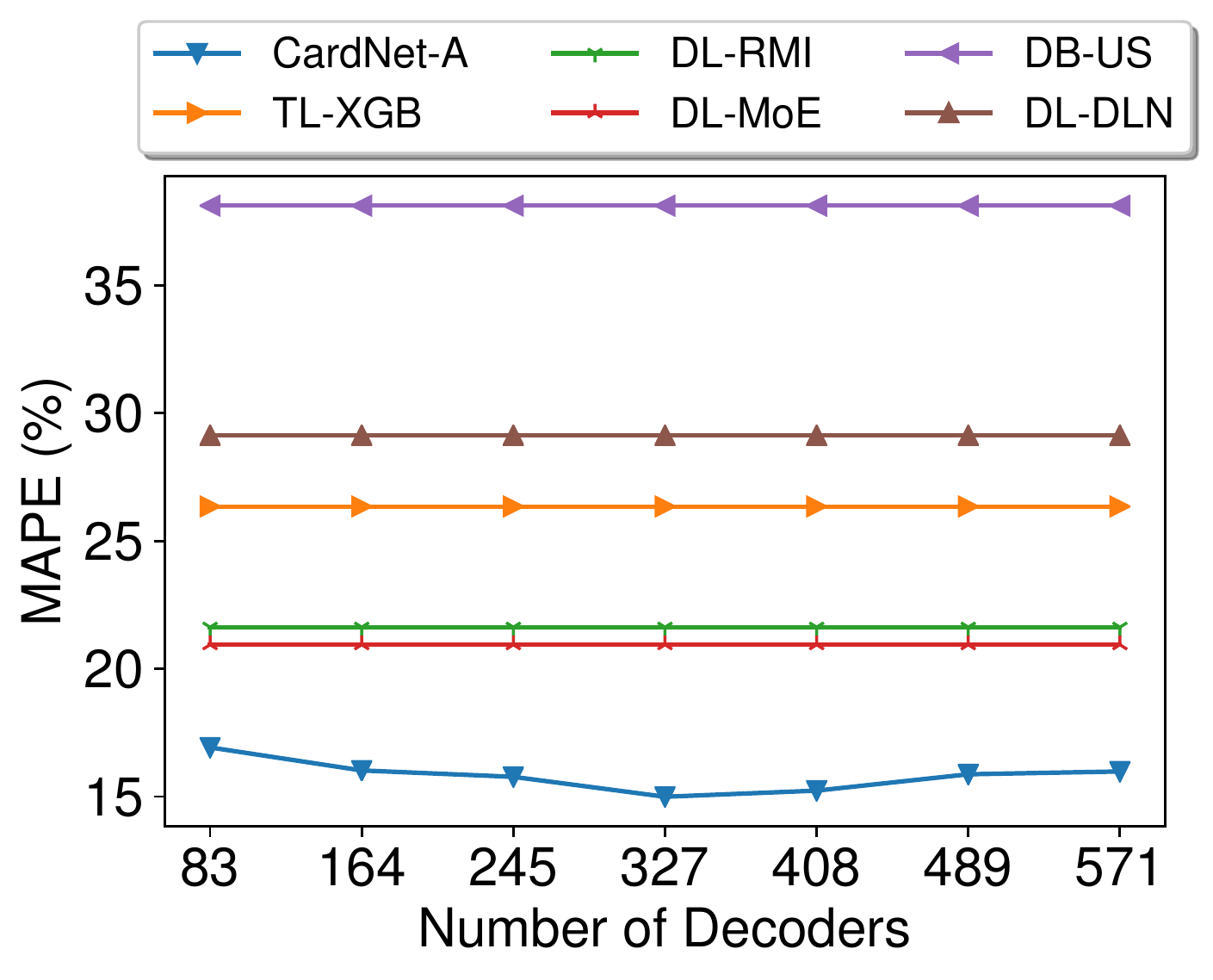}
    \label{fig:exp-granu-mape-youtube}
  }
  \subfigure[\textsf{MSE}, \gistlarge]{
    \includegraphics[width=0.46\linewidth]{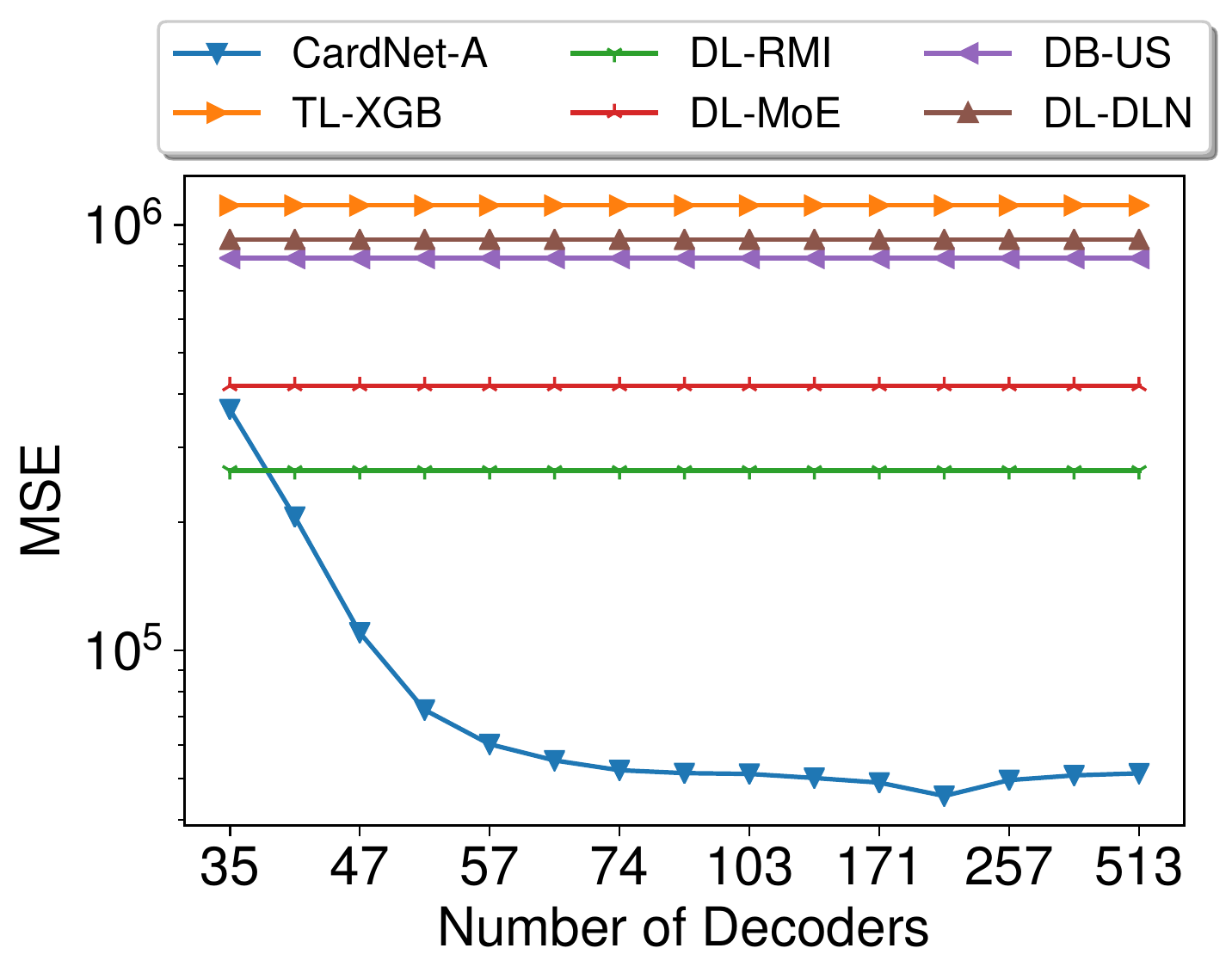}
    \label{fig:exp-granu-mse-gist}
  }
  \subfigure[\textsf{MAPE}, \gistlarge]{
    \includegraphics[width=0.46\linewidth]{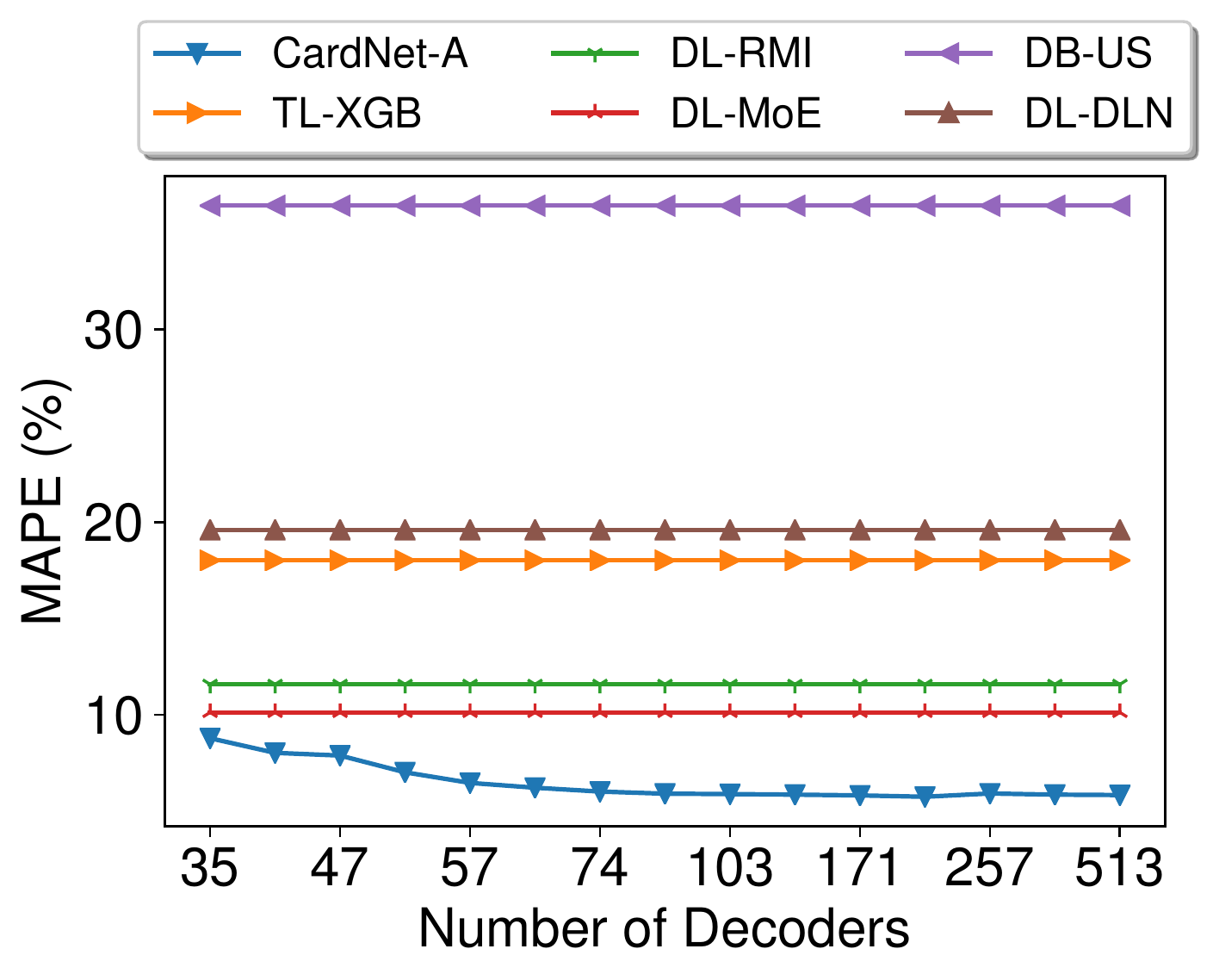}
    \label{fig:exp-granu-mape-gist}
  }
  \subfigure[\textsf{MSE}, \dblped]{
    \includegraphics[width=0.46\linewidth]{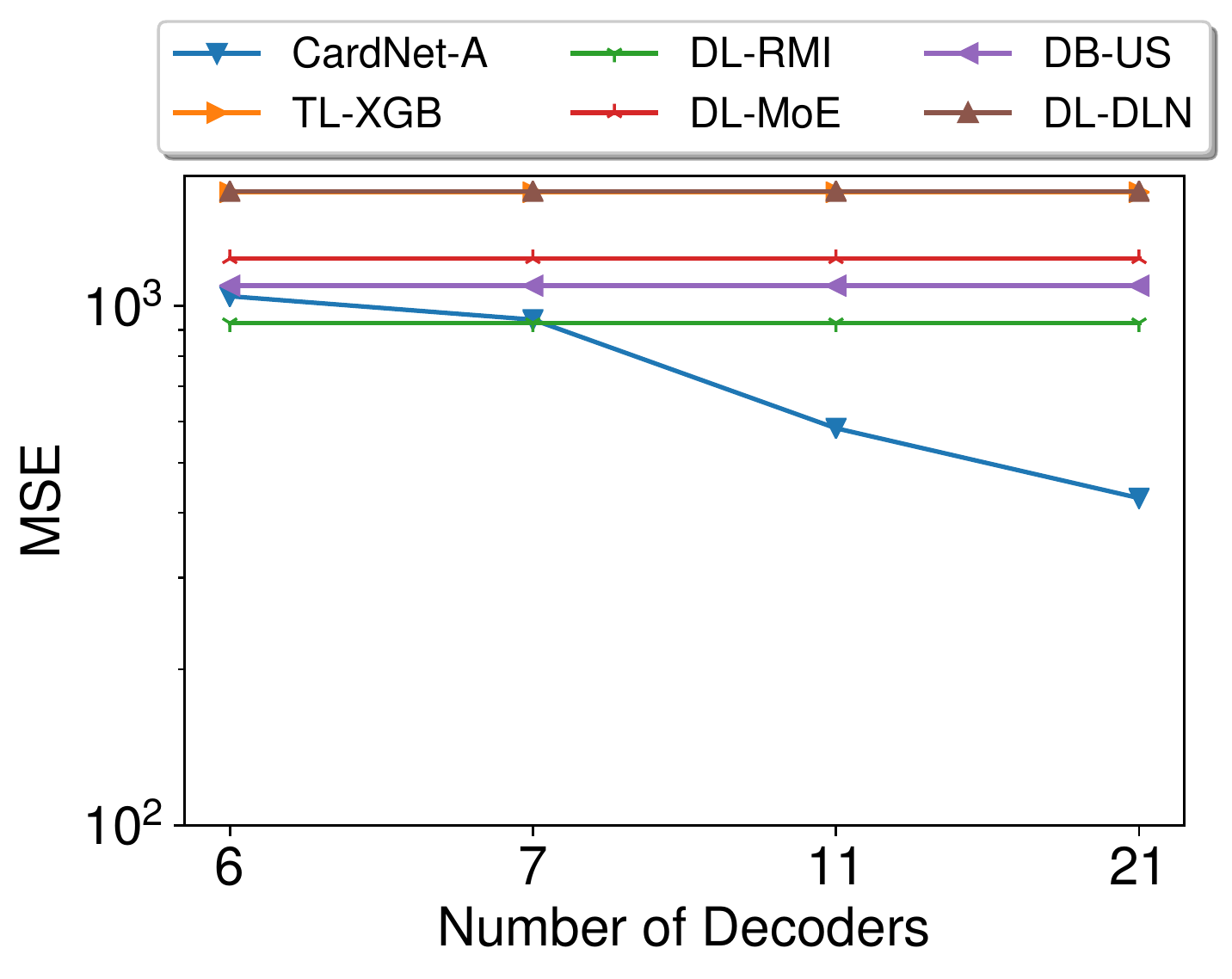}
    \label{fig:exp-granu-mse-wiki-ol}
  }
  \subfigure[\textsf{MAPE}, \dblped]{
    \includegraphics[width=0.46\linewidth]{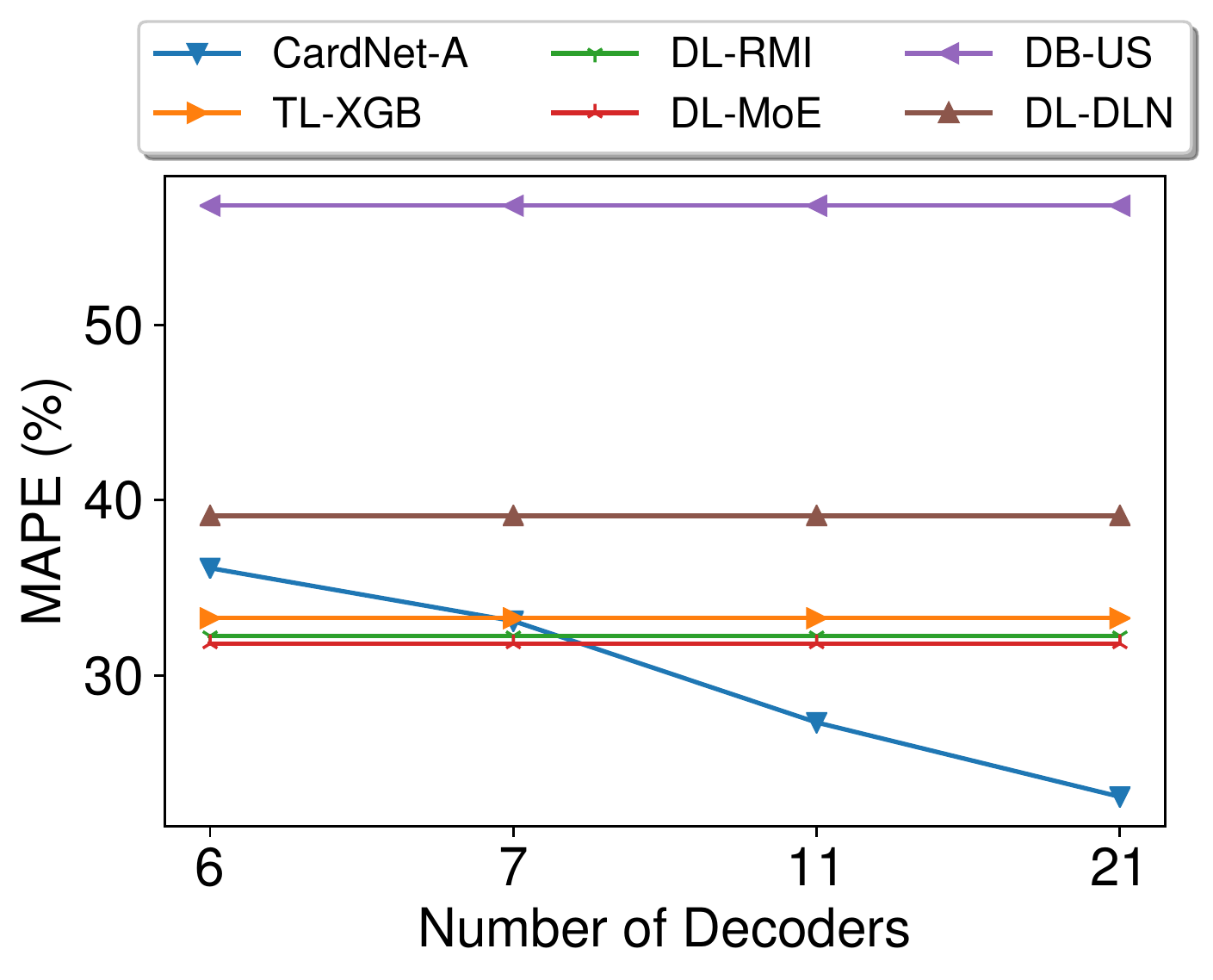}
    \label{fig:exp-granu-mape-wiki-ol}
  }
  \subfigure[\textsf{MSE}, \wikijacc]{
    \includegraphics[width=0.46\linewidth]{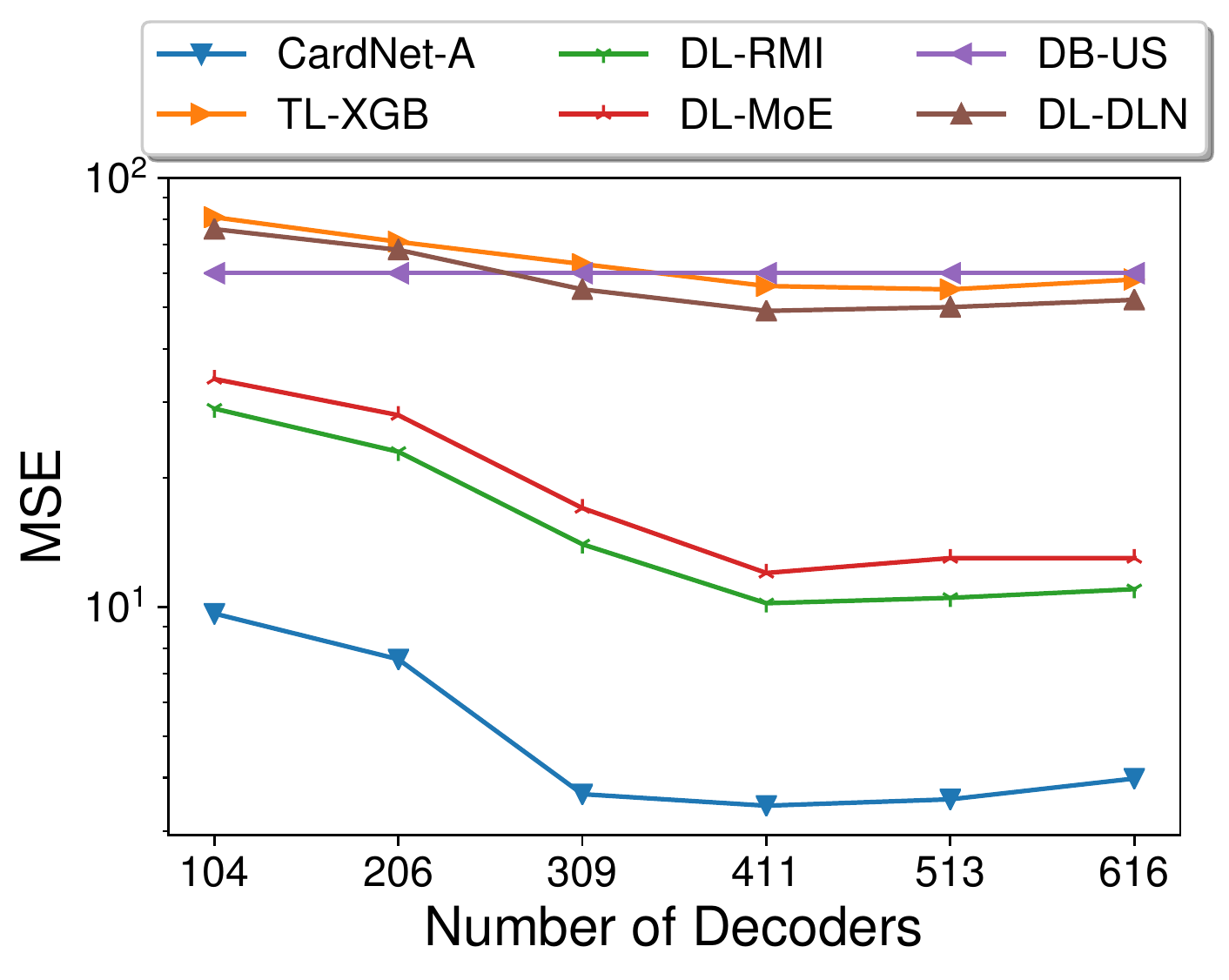}
    \label{fig:exp-granu-mse-wiki-ol}
  }
  \subfigure[\textsf{MAPE}, \wikijacc]{
    \includegraphics[width=0.46\linewidth]{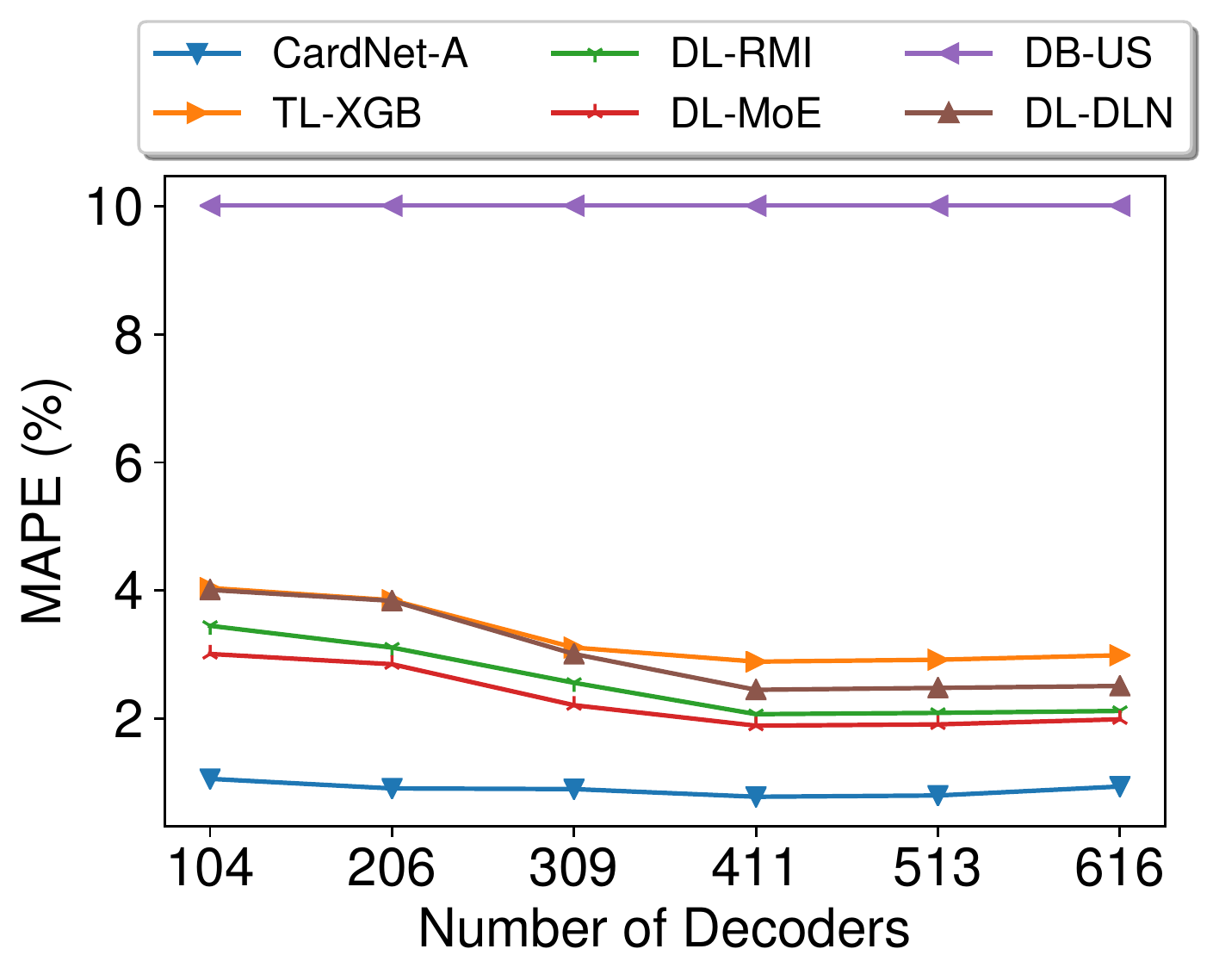}
    \label{fig:exp-granu-mape-wiki-ol}
  }
  \caption{Accuracy v.s. number of decoders.}
  \label{fig:exp-granu}
\end{figure}
}

\subsection{Model Size}
Table~\ref{tab:modelsize} shows the storage sizes of the competitors\fullversion{ on the four default
datasets}. \usexp does not need any storage and thus shows zero model size. \kdeexp has 
the smallest model size among the others, because it only stores the kernel instances 
for estimation. For deep learning models, \dnnexp has the smallest model size. Our model 
sizes range from 10 to 55 MB, smaller than the other deep models except \dnnexp. 

\subsection{Evaluation of Training}

\subsubsection{Training Time}
Table~\ref{tab:traintime} shows the training times of various models\fullversion{ on the four default
datasets}. Traditional learning models are faster to train. Our models spend 2 -- 4 hours, 
similar to other deep models. 
\dnnsexp is the slowest since its has $(\tau_{max} + 1)$ independently learned deep neural networks. 

\subsubsection{Varying the Size of Training Data}
\begin{figure} [t]
  \centering
  \subfigure[\textsf{MSE}, \imagenet]{
    \includegraphics[width=0.46\linewidth]{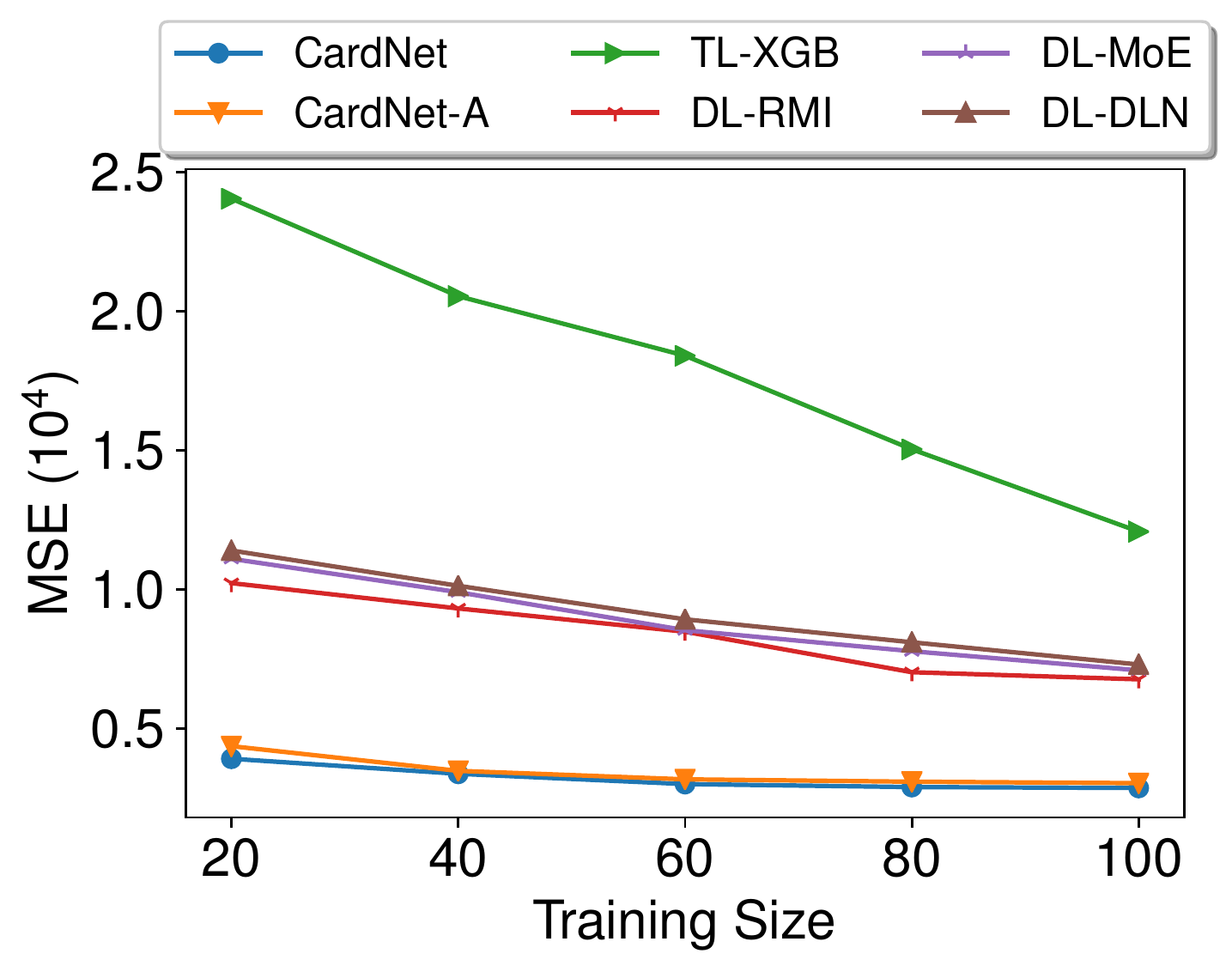}
    \label{fig:exp-imagenet-mse-trainsize}
  }
  \subfigure[\textsf{MSE}, \aminer]{
    \includegraphics[width=0.46\linewidth]{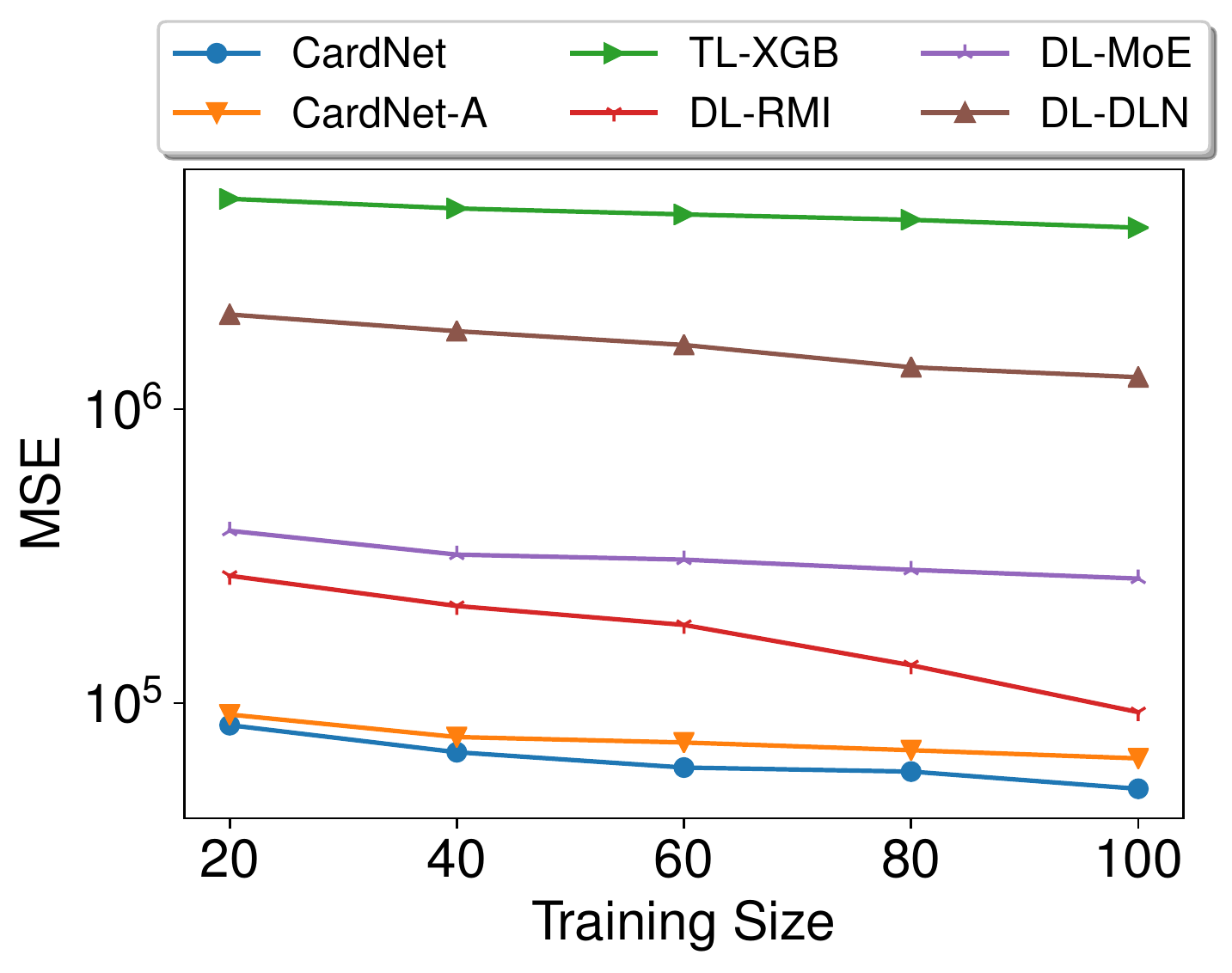}
    \label{fig:exp-aminer-mse-trainsize}
  }
  \subfigure[\textsf{MSE}, \bmsjacc]{
    \includegraphics[width=0.46\linewidth]{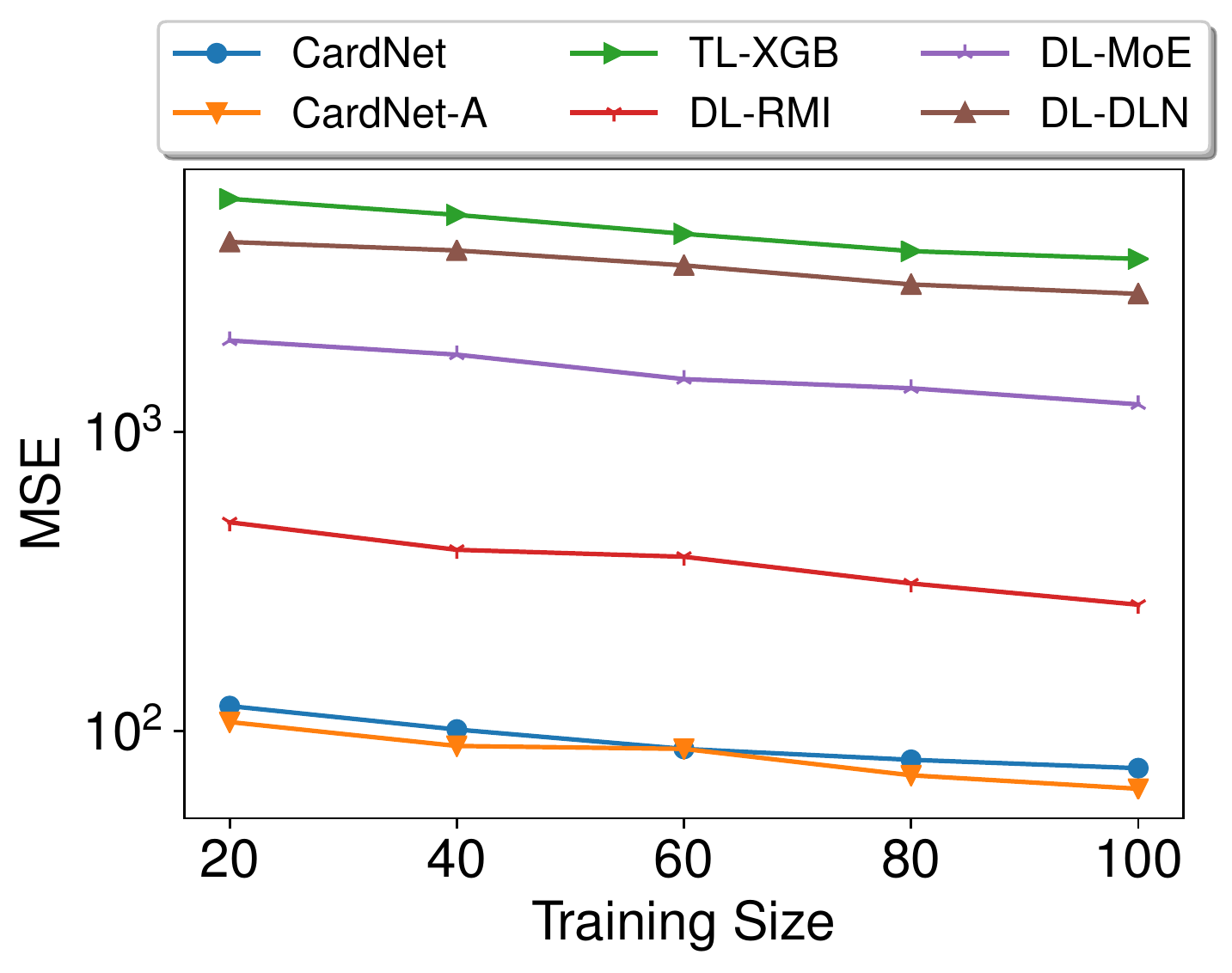}
    \label{fig:exp-bms-mse-trainsize}
  }
  \subfigure[\textsf{MSE}, \glovetwo]{
    \includegraphics[width=0.46\linewidth]{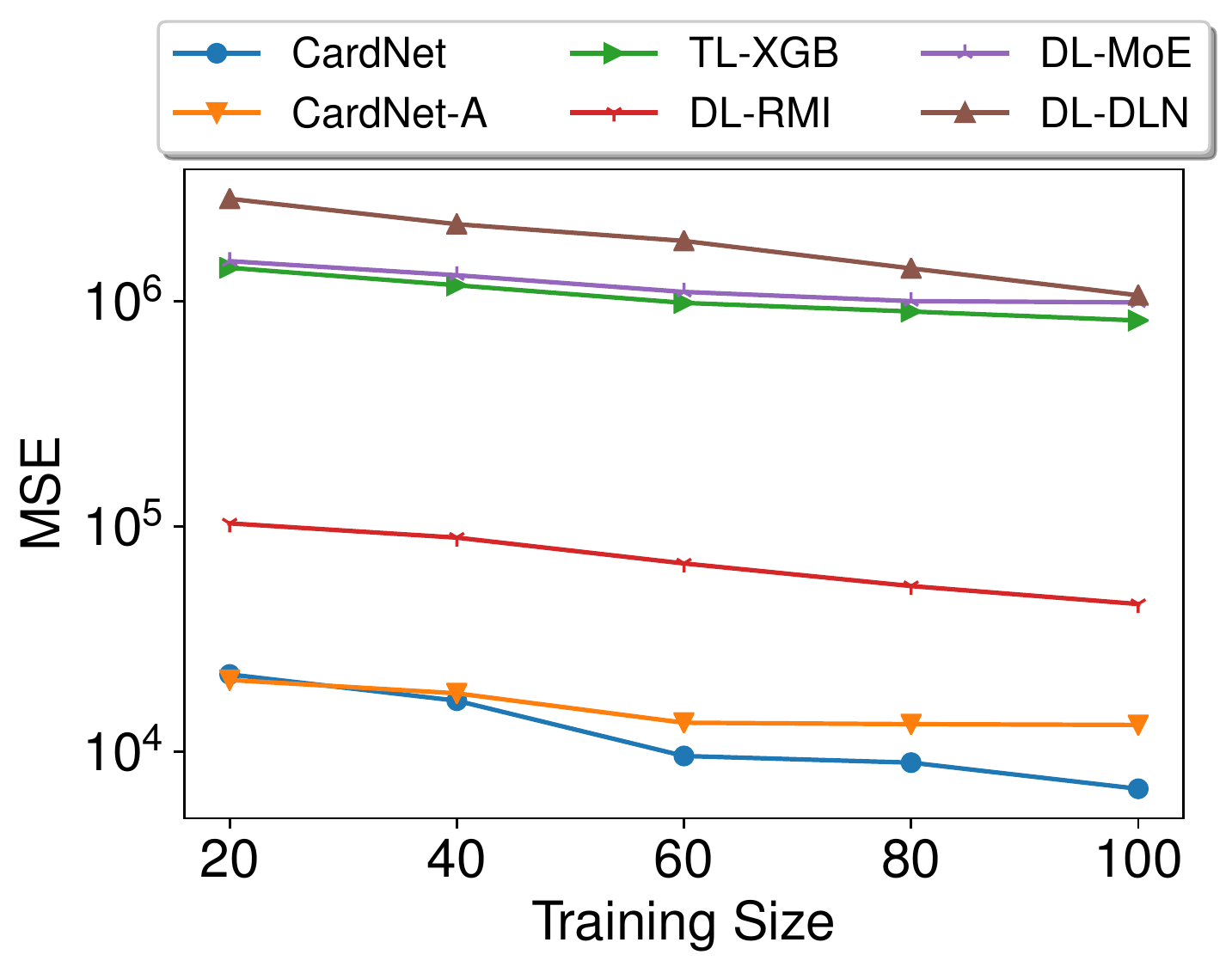}
    \label{fig:exp-glove-mse-trainsize}
  }
  \caption{Accuracy v.s. training data size.}
  \label{fig:exp-trainsize}
\end{figure}

In Figure~\ref{fig:exp-trainsize}, we show the performance of different models by varying 
the scale of training examples from 20\% to 100\% of the original training data. We only 
plot \mse due to the page limitation. All the models have worse performance with fewer 
training data, but our models are more robust, showing moderate accuracy loss. 

\begin{table} [t]
  \small
  \caption{Model size (MB).}
  \label{tab:modelsize}
  \centering
  \resizebox{\linewidth}{!}{  
  \begin{tabular}[b]{| l | c | c | c | c |}
    \hline%
    Model & \imagenet & \aminer & \bmsjacc & \glovetwo \\
    \hline%
    \spestexp & 10.4 & 31.2 & 39.4 & 86.2 \\
    \usexp & 0.6 & 0.5 & 0.6 & 3.4 \\
    \xgbexp & 36.4 & 36.4 & 48.8 & 63.2 \\
    \lightgbmexp & 32.6 & 32.8 & 45.4 & 60.4 \\
    \kdeexp & 4.5 & 1.5 & 3.6 & 18.1 \\
    \dlnexp & 28.4 & 75.4 & 28.6 & 64.4 \\
    \moeexp & 16.8 & 52.5 & 35.4 & 52.5 \\
    \hierdnnexp & 57.7 & 84.8 & 54.6 & 66.1 \\
    \dnnexp & 5.0 & 14.5 & 8.7 & 9.8 \\
    \dnnsexp & 105.4 & 154.2 & 183.2 & 158.4 \\
    \textbf{\modelone} & 9.6 & 40.2 & 16.4 & 23.8 \\
    \textbf{\modeltwo} & 16.2 & 54.5 & 22.8 & 35.3 \\
    \hline%
  \end{tabular}
  }
\end{table}

\begin{table} [t]
  \small
  \caption{Training time (hours).}
  \label{tab:traintime}
  \centering
  \resizebox{\linewidth}{!}{  
  \begin{tabular}[b]{| l | c | c | c | c |}
    \hline%
    Model & \imagenet & \aminer & \bmsjacc & \glovetwo \\
    \hline%
    \xgbexp  & 0.8 & 0.8 & 1.0 & 1.2  \\
    \lightgbmexp & 0.6 & 0.5 & 0.7 & 0.6 \\
    \kdeexp & 0.3 & 0.3 & 0.5 & 0.6 \\
    \dlnexp & 4.1 & 4.6 & 4.5 & 4.9 \\    
    \moeexp & 2.7 & 3.5 & 3.8 & 3.8 \\
    \hierdnnexp & 3.2 & 3.7 & 3.9 & 4.4 \\
    \dnnexp & 1.2 & 1.5 & 1.7 & 1.8 \\
    \dnnsexp & 15 & 12 & 12 & 20 \\
    \textbf{\modelone} & 3.3 & 3.4 & 4.1 & 4.2 \\
    \textbf{\modeltwo} & 1.7 & 2.2 & 2.4 & 2.6 \\
    \hline%
  \end{tabular}
  }
\end{table}

\subsection{Evaluation of Updates}
\begin{figure} [t]
  \centering
  \subfigure[\textsf{MSE}, \imagenet]{
    \includegraphics[width=0.46\linewidth]{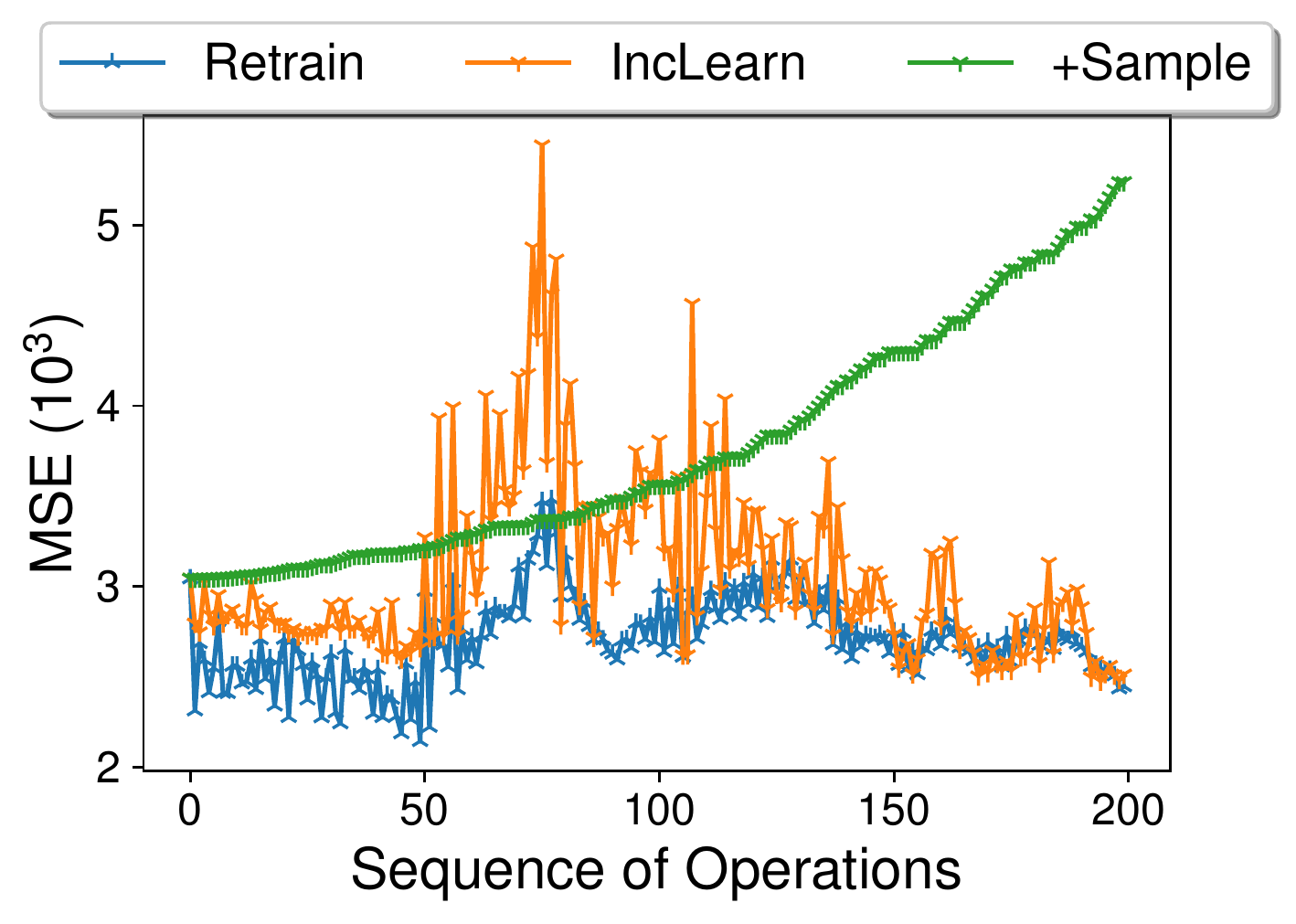}
    \label{fig:exp-imagenet-mse-update}
  }
  \subfigure[\textsf{MSE}, \glovetwo]{
    \includegraphics[width=0.46\linewidth]{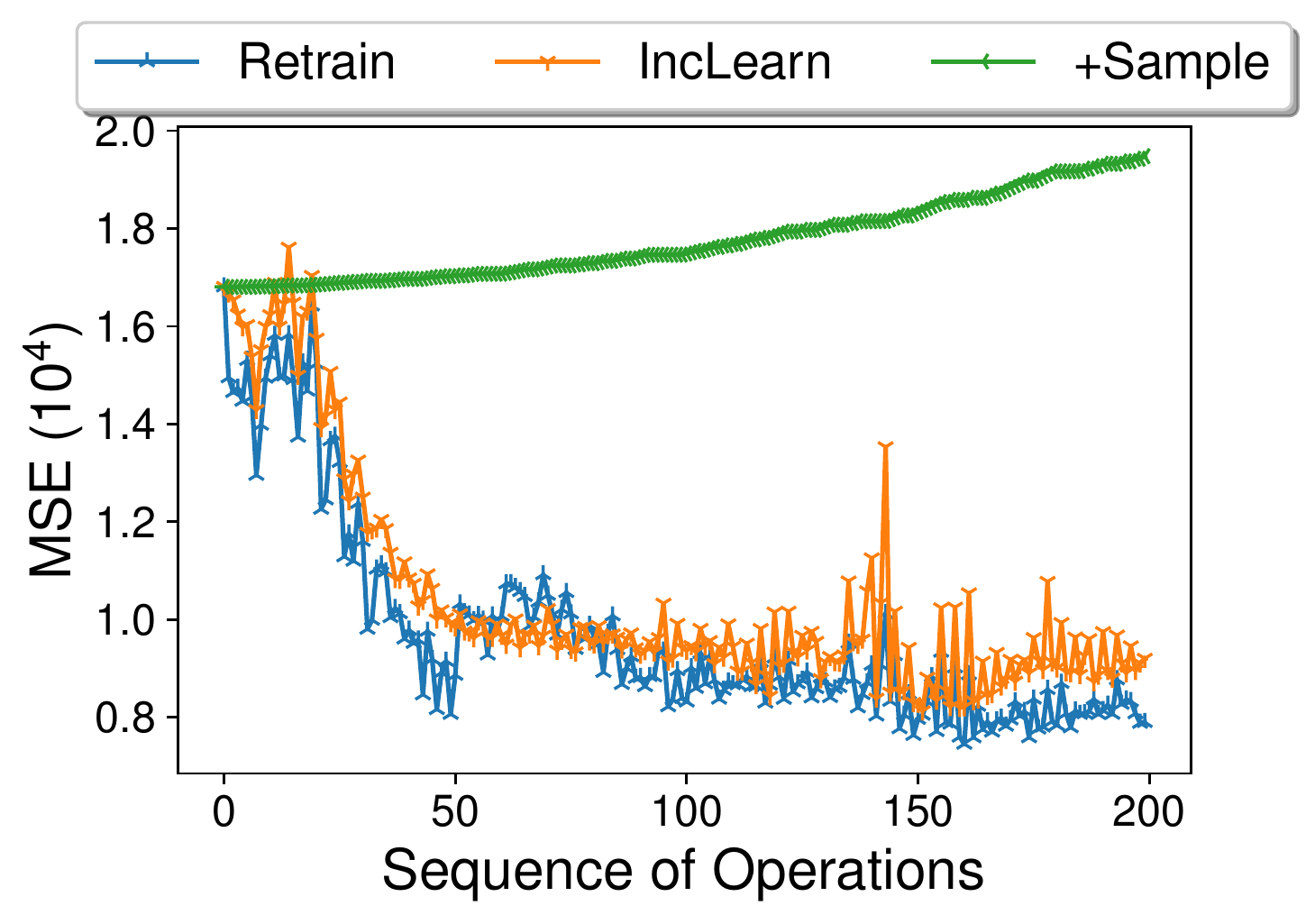}
    \label{fig:exp-glove-mse-update}
  }
  \caption{Evaluation of updates.}
  \label{fig:exp-update}
\end{figure}



We generate a stream of \revise{200 update operations, each with an insertion or deletion of 
5 records}. 
We compare three methods: \textsf{IncLearn} that utilizes incremental learning on \modeltwo, 
\textsf{Retrain} that retrains \modeltwo for each operation, and \textsf{+Sample} that performs 
sampling (\usexp) on the updated data and add the result to that of \modeltwo on the original data. 
Figure~\ref{fig:exp-update} 
plots the \mse on \imagenet and \glovetwo. We observe that in most cases, \textsf{IncLearn} has 
similar performance to \textsf{Retrain} and performs better than \textsf{+Sample}, especially when 
there are more updates. Compared to \textsf{Retrain} that spends several hours to retrain 
the model (Table~\ref{tab:traintime}), \textsf{IncLearn} only needs 1.2 -- 1.5 minutes to perform 
incremental learning. 

\mycomment{In Figure~\ref{fig:exp-update}, we test the accuracy of scaling methods and non-scaling ones
with insertions and deletions of 2\%, 4\%, 8\% and 16\% records, e.g., +2\% means inserting 2\%
records into the original dataset $\mathcal{D}$, and -2\% means deleting 2\% records. 
Records used for deletion and insertion are sampled from the original 
dataset with adding a few noise.
The experimental results show that scaling methods can still have relatively good prediction
(better than non-scaling),
when a small number of
modifications are operated on $\mathcal{D}$, e.g., competitive \mse and \mape in
the range [-4\%, +4\%] in \bmsjacc dataset. Thus, when a small amount of insertions and deletions
incur, we can adopt scaling strategy, and do not need to re-train our models. However,
when there are large updates, e.g., inserting more than 4\% records into dataset, we have to
re-train our models with updated training data.}

\begin{figure} [t]
  \centering
  \subfigure[\textsf{MSE}, \imagenet]{
    \includegraphics[width=0.46\linewidth]{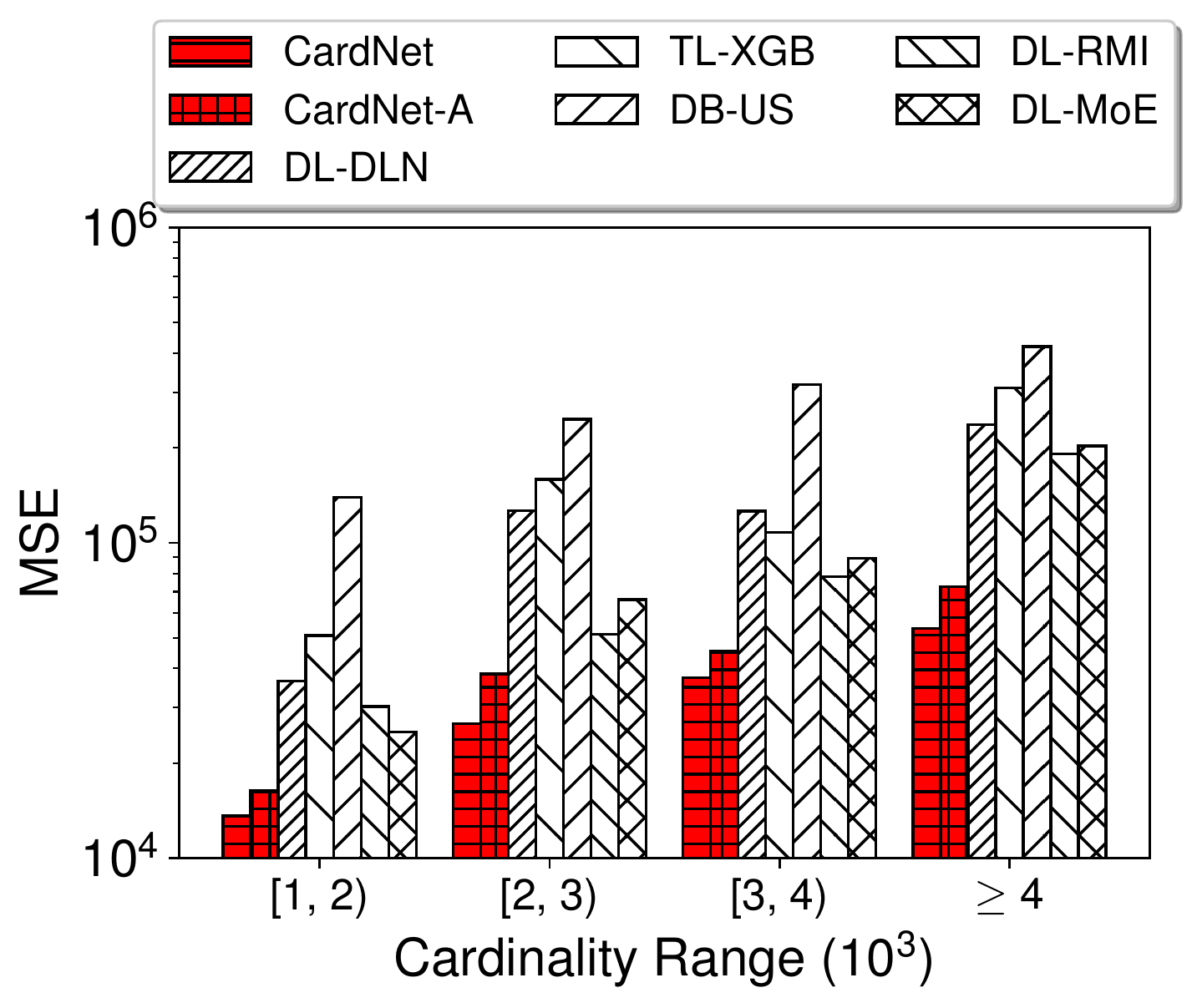}
    \label{fig:exp-imagenet-mse-long-tail}
  }
  \subfigure[\textsf{MSE}, \aminer]{
    \includegraphics[width=0.46\linewidth]{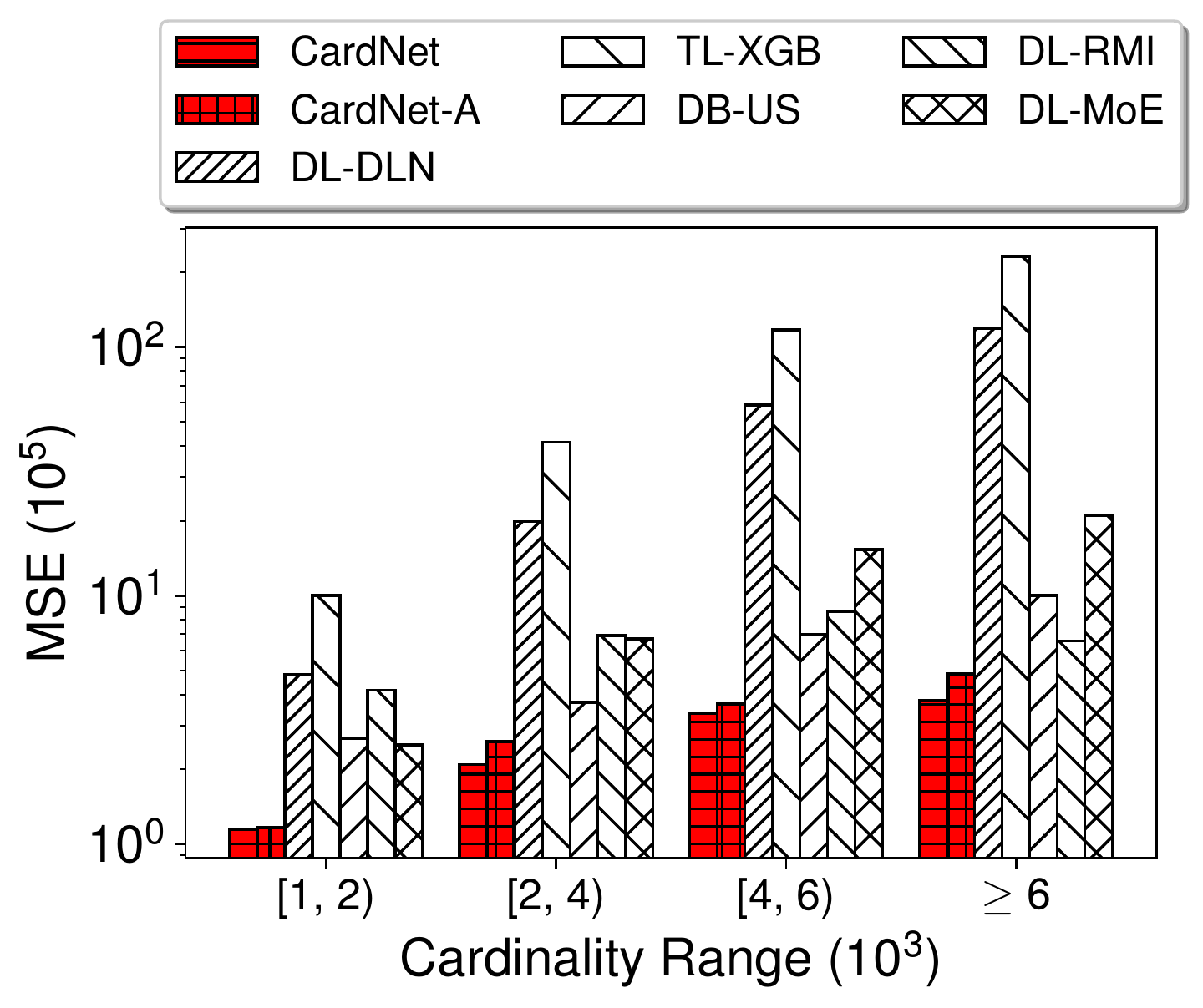}
    \label{fig:exp-aminer-mse-long-tail}
  }  
  \subfigure[\reviseR{\textsf{MSE}, \bmsjacc}]{
    \includegraphics[width=0.46\linewidth]{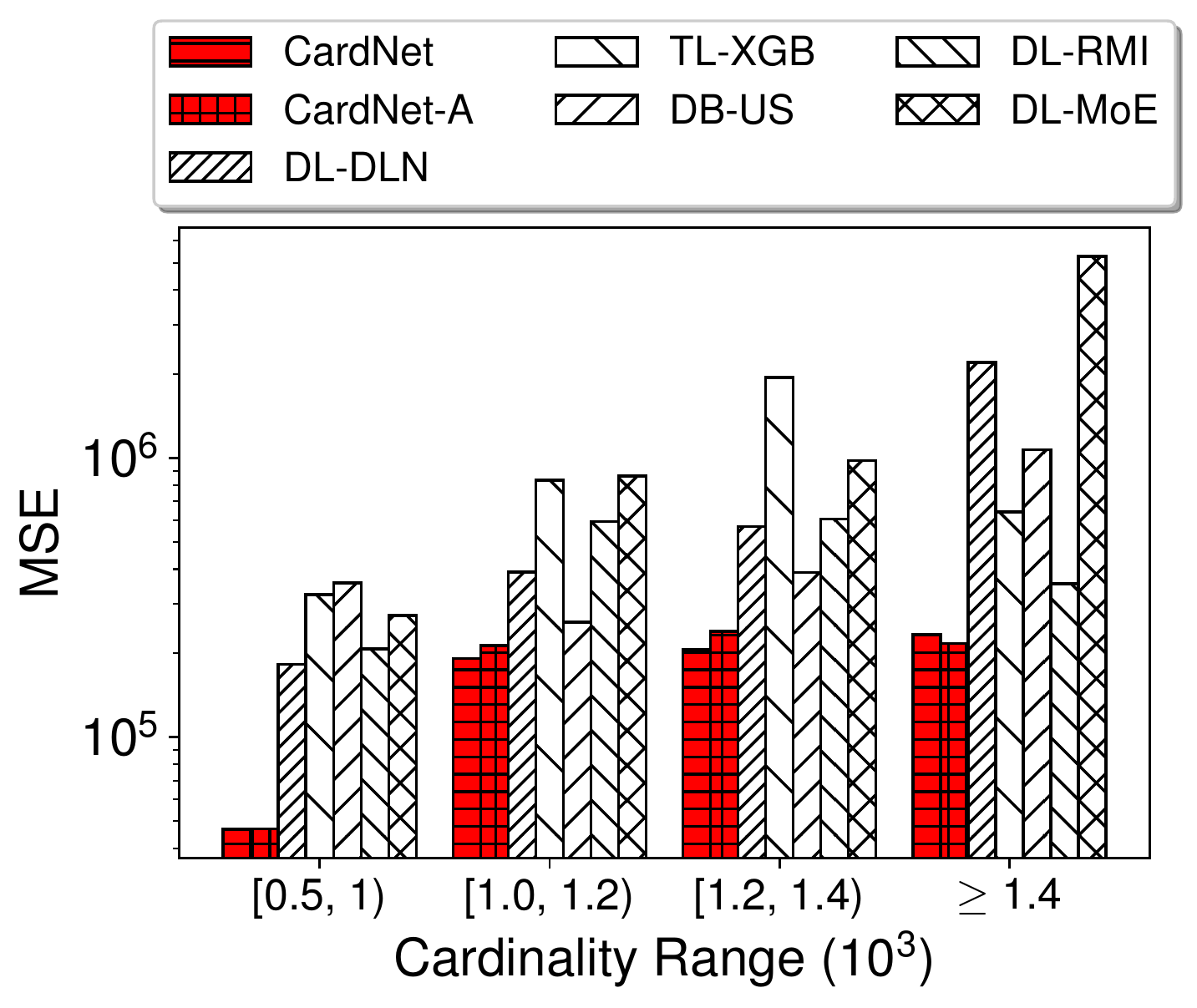}
    \label{fig:exp-bms-mse-long-tail}
  }
  \subfigure[\textsf{MSE}, \glovetwo]{
    \includegraphics[width=0.46\linewidth]{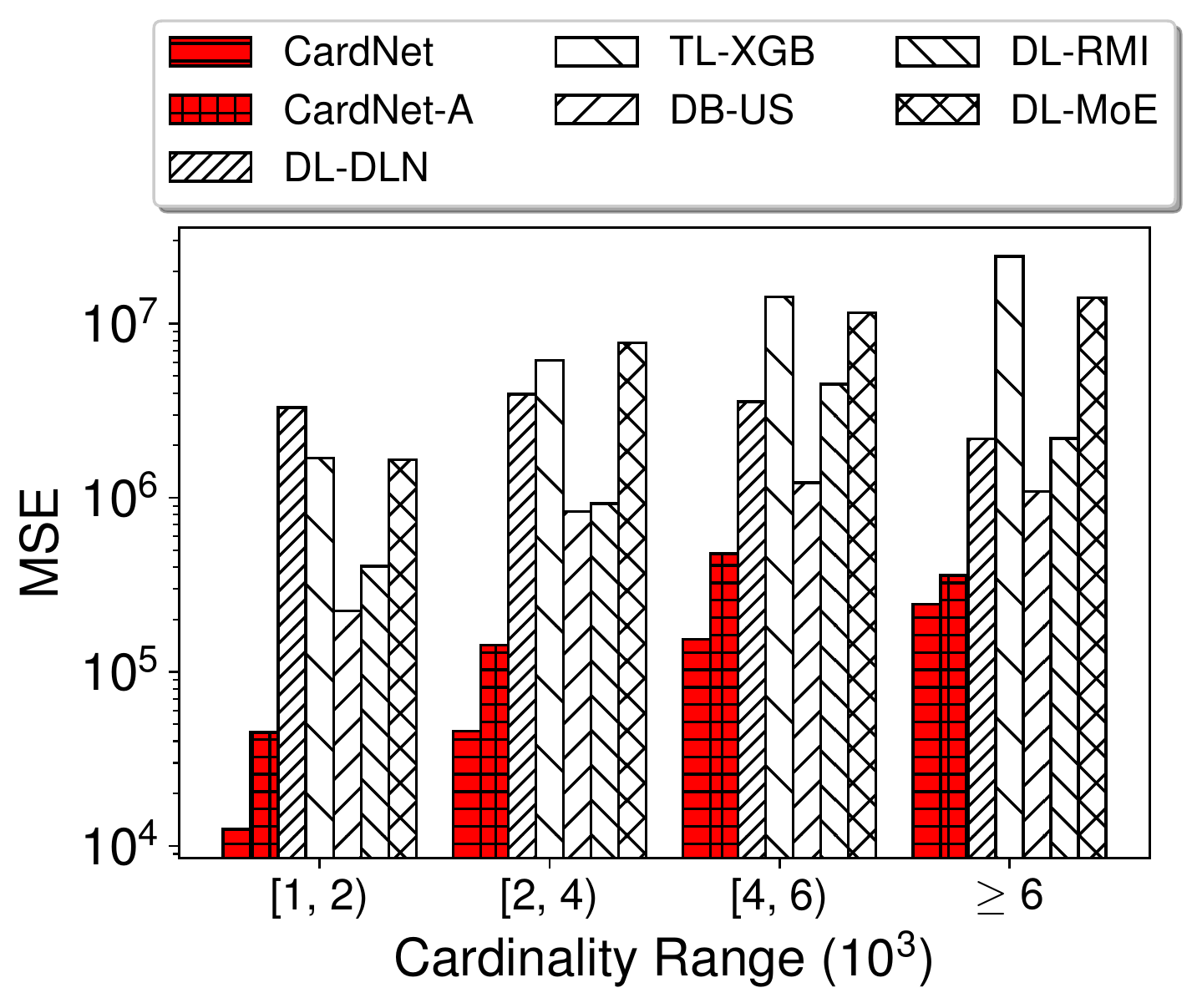}
    \label{fig:exp-glovetwo-mse-long-tail}
  }
  \caption{\reviseR{Evaluation of long-tail queries.}}
  \label{fig:exp-long-tail}
\end{figure}

\begin{figure} [t]
  \centering
  \subfigure[\textsf{MSE}, \imagenet]{
    \includegraphics[width=0.46\linewidth]{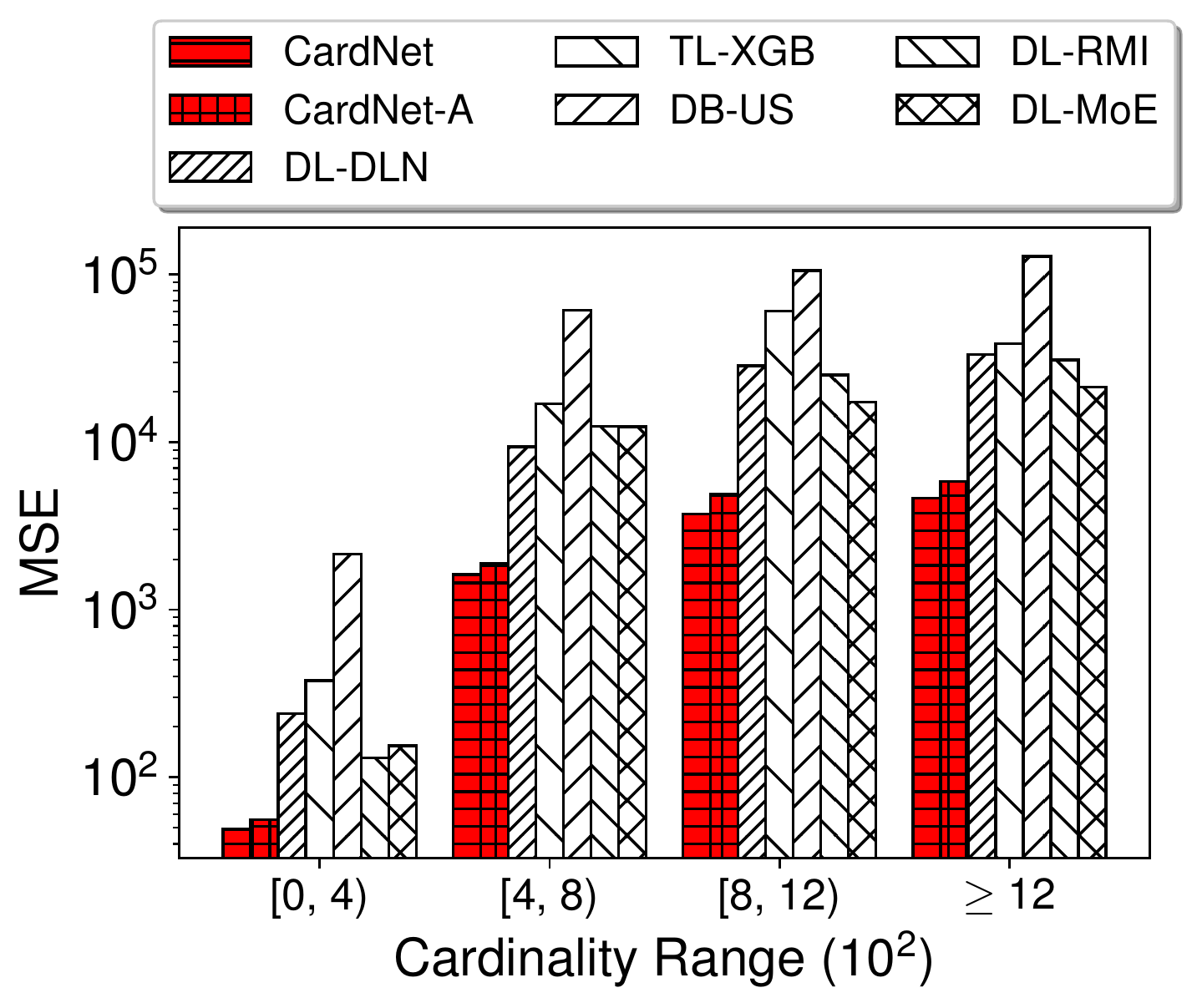}
    \label{fig:exp-imagenet-mse-unseen-queries}
  }
  \subfigure[\textsf{MSE}, \aminer]{
    \includegraphics[width=0.46\linewidth]{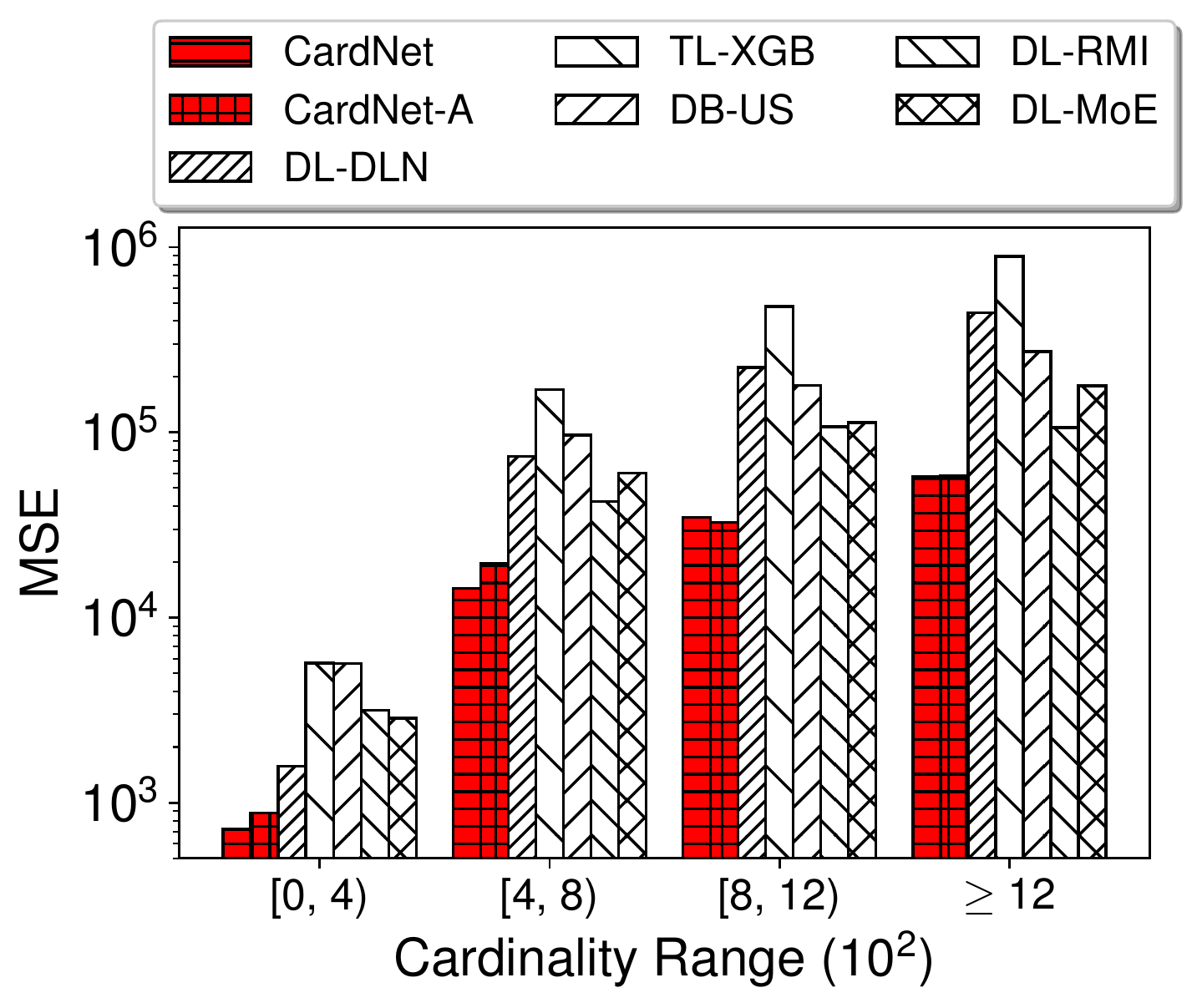}
    \label{fig:exp-aminer-mse-unseen-queries}
  }  
  \subfigure[\reviseR{\textsf{MSE}, \bmsjacc}]{
    \includegraphics[width=0.46\linewidth]{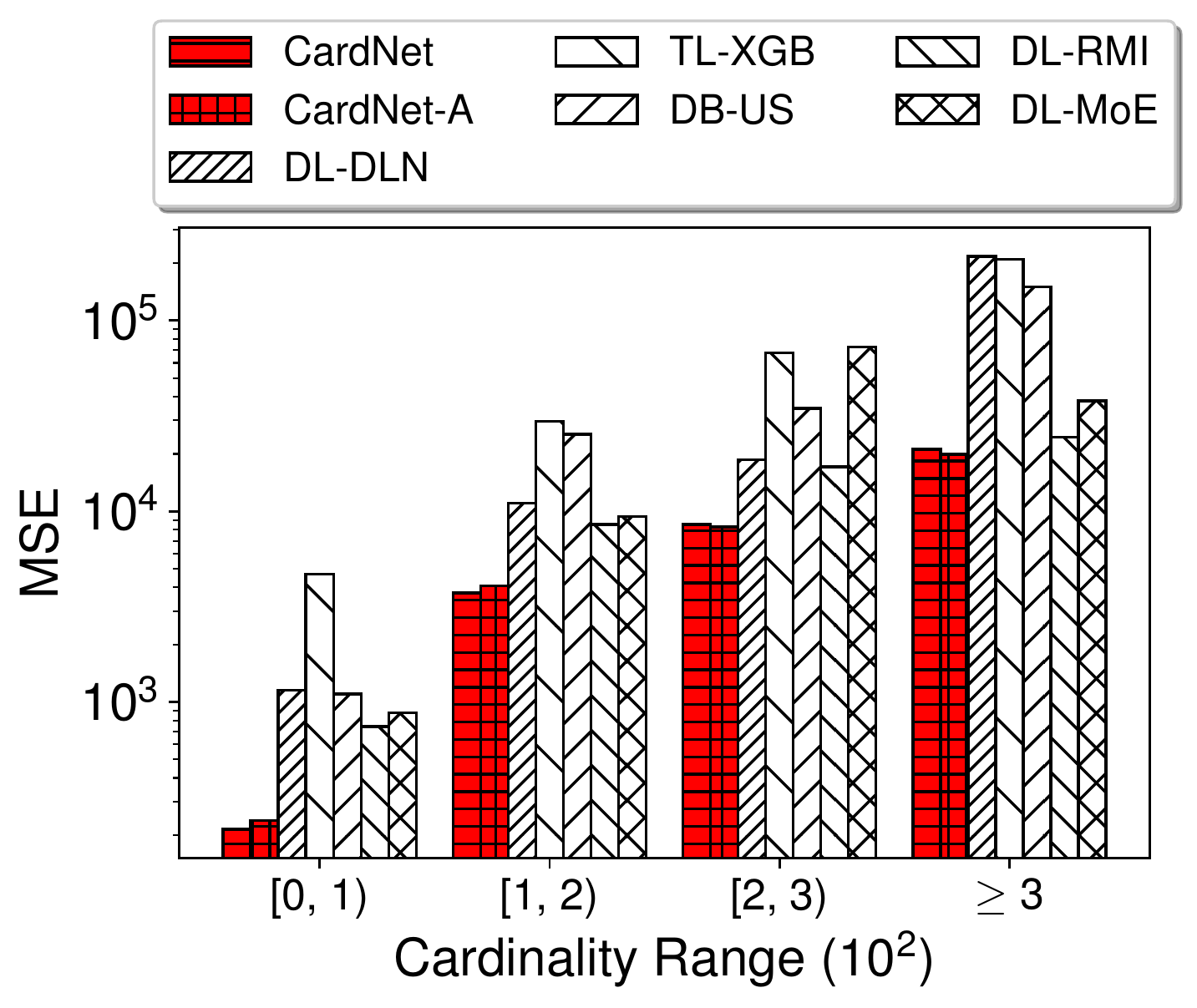}
    \label{fig:exp-bms-mse-unseen-queries}
  }
  \subfigure[\textsf{MSE}, \glovetwo]{
    \includegraphics[width=0.46\linewidth]{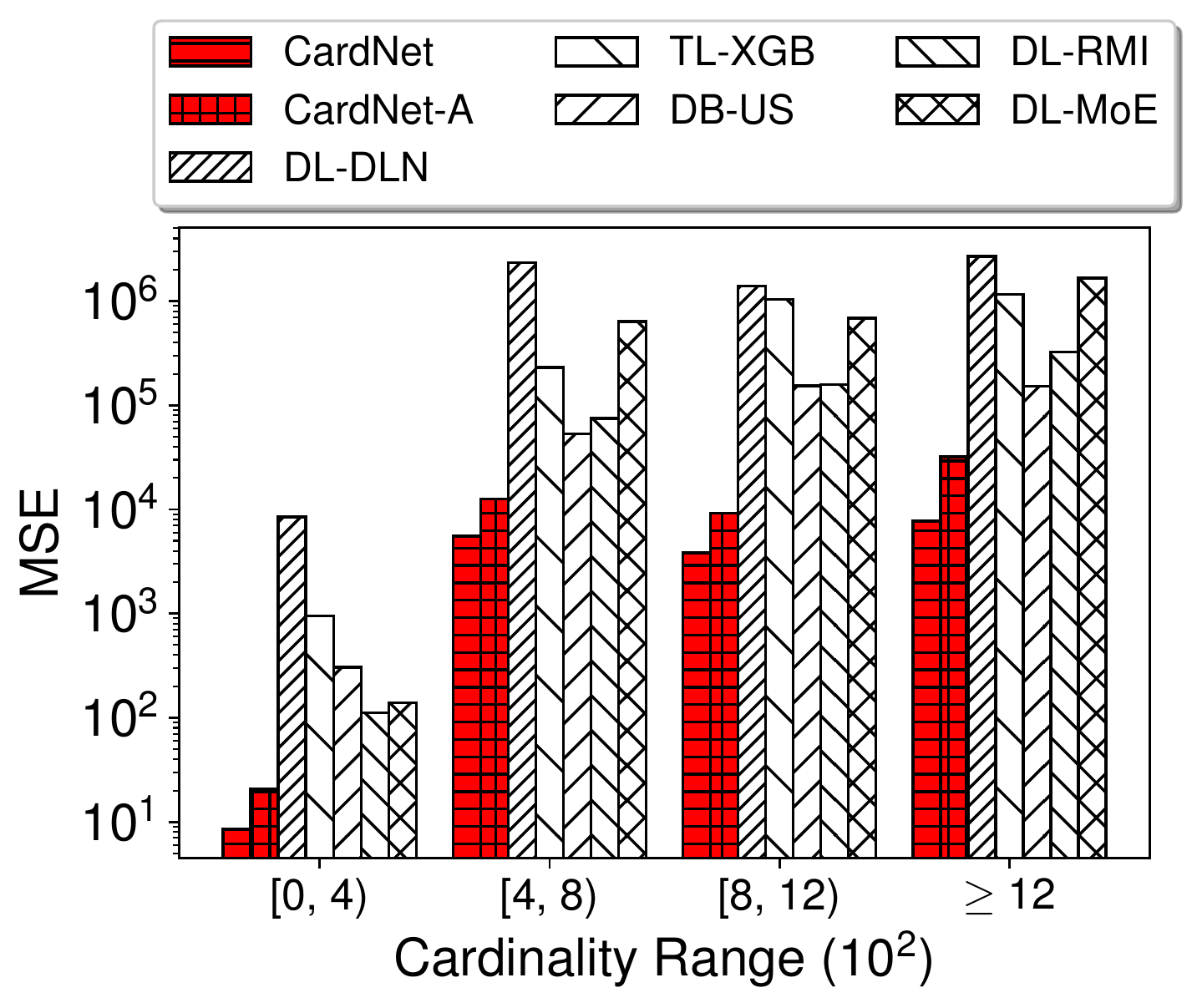}
    \label{fig:exp-glovetwo-mse-unseen-queries}
  }
  \caption{\reviseR{Generalizability.}}
  \label{fig:exp-out-of-dataset}
\end{figure}


\subsection{Evaluation of Long-tail Queries}
We compare the performance on long-tail queries, i.e., those having exceptionally large 
cardinalities ($\geq$ 1000). They are outliers and a hard case of estimation. 
We divide queries into different cardinality groups by every 
thousand. Figure~\ref{fig:exp-long-tail} shows the \mse \fullversion{on the four default datasets} 
by varying cardinality groups. The \mse increases with cardinality for all the methods. 
This is expected since the larger the cardinality is, the more exceptional is the query 
(see Figure~\ref{fig:pct-card}). Our models outperform the others by 1 to 3 orders of 
magnitude. Moreover, the \mse growth rates of our models w.r.t. cardinality are smaller 
than the others, suggesting that our models are more robust against long-tail queries. 

\subsection{Generalizability} \label{sec:exp-generalizability}
To show the generalizability of our models, we evaluate the performance on the queries that 
significantly differ from the records in the dataset and the training data. To prepare 
such queries, we first perform a $k$-medoids clustering on the dataset, and then randomly generate 
10,000 out-of-dataset queries \confversion{(e.g., for binary vectors, we generate a query such that 
$q[i] \sim \text{uniform}\set{0, 1}$ and accept it only if it is not in $\mathcal{D}$)} and pick 
the top-2,000 ones having the largest sum of squared distance to the $k$ centroids. 
\fullversion{
To prepare an out-of-dataset query, we generate a random query $q$ and accept it only if it is 
not in the dataset $\mathcal{D}$. 
Specifically, 
\begin{inparaenum} [(1)]
  \item for binary vectors, $q[i] \sim \text{uniform}\set{0, 1}$; 
  \item for strings, since AMiner and DBLP both contain author names, we take a 
  random author name from the set $(\text{DBLP} \backslash \text{AMiner})$; 
  \item for sets, we generate a length $l \sim \text{uniform}[l_{\min}, l_{\max}]$, 
  where $l_{\min}$ and $l_{\max}$ are the minimum and maximum set sizes in $\mathcal{D}$, 
  respectively, and then generate a random set of length $l$ sampled from the universe of 
  all the elements in $\mathcal{D}$; 
  \item for real vectors, $q[i] \sim \text{uniform}[-1, 1]$.
\end{inparaenum}
}
Figure~\ref{fig:exp-out-of-dataset} shows the \mse \fullversion{on the 
four default datasets} by varying cardinality groups. The same trend is witnessed as we have 
seen for long-tail queries. Due to the use of \vae and dynamic training, our models always perform 
better than the other methods, especially for Jaccard distance. The results demonstrate that our 
models generalize well for out-of-dataset queries. 

\begin{table*} [t]
  \small
  \caption{\revise{Statistics of datasets for conjunctive query optimizer.}}
  \label{tab:dataset-optimizer-conj}
  \centering
  \begin{tabular}[b]{| l | c | c | c | c | c |}
    \hline%
    \texttt{Dataset} & \texttt{Source} & \texttt{Attributes} 
    & \texttt{\# Records} & $\theta_{\min}$ & $\theta_{\max}$ \\
    \hline%
    \textsf{AMiner-Publication} & \cite{URL:aminer} & title, authors, affiliations, venue, abstract & 2,092,356 & 0.2 & 0.5 \\
    \hline%
    \textsf{AMiner-Author} & \cite{URL:aminer} & name, affiliations, research interests & 1,712,433 & 0.2 & 0.5 \\
    \hline%
    \textsf{IMDB-Movie} & \cite{URL:imdb} & title type, primary title, original title, genres & 6,250,486 & 0.2 & 0.5 \\
    \hline%
    \textsf{IMDB-Actor} & \cite{URL:imdb} & primary name, primary profession & 9,822,710 & 0.2 & 0.5 \\    
    \hline%
  \end{tabular}
\end{table*}

\begin{figure} [t]
  \centering
  \subfigure[Time, \textsf{AMiner-Publication}]{
    \includegraphics[width=0.46\linewidth]{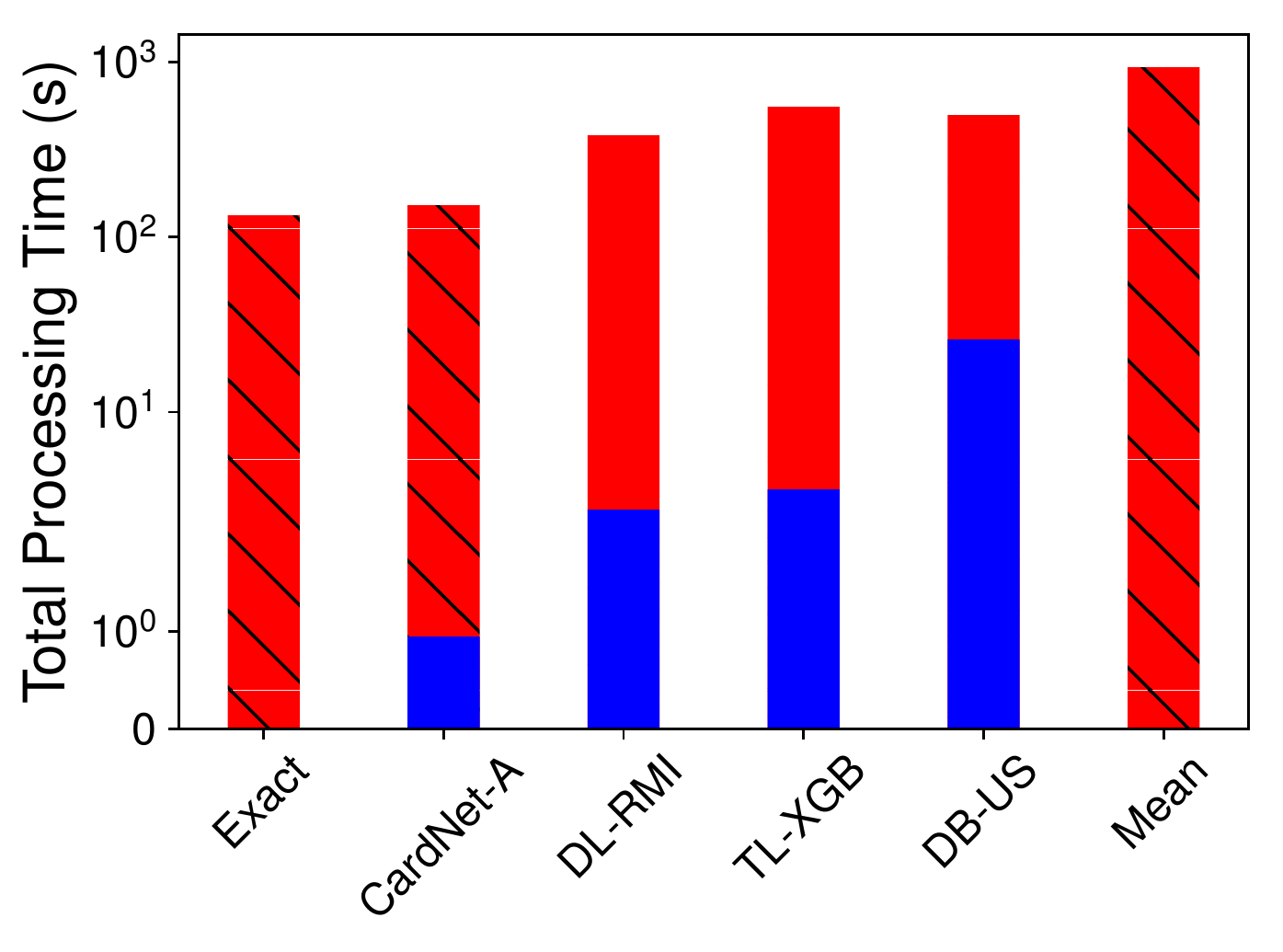}
    \label{fig:exp-aminer-paper-optimizer-time}
  }  
  \subfigure[Time, \textsf{AMiner-Author}]{
    \includegraphics[width=0.46\linewidth]{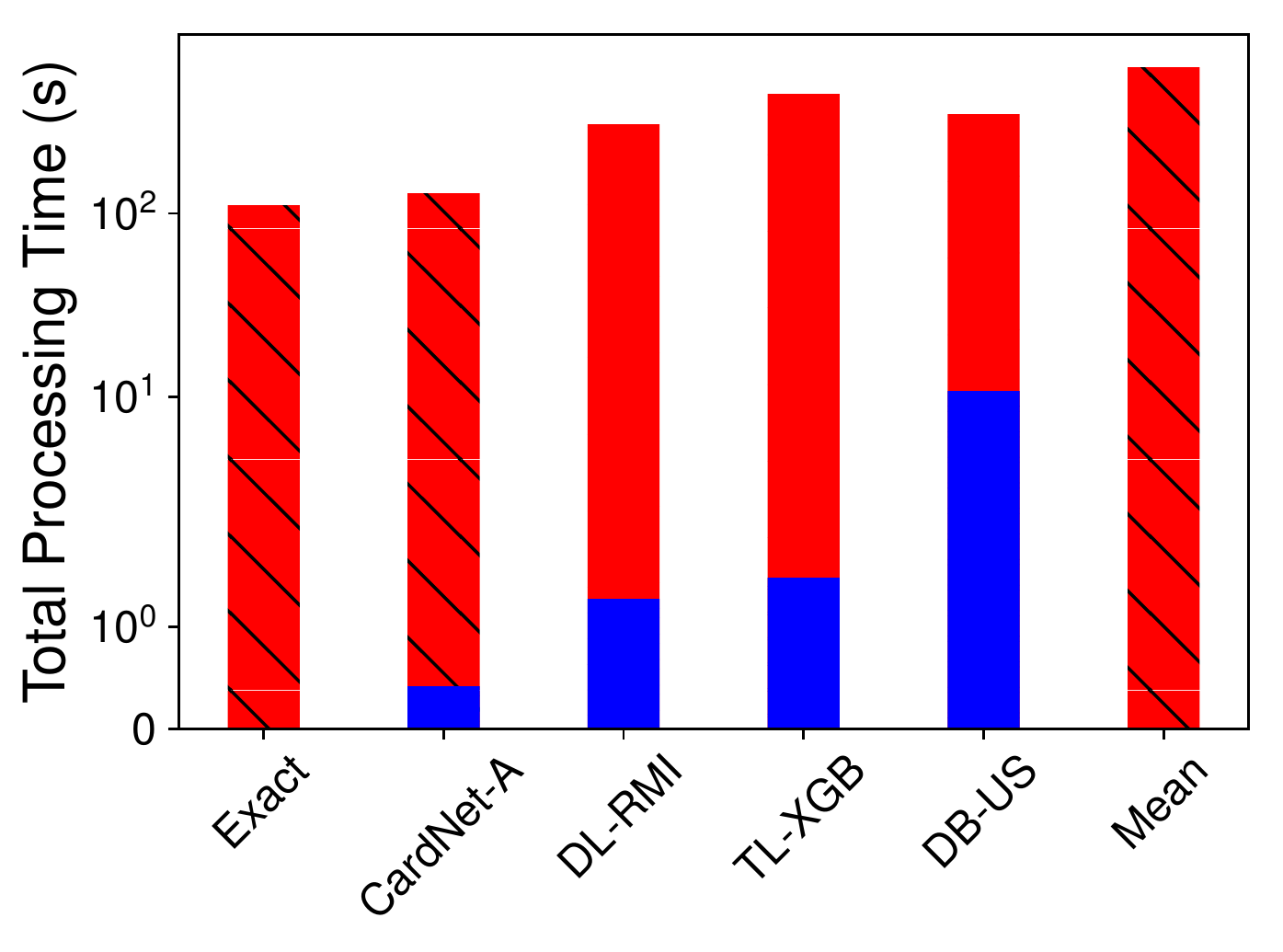}
    \label{fig:exp-aminer-author-optimizer-time}
  }
  \subfigure[Time, \textsf{IMDB-Movie}]{
    \includegraphics[width=0.46\linewidth]{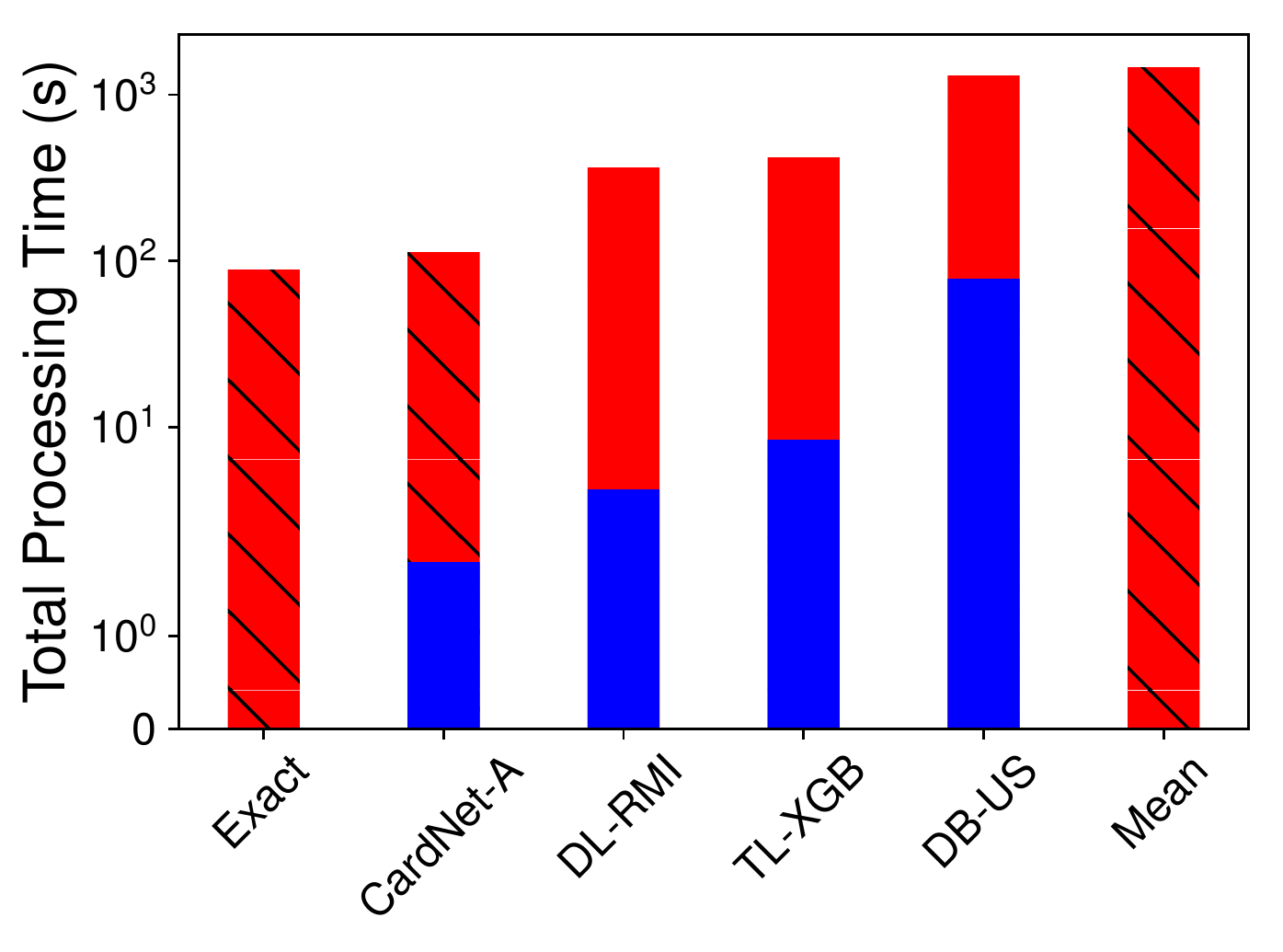}
    \label{fig:exp-imdb-title-optimizer-time}
  }
  \subfigure[Time, \textsf{IMDB-Actor}]{
    \includegraphics[width=0.46\linewidth]{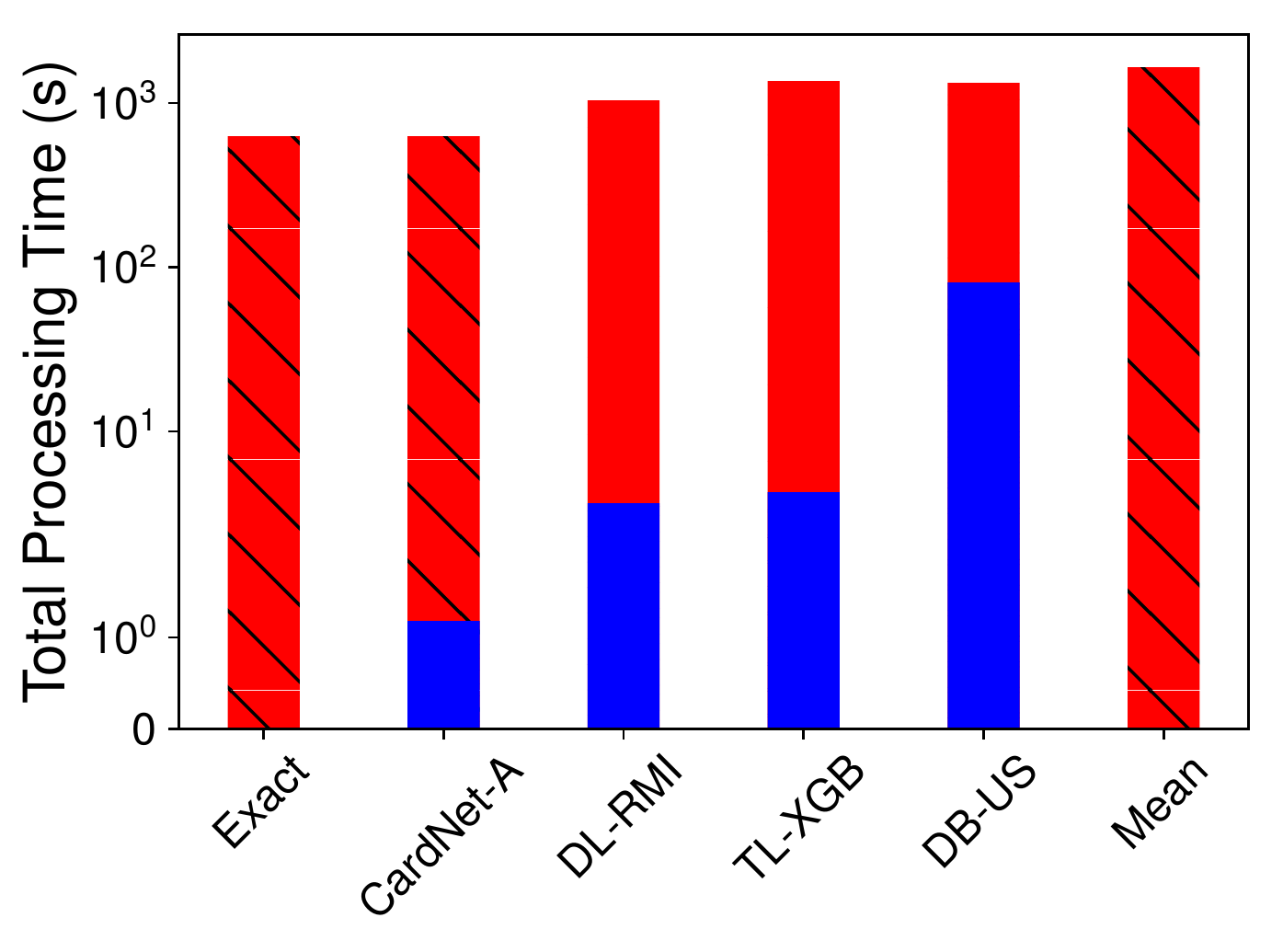}
    \label{fig:exp-imdb-actor-optimizer-time}
  }  
  \confversion{\caption{\revise{Query processing time.}}}
  \fullversion{\caption{\revise{Conjunctive euclidean distance query -- query processing time.}}}
  \label{fig:optimizer-time-conjunctive}
\end{figure}

\begin{figure} [t]
  \centering
  \subfigure[Time, \textsf{AMiner-Publication}]{
    \includegraphics[width=0.46\linewidth]{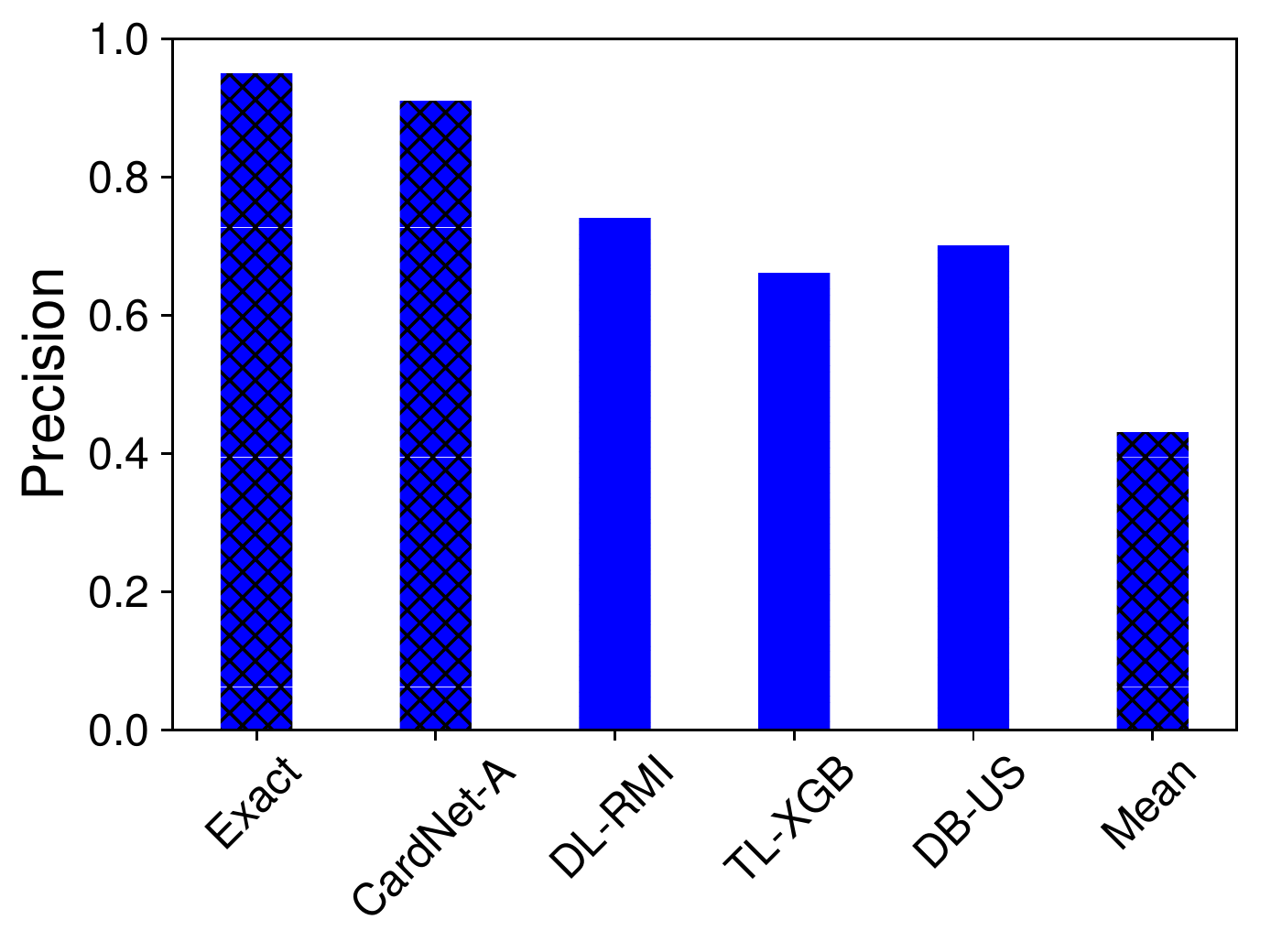}
    \label{fig:exp-aminer-paper-optimizer-time-ratio}
  }  
  \subfigure[Time, \textsf{AMiner-Author}]{
    \includegraphics[width=0.46\linewidth]{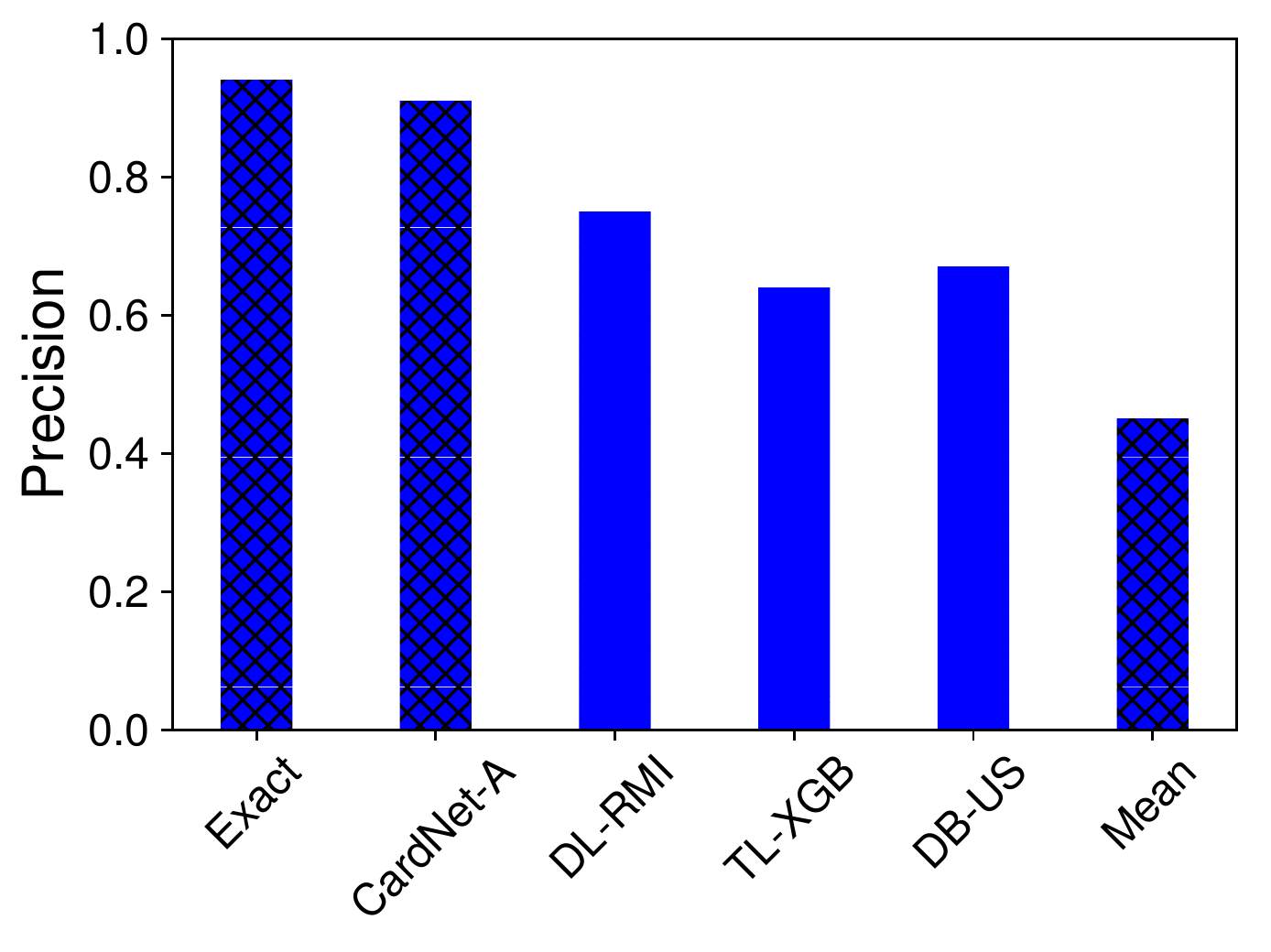}
    \label{fig:exp-aminer-author-optimizer-time-ratio}
  }
  \subfigure[Time, \textsf{IMDB-Movie}]{
    \includegraphics[width=0.46\linewidth]{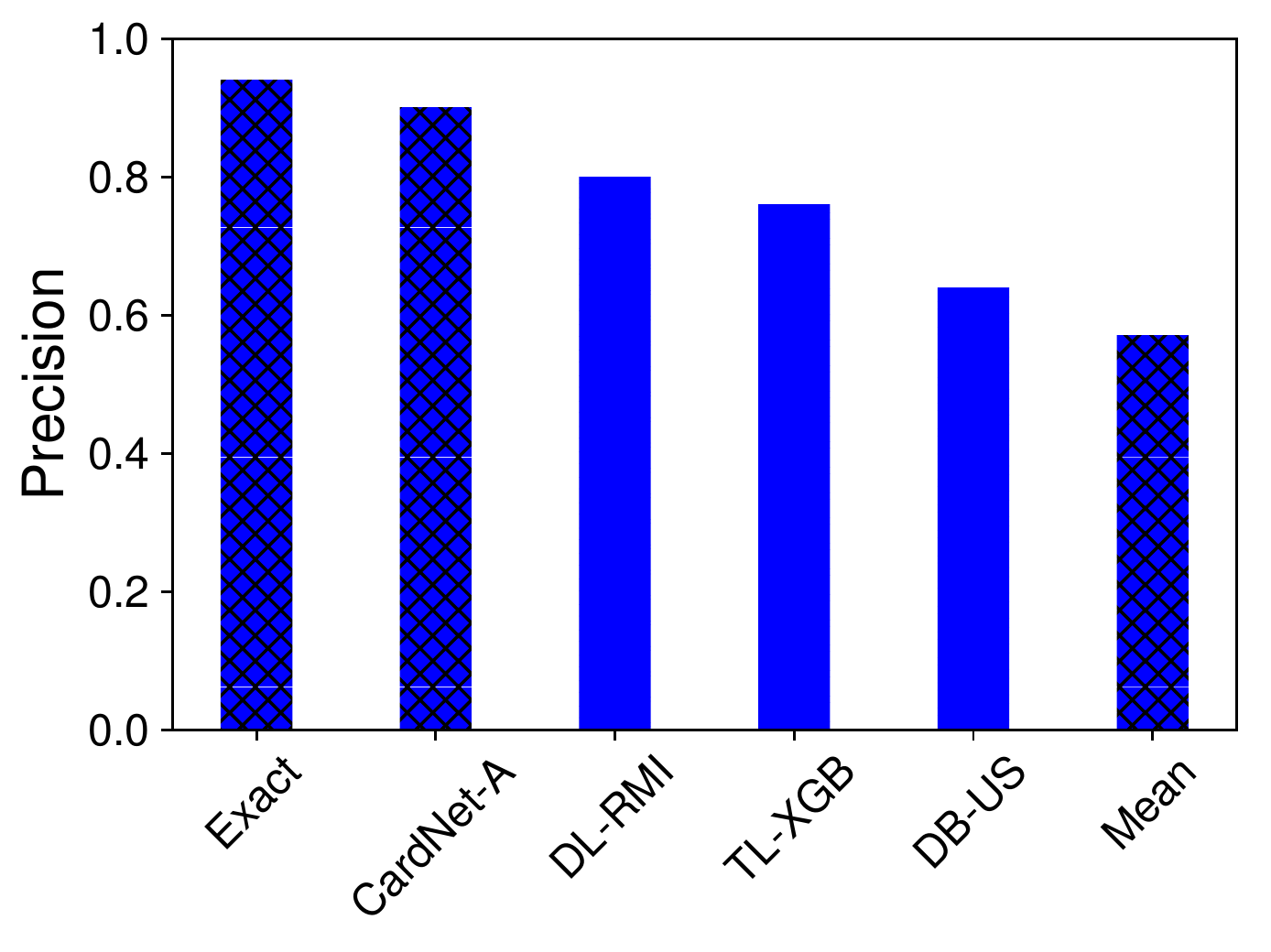}
    \label{fig:exp-imdb-title-optimizer-time-ratio}
  }
  \subfigure[Time, \textsf{IMDB-Actor}]{
    \includegraphics[width=0.46\linewidth]{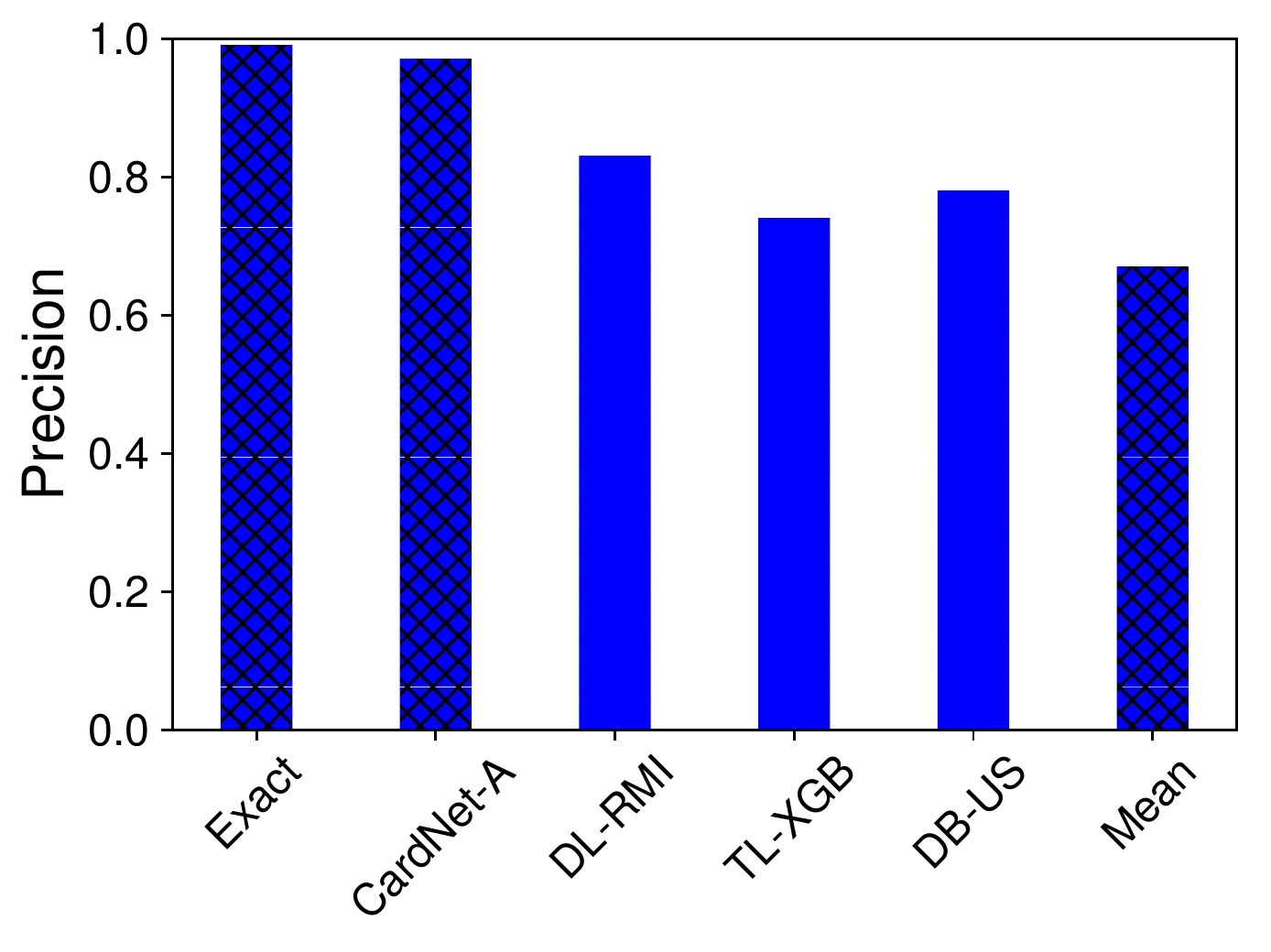}
    \label{fig:exp-imdb-actor-optimizer-time-ratio}
  }  
  \confversion{\caption{\revise{Query planning precision.}}}
  \fullversion{\caption{\revise{Conjunctive euclidean distance query -- query planning precision.}}}  
  \label{fig:optimizer-time-conjunctive-ratio}
\end{figure}

\revise{
\subsection{Performance in a Query Optimizer}
\fullversion{\subsubsection{Conjunctive Euclidean Distance Query}}
We consider a case study of conjunctive queries. Four textual datasets with multiple 
attributes are used (statistics shown in Table~\ref{tab:dataset-optimizer-conj}). Given a dataset, 
we convert each attribute to a word embedding (768 dimensions) by Sentence-BERT~\cite{DBLP:conf/emnlp/ReimersG19}. 
A query is a conjunction of Euclidean distance predicates (a.k.a. high-dimensional range 
predicates~\cite{DBLP:journals/vldb/LinJF94}) on normalized word 
embeddings, with thresholds uniformly sampled from $[0.2, 0.5]$; e.g., ``$\textsf{EU}(\text{name}) 
\leq 0.25$ \textsf{AND} $\textsf{EU}(\text{affiliations}) \leq 0.4$ \textsf{AND} 
$\textsf{EU}(\text{research interests}) \leq 0.45$'', where $\textsf{EU}()$ measures the Euclidean 
distance between the embeddings of a query and a database record. Such queries can be used for 
entity matching 
as blocking rules~\cite{DBLP:conf/sigmod/GokhaleDDNRSZ14,DBLP:conf/sigmod/DasCDNKDARP17}.  

To process a query, we first find the records that satisfy one predicate by index lookup (by a 
cover tree~\cite{fastercovertree}), and then check other predicates on the fly. We estimate for each 
predicate and pick the one with the smallest cardinality for index lookup. We compare \modeltwo with: 
\begin{inparaenum} [(1)]
  \item \usexp, sampling ratio tuned for fastest query processing speed; 
  \item \xgbexp; 
  \item \hierdnnexp; 
  \item \mean, which returns the same cardinality for a given threshold;   
  each threshold is quantized to an integer in $[0, 255]$ using the threshold transformation in 
  Section~\ref{sec:featext-eu}, and then we offline generate 10,000 random queries for each integer 
  in $[0, 255]$ and take the mean; and 
  \item \exact, an oracle that instantly returns the exact cardinality. 
\end{inparaenum}

Figure~\ref{fig:optimizer-time-conjunctive} reports the processing time of 1,000 queries. The time 
is broken down to cardinality estimation (in blue) and postprocessing (in red, including index 
lookup and on-the-fly check). We observe: 
\begin{inparaenum} [(1)]
  \item more accurate cardinality estimation (as we have seen in Section~\ref{sec:exp-accuracy}) 
  contributes to faster query processing speed; 
  \item cardinality estimation spends much less time than postprocessing; 
  \item uniform estimation (\mean) has the slowest overall speed; 
  \item deep learning performs better than database and traditional learning methods in 
  both estimation and overall speeds; 
  \item except \exact, \modeltwo is the fastest and most accurate in estimation, and its overall 
  speed is also the fastest (by 1.7 to 3.3 times faster than the runner-up \hierdnnexp) and even 
  close to \exact. 
\end{inparaenum}


In Figure~\ref{fig:optimizer-time-conjunctive-ratio}, we show the precision of query planning, i.e., 
the percentage of queries on which a method picks the fastest (excluding estimation time) plan. 
The result is in accord with what we have observed in Figure~\ref{fig:optimizer-time-conjunctive}. 
The precision of \modeltwo ranges from 90\% to 96\%, second only to \exact. The gap between \modeltwo 
and \hierdnnexp is within 20\%, but results in the speedup of 1.7 to 3.3 times, showcasing the effect 
of correct query planning. We also observe that \exact is not 100\% precise, though very close, 
indicating that smallest cardinality does not always yield the best query plan. Future work on cost 
estimation may further improve query processing. 
}


\fullversion{
\begin{table*} [t]
  \small
  \caption{\revise{Statistics of datasets for Hamming distance query optimizer.}}
  \label{tab:dataset-optimizer-gph}
  \centering
  \begin{tabular}[b]{| l | c | c | c | c | c | c |}
    \hline%
    \texttt{Dataset} & \texttt{Source} & \texttt{Process} & 
    \texttt{Domain} & \texttt{\# Records} & $\ell$ & $\theta_{max}$ \\
    \hline%
    \pubchem & \cite{URL:pubchem} 
    & - & biological sequence & 1,000,000 & 881 & 32 \\
    \hline%
    \video & \cite{DBLP:journals/tmm/SongYHSL13} 
    & multiple feature hashing~\cite{DBLP:journals/tmm/SongYHSL13} & video embedding & 1,000,000 & 128 & 12 \\
    \hline%
    \fasttext & \cite{URL:fasttext} 
    & spectral hashing~\cite{weiss2009spectral} & word embedding & 999,999 & 256 & 24 \\
    \hline%
    \mnist & \cite{DBLP:conf/ijcnn/CohenATS17} 
    & image binarization & image pixel & 814,255 & 784 & 32 \\    
    \hline%
  \end{tabular}
\end{table*}

\begin{figure} [t]
  \centering
  \subfigure[Time, \pubchem]{
    \includegraphics[width=0.46\linewidth]{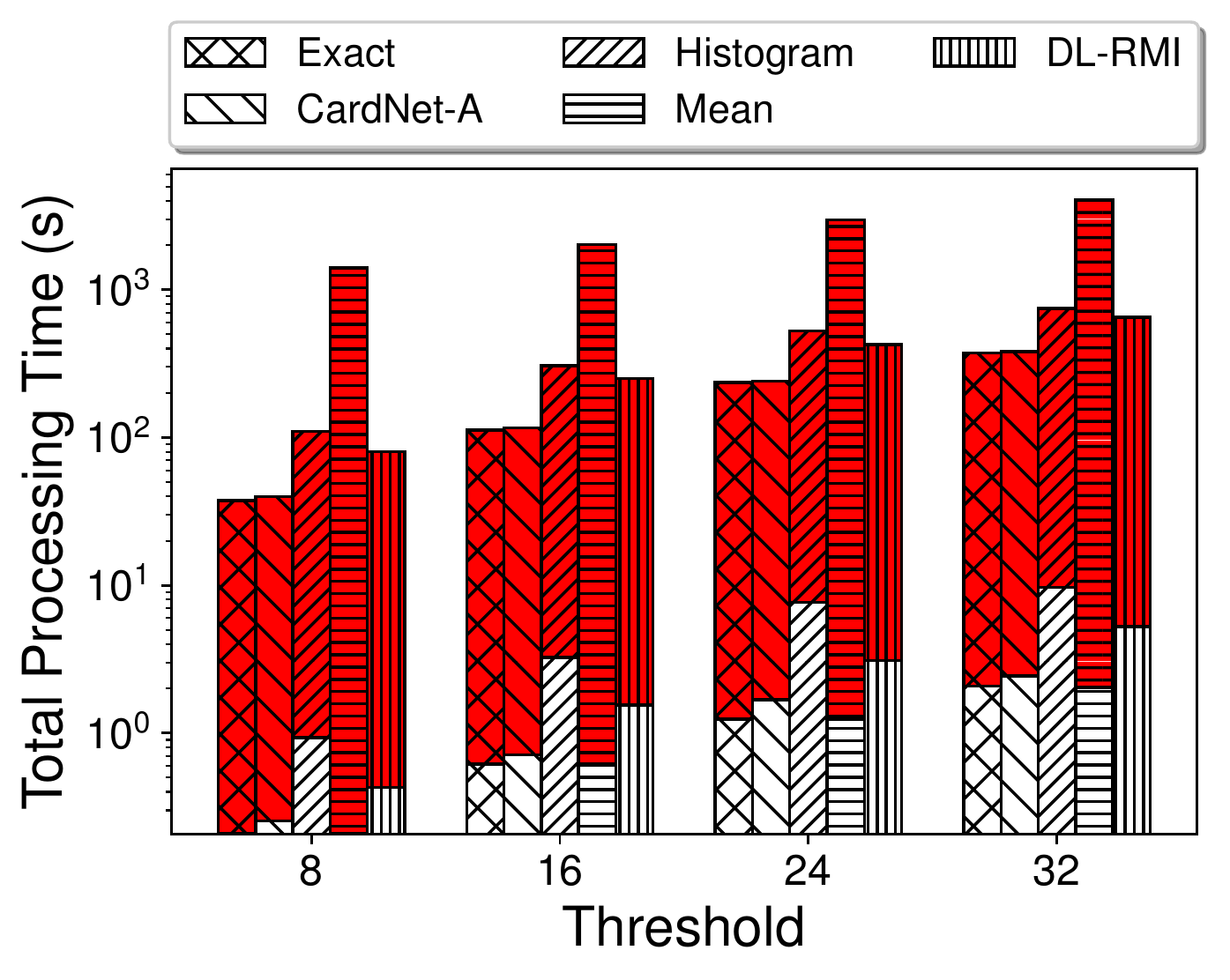}
    \label{fig:exp-pubchem-optimizer-time-threshold}
  }  
  \subfigure[Time, \video]{
    \includegraphics[width=0.46\linewidth]{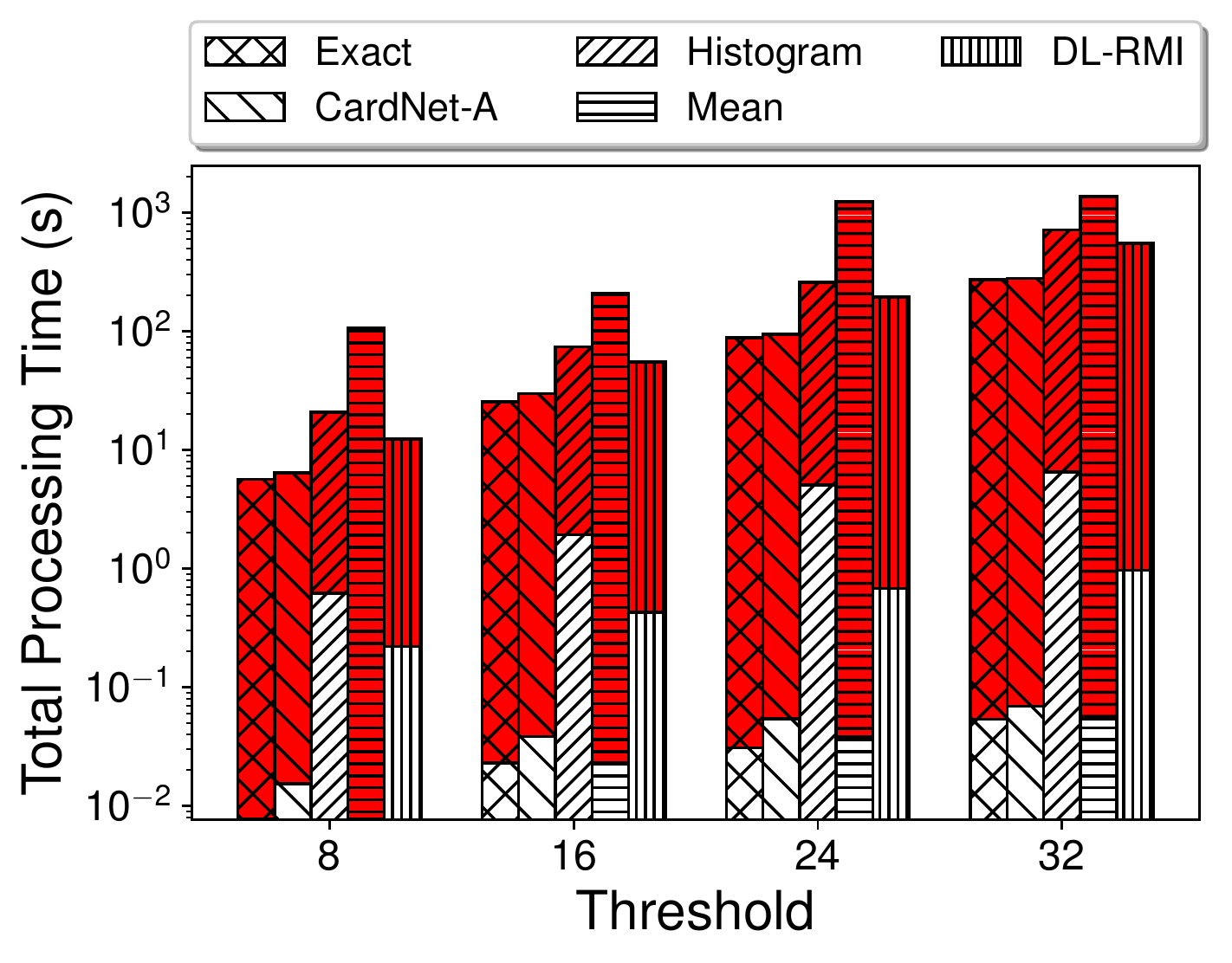}
    \label{fig:exp-youtube-optimizer-time-threshold}
  }
  \subfigure[Time, \fasttext]{
    \includegraphics[width=0.46\linewidth]{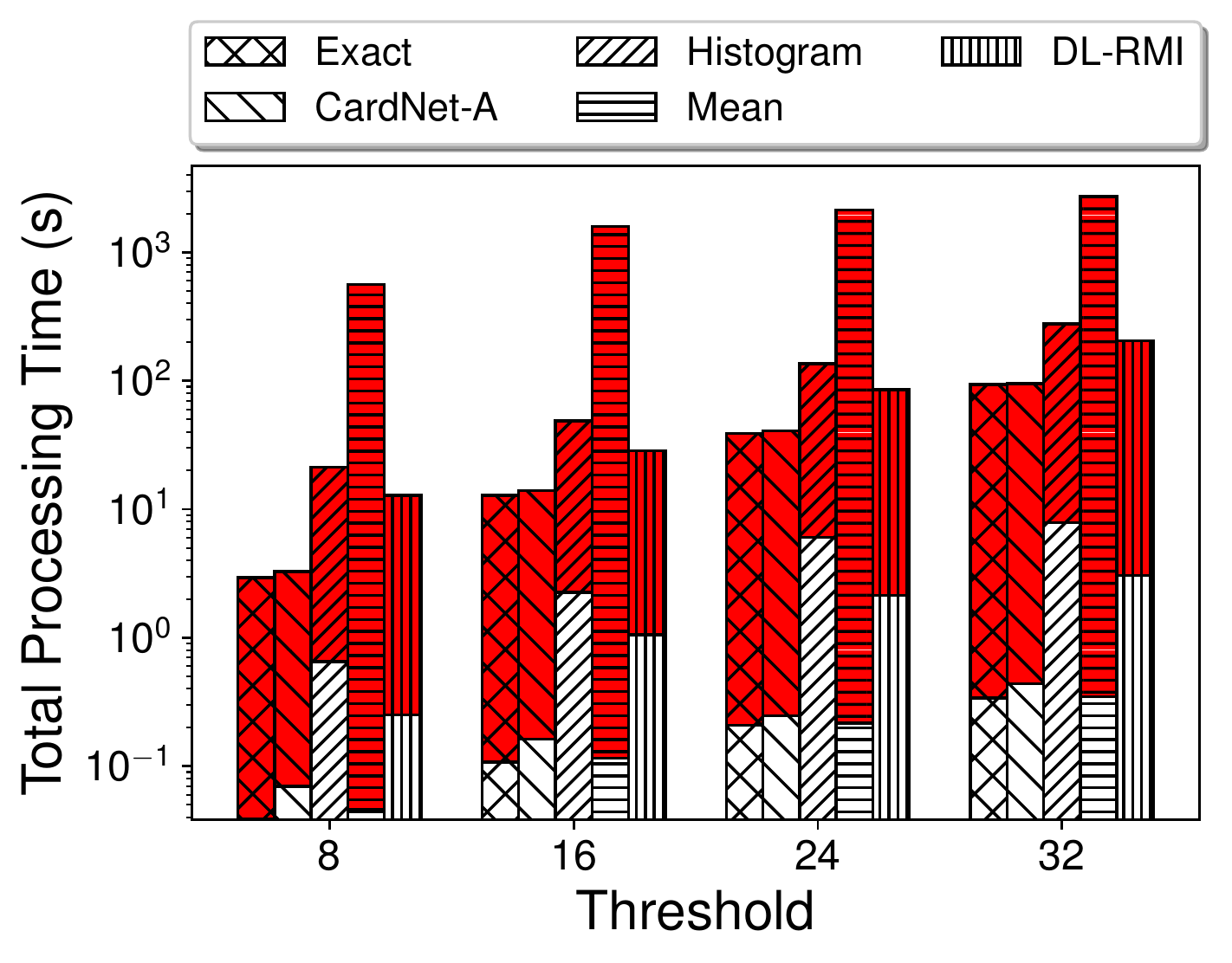}
    \label{fig:exp-fasttext-optimizer-time-threshold}
  }
  \subfigure[Time, \mnist]{
    \includegraphics[width=0.46\linewidth]{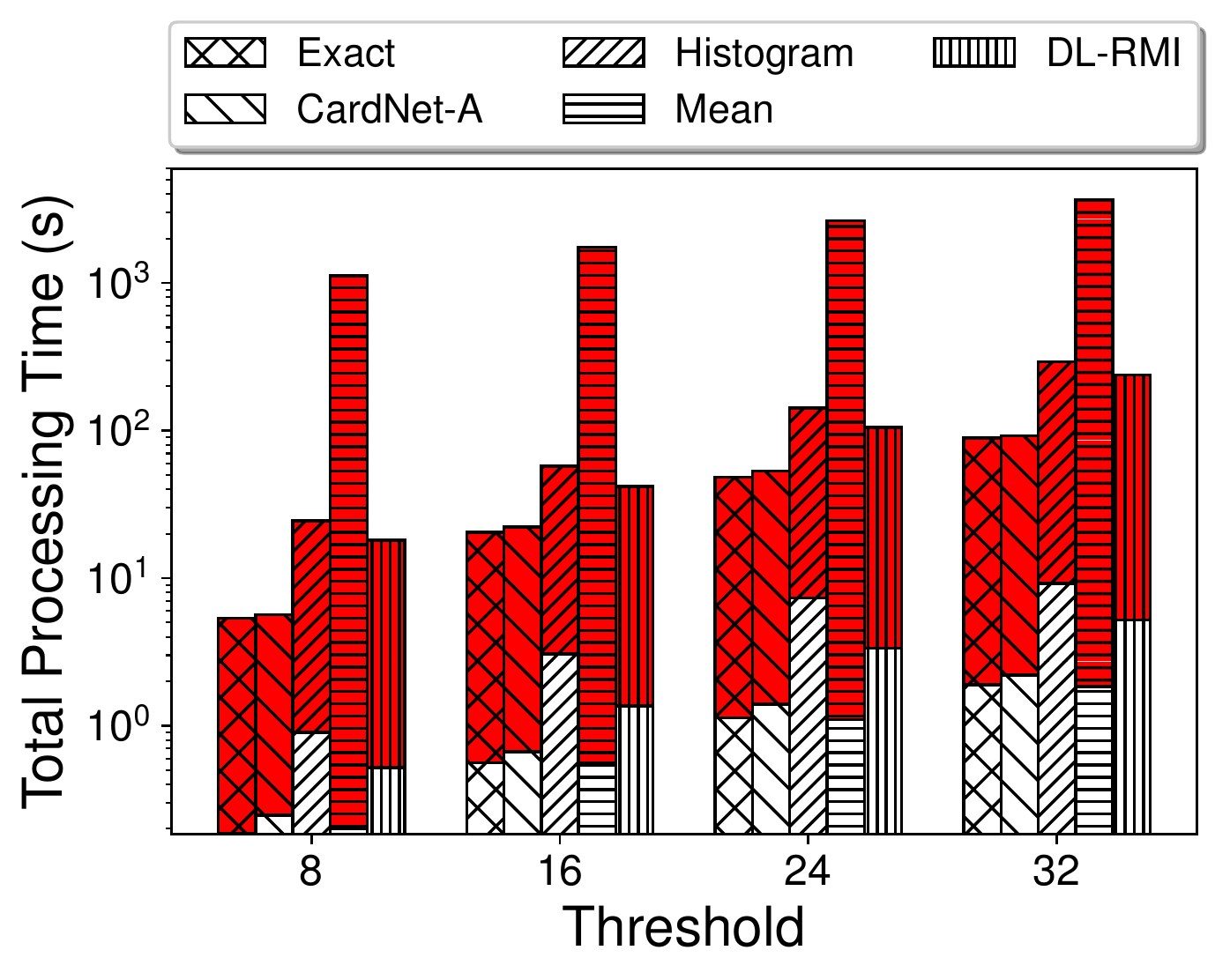}
    \label{fig:exp-mnist-optimizer-time-threshold}
  }  
  \caption{\revise{Hamming distance query -- query processing time.}}
  \label{fig:optimizer-time-threshold}
\end{figure}

\begin{figure} [t]
  \centering
  \subfigure[Time, \pubchem]{
    \includegraphics[width=0.46\linewidth]{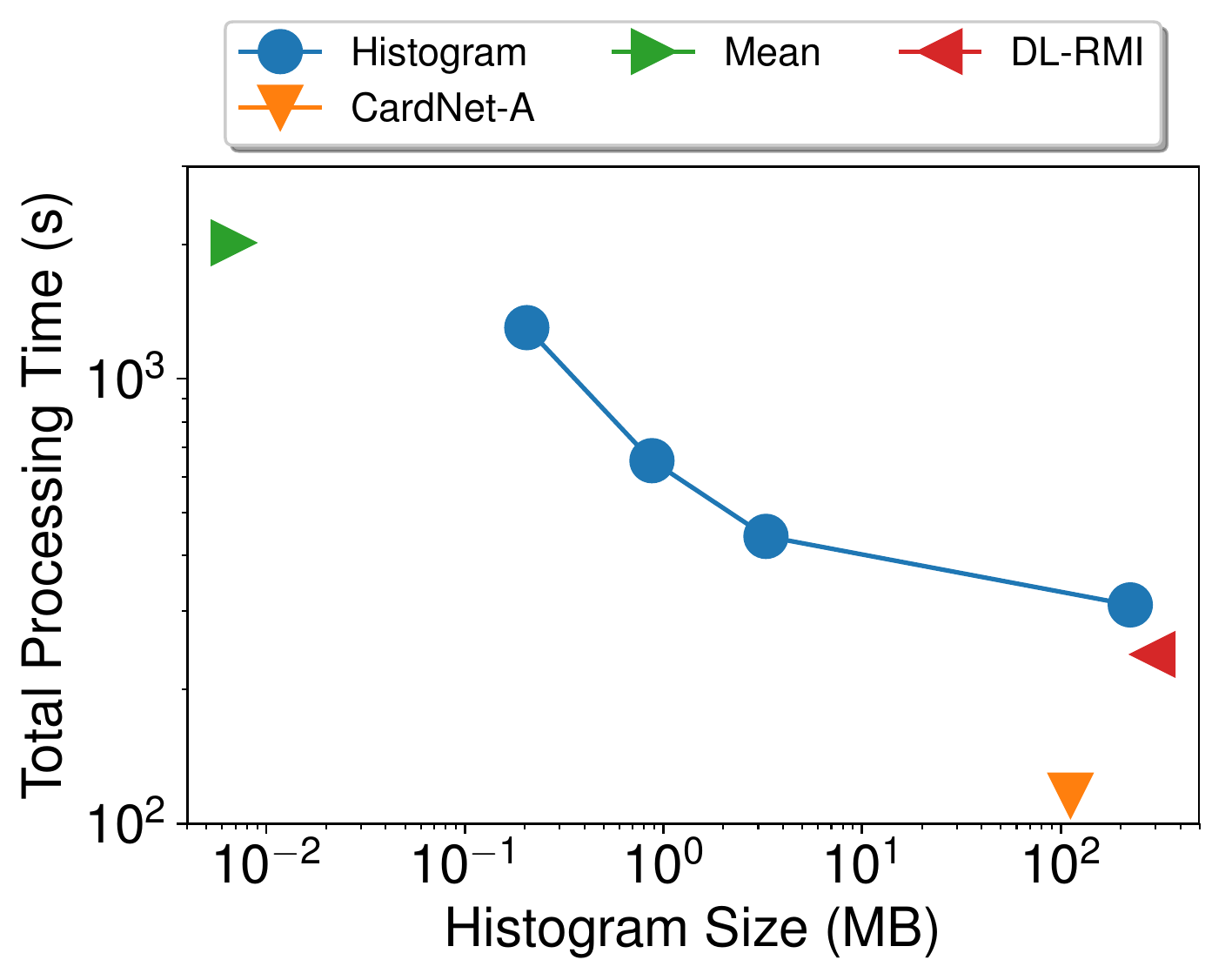}
    \label{fig:exp-pubchem-optimizer-time-size}
  }  
  \subfigure[Time, \video]{
    \includegraphics[width=0.46\linewidth]{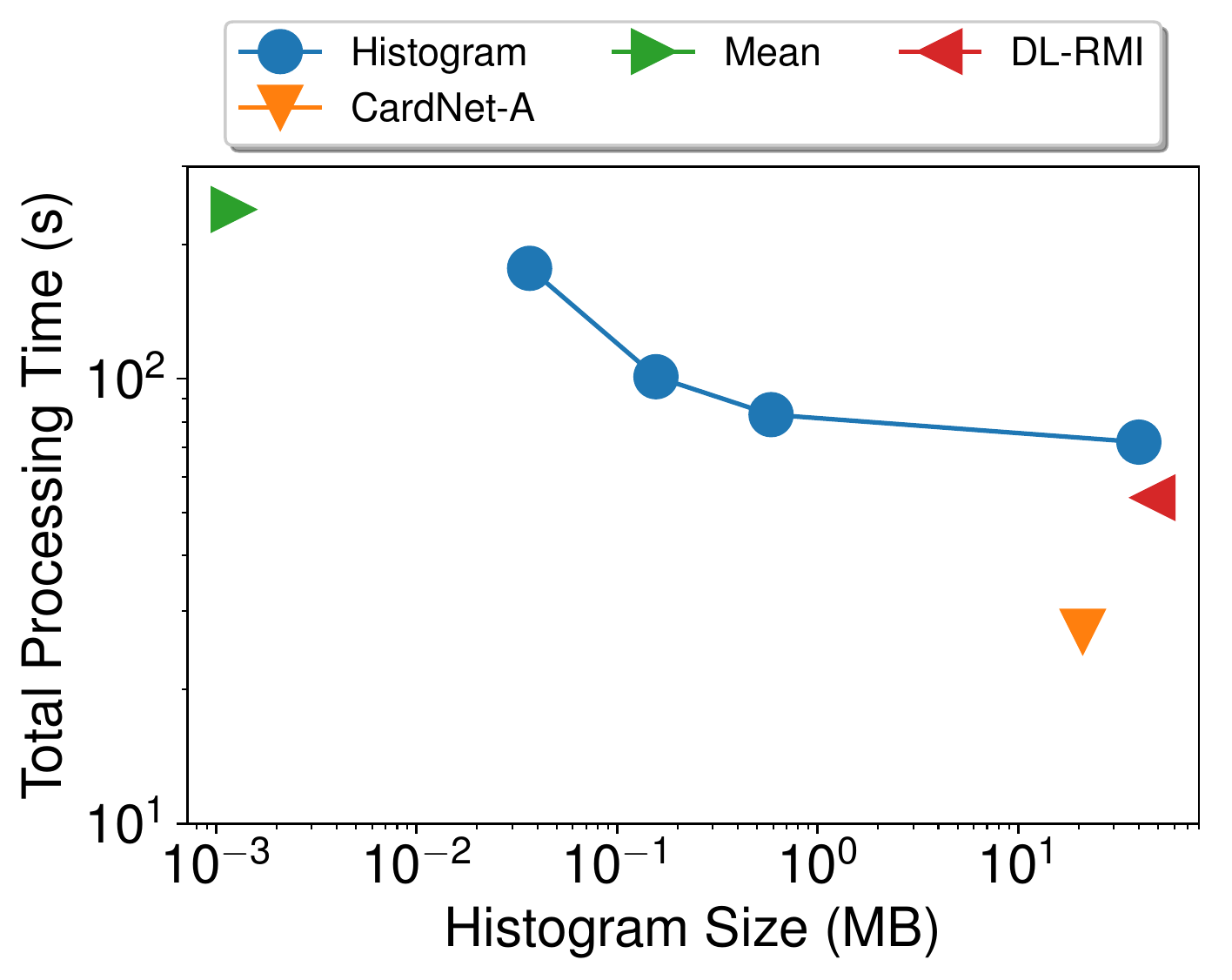}
    \label{fig:exp-youtube-optimizer-time-size}
  }
  \subfigure[Time, \fasttext]{
    \includegraphics[width=0.46\linewidth]{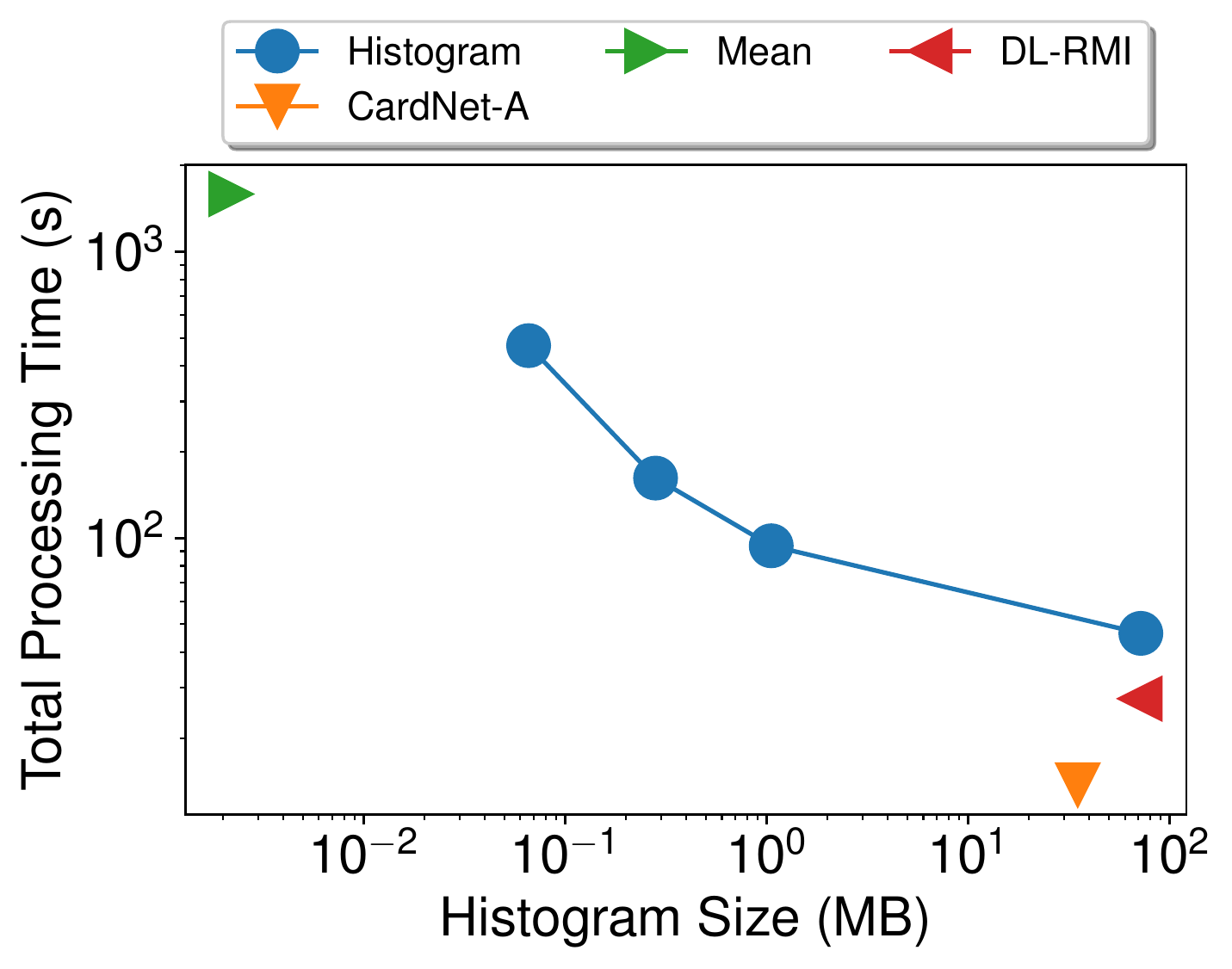}
    \label{fig:exp-fasttext-optimizer-time-size}
  }
  \subfigure[Time, \mnist]{
    \includegraphics[width=0.46\linewidth]{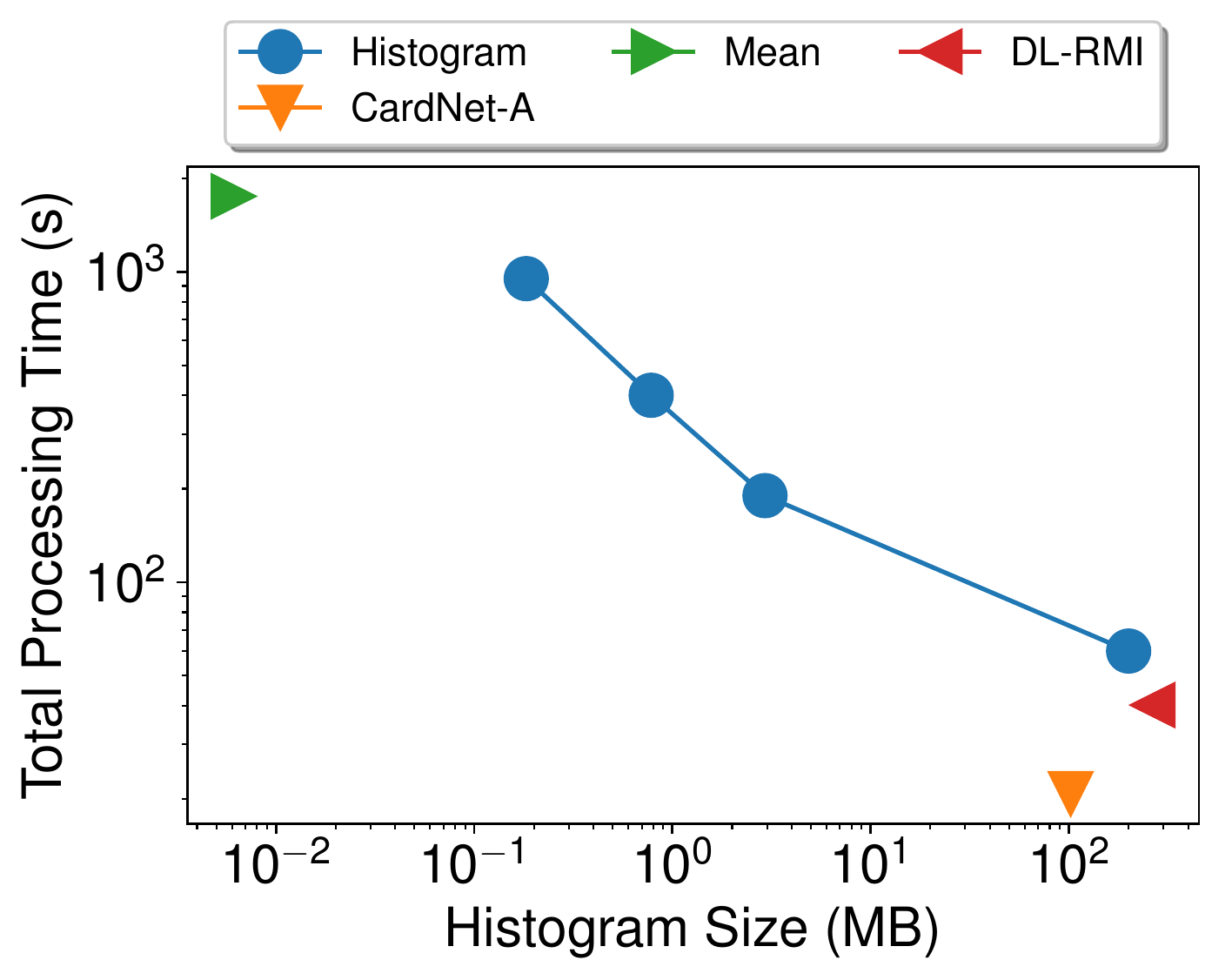}
    \label{fig:exp-mnist-optimizer-time-size}
  }  
  \caption{\revise{Hamming distance query -- varying model size.}}
  \label{fig:optimizer-time-size}
\end{figure}

\subsubsection{Hamming Distance Query}
We also consider a case study of the \gph algorithm~\cite{qin2018gph}, 
which processes Hamming distance queries 
over high dimensional vectors through a query optimizer. To cope with the high dimensionality, 
the algorithm answers a query $q$ by dividing it into $m$ non-overlapping parts and 
allocating a threshold (with dynamic programming) to each part using the pigeonhole principle. 
Each part itself is a 
Hamming distance selection query that can be answered by bit enumeration and index lookup. 
The union of the answers of the $m$ parts are the candidates of $q$. To allocate thresholds 
and hence to achieve small query processing cost, a query optimizer is used to minimize the 
sum of estimated cardinalities of the $m$ parts. 
We compare \modeltwo with the following options: 
\begin{inparaenum} [(1)]
  \item \hist, the histogram estimator in \cite{qin2018gph}; 
  \item \hierdnnexp; 
  \item \mean, an naive estimator that returns the same cardinality for a given threshold 
  (for each threshold, we offline generate 10,000 random queries and take the mean); and 
  \item \exact, an oracle that instantly returns the exact cardinality.
\end{inparaenum}
We use four datasets. The statistics is given in Table~\ref{tab:dataset-optimizer-gph}, 
where $\ell$ denotes the dimensionality. 
The records are converted to binary vectors as per the process in the table. 
We set each part to 32 bits (the last part is smaller if not divisible).



Figure~\ref{fig:optimizer-time-threshold} reports the processing time of 1,000 queries by varying
thresholds from 8 to 32 on the four datasets. The time is broken down to threshold allocation 
time (in white, which contains cardinality estimation) and postprocessing time (in red). 
The performance of \modeltwo is very close to \exact and faster than \hist by 1.6 to 4.9 times 
speedup. \hierdnnexp is slightly faster than \hist. 
\mean is much slower (typically one order of magnitude) than other 
methods, suggesting that cardinality estimation is important for this application. The threshold 
allocation (including cardinality estimation) spends less time than the subsequent query 
processing. \modeltwo also performs better than \hist in threshold allocation. This is because 
\begin{inparaenum} [(1)]
  \item \modeltwo itself is faster in estimation, and 
  \item the more accuracy makes the dynamic programming-based allocation terminate earlier. 
\end{inparaenum}

Next we fix the threshold at 50\% of the maximum threshold in Figure~\ref{fig:optimizer-time-threshold} 
and vary the size of the histogram. 
Figure~\ref{fig:optimizer-time-size} reports the average query processing time. The positions of 
other methods (except \exact) are also marked in the figure. As expected, the query processing time reduces 
when using larger histograms. However, the speed is still 1.6 to 2.6 times slower than \modeltwo 
even if the size of the histogram exceeds the model size of \modeltwo (see the rightmost point of 
\hist). This result showcases the superiority of our model compared to the traditional database 
method in an application of cardinality estimation of similarity selection. 
}

\fullversion{
}


\section{Conclusion} \label{sec:concl}
We investigated utilizing deep learning for cardinality estimation of similarity 
selection. Observing the challenges of this problem and the advantages of using 
deep learning, we designed a method composed of two components. The feature 
extraction component transforms original data and threshold to Hamming space, 
hence to support any data types and distance functions. The regression component 
estimates the cardinality in the Hamming space based on a deep learning model. We 
exploited the incremental property of cardinality to output monotonic results and 
devised a set of encoder and decoders that estimates the cardinality for each 
distance value. We developed a training strategy tailored to our model and 
proposed optimization techniques to speed up estimation. We discussed incremental 
learning for updates. The experimental results demonstrated the accuracy, 
efficiency, and generalizability of the proposed method as well as the 
effectiveness of integrating our method to a query optimizer. 

\section*{Acknowledgments}
This work was supported by JSPS 16H01722, 17H06099, 18H04093, and 19K11979, NSFC 61702409, 
CCF DBIR2019001A, NKRDP of China 2018YFB1003201, ARC DE190100663, DP170103710, and 
DP180103411, and D2D CRC DC25002 and DC25003. The Titan V was donated by Nvidia. We 
thank Rui Zhang (the University of Melbourne) for his precious comments. 

\balance
\bibliographystyle{abbrv}
\bibliography{citation}


\end{document}